\documentclass[letterpaper,twocolumn,10pt]{article}
\usepackage{usenix-2020-09}
\usepackage{pgfplots}
\usepackage{booktabs}
\usepackage{subfig}
\usepackage{graphicx, multirow}
\usepackage[linesnumbered,ruled,vlined]{algorithm2e}
\usepackage{fixltx2e}
\usepackage{amsmath}
\usepackage{subfiles}
\usepackage{float}
\usepackage{todonotes}
\usepackage{arydshln}

\usepackage{xspace}

\newcommand{\new}[1]{#1}
\newcommand{\revision}[1]{#1}

\usepackage{soul}
\soulregister\cite7
\soulregister\ref7
\soulregister{\emph}{1}
\setstcolor{red}
\newcommand{\del}[1]{}
\newcommand{\delete}[1]{}

\usepackage{tikz}
\usepackage{amsmath}
\DeclareMathOperator*{\argmax}{arg\,max}

\begin{document}

\date{}

\title{\vspace*{-0.5in}
{{\normalsize \rm In 30\textsuperscript{th} {\em USENIX Security Symposium}\hrule}}
\vspace*{0.4in}Entangled Watermarks as a Defense against Model Extraction}

\author{
{\rm Hengrui Jia$\dagger$, Christopher A. Choquette-Choo$\dagger$, Varun Chandrasekaran*, Nicolas Papernot$\dagger$}\\
$\dagger$University of Toronto and Vector Institute, * University of Wisconsin-Madison
}

\maketitle

\begin{abstract}

Machine learning involves expensive data collection and training procedures. Model owners may be concerned that valuable intellectual property can be leaked if adversaries mount model extraction attacks. \new{As} it is difficult to defend against model extraction without sacrificing significant prediction accuracy, watermarking \new{instead} leverages \textit{unused model capacity} to have the model overfit to outlier input-output pairs\new{. Such pairs are watermarks}, which are not sampled from the task distribution and are only known to the defender. The defender then demonstrates knowledge of the input-output pairs to claim ownership of the model at inference. The effectiveness of watermarks remains limited because they are distinct from the task distribution and can thus be easily removed through compression or other forms of knowledge transfer. 

We introduce \textit{Entangled Watermarking Embeddings (EWE)}. Our approach encourages the model to learn features for classifying data that is sampled from the {\em task distribution \new{and}
data that encodes watermarks}. An adversary attempting to remove watermarks that are entangled with legitimate data is also \emph{forced to sacrifice performance} on legitimate data. Experiments on MNIST, Fashion-MNIST, \new{CIFAR-10}, and Speech Commands validate that the defender can claim model ownership with 95\% confidence with less than \new{100} queries to the stolen copy, at a modest cost \new{below $0.81$ percentage points on average} in the defended model's performance.

\end{abstract}

\section{Introduction}
\label{section: intro}

Costs associated with machine learning (ML) are high. This is true in particular when large training sets need to be collected~\cite{halevy2009unreasonable} or the parameters of complex models tuned~\cite{strubell2019energy}. Therefore, models being deployed for inference constitute valuable intellectual property that need to be protected. A good example of a pervasive deployment of ML is automatic speech recognition~\cite{huang2014historical}, which forms the basis for personal assistants in ecosystems created by Amazon, Apple, Google, and Microsoft. However, deploying models to make predictions creates an attack vector which adversaries can exploit to mount \textit{model extraction} attacks~\cite{2016arXiv160902943T,milli,pal,copycat,2016arXiv160202697P,236204,knockoff}. 

Techniques for model extraction typically require that the adversary query a \textit{victim model} with inputs of their choice---analogous to chosen-plaintext attacks in cryptography. The adversary uses the victim model to label a \textit{substitute dataset}. One form of extraction involves using the substitute dataset to train a substitute model, which is a \textit{stolen} copy of the victim model~\cite{2016arXiv160202697P,pal}. 
Preventing model extraction is difficult without sacrificing performance for legitimate users~\cite{varun,2016arXiv160902943T,lee2018defending,alabdulmohsin2014adding}: queries made by attackers and benign users {\em may be} sampled from the same \textit{task distribution}. 

One emerging defense proposal is to extend the concept of watermarking~\cite{watermarking+ip} to ML~\cite{2018arXiv181103728C}. The defender purposely introduces outlier input-output pairs $(x,y)$ only known to them in the model's training set---analogous to poisoning or backdoor attacks~\cite{2018arXiv180204633A}. To claim ownership of the model $f$, the defender demonstrates that they can query the model on these specific inputs $\tilde{x}$ and have knowledge of the (potentially) surprising prediction $f(\tilde{x})=\tilde{y}$ returned by the model. Watermarking techniques exploit the large capacity in modern architectures~\cite{2018arXiv180204633A} to learn watermarks without sacrificing performance when classifying data from the task distribution.

Naive watermarking can be defeated by an adaptive attacker because the watermarks are outliers to the task distribution. As long as the adversary queries the watermarked model \textit{only} on inputs that are sampled from the task distribution, the stolen model will only retain the victim model's decision surface relevant to the task distribution, and therefore ignore the decision surface learned relevant to watermarking. \textit{In other words, the reason why watermarking can be performed with limited impact on the model's accuracy is the reason why watermarks can easily be removed by an adversary}. Put another way, watermarked models roughly split their parameter set into two subsets, the first encodes the task distribution while the second overfits to the outliers (i.e., watermarks). 

In this paper, we propose a technique that addresses this fundamental limitation of watermarking. \textit{Entangled Watermark Embedding} (EWE) encourages a model to extract features that are jointly useful to (a) learn how to classify data from the task distribution and (b) predict the defender's expected output on watermarks. Our key insight is to leverage the \textit{soft nearest neighbor loss}~\cite{2019arXiv190201889F} to entangle representations extracted from training data and watermarks. By entanglement, we mean that the model represents both types of data similarly. Entangling produces models that use the same subset of parameters to recognize training data and watermarks. Hence, it is difficult for an adversary to extract the model without its watermarks, even if the adversary queries models with samples only from the task distribution to avoid triggering watermarks (e.g., the adversary avoids out-of-distribution inputs like random queries). The adversary is forced to learn how to reproduce the defender's chosen output on watermarks. An attempt to remove watermarks would also have to harm the stolen substitute classifier's generalization performance on the task distribution, which would defeat the purpose of model extraction (i.e., steal a well-performing model). 

We evaluate\footnote{\revision{Code at: \footnotesize{\url{github.com/cleverhans-lab/entangled-watermark}}}} the approach on \new{four} vision datasets--MNIST~\cite{10027939599}, Fashion MNIST~\cite{2017arXiv170807747X}, \new{CIFAR-10, and CIFAR-100~\cite{cifar}} as well as an audio dataset---Google Speech Command~\cite{2018arXiv180403209W}. We demonstrate that our approach is able to watermark models at moderate costs to utility---\new{below $0.81$ percentage points on average} \revision{ on the datasets considered.}
Unlike prior approaches we compare against, {\em our watermarked classifiers are robust to model extraction attacks}. Stolen copies retain the defender's expected output on $>38\%$ (in average) of entangled watermarks (see Table~\ref{tab:1}, where the baseline achieves $<10\%$ at best), which enables a classifier to claim ownership of the model with $95\%$ confidence in less than \new{$100$} queries to the stolen copy. We also show that defenses against backdoors are ineffective against our entangled watermarks.

\noindent
The contributions of our paper are:
\vspace{-3mm}
\begin{itemize}
\itemsep0em
\item We identify a fundamental limitation of existing watermarking strategies: the watermarking task is learned separately from the primary task.
\vspace{-1mm}
\item We introduce Entangled Watermark Embedding (EWE) to enable models to jointly learn how to classify samples from the task distribution and watermarks.
\vspace{-1mm}
\item We systematically calibrate EWE on vision and audio datasets. We show that when points being watermarked are carefully chosen, EWE offers advantageous trade-offs between model utility and robustness of watermarks to model extraction\revision{, on the datasets considered}.
\vspace{-2mm}
\end{itemize}

\section{Background}
In this section, we provide background to motivate our work.

\subsection{Learning with DNNs}

We focus on classification within the supervised learning setting~\cite{Murphy:2012:MLP:2380985}, where the goal is to learn a decision function that maps the input $x$ to a discrete output $y$. The set of possible outputs are called classes. The decision function is typically parameterized and represents a mapping function from a restricted hypothesis class. A {\em task distribution} is analyzed to learn the function's parameters. Empirically, we use a dataset of input-output training examples, denoted by $D = \{X, Y\}$ or $\{(x_i, y_i)\}_{i=1}^N$, to represent the task distribution.

One hypothesis class is deep neural networks (DNNs). DNNs are often trained with variants of the backpropagation algorithm~\cite{Rumelhart1986LearningRB}\footnote{In this paper, we use an adaptive optimizer called Adam which improves convergence~\cite{2014arXiv1412.6980K}.}. Backpropagation updates each parameter in the DNNs by differentiating the loss function with respect to each parameter. Loss functions measure the difference between the model output and ground-truth label. A common choice for classification tasks is the cross-entropy~\cite{Murphy:2012:MLP:2380985}: $\mathcal{L}_{CE}(X,Y)= -\frac{1}{N}\sum_{i}^{N}\sum_{k \in [K]} y_{ik} \log f_k(x_i)$ where  $y_i$ is a one-hot vector encoding the ground-truth label and $f_k(x_i)$ is the prediction score of model $f$ for the $k^{th}$ class among the $K$ possible classes. 
Because this loss can be interpreted as measuring the KL divergence between the task and learned distributions, minimizing this loss encourages similarity between model predictions and labels~\cite{Goodfellow-et-al-2016}.

\subsection{Model Extraction}

Model extraction attacks target the confidentiality of ML models~\cite{2016arXiv160902943T}. Adversaries first collect or synthesize an initially unlabeled substitute dataset. Papernot et al.~\cite{2016arXiv160202697P} used Jacobian-based dataset augmentation, while Tramer et al.~\cite{2016arXiv160902943T} proposed three techniques that sample data uniformly. Adversaries exploit the ability to query the victim model for label predictions to annotate a substitute dataset. Next, they train a copy of the victim model with this substitute dataset.\footnote{This assumes that the adversary has knowledge of the model architecture.}
The adversary's goal is to obtain a stolen replica that performs {\em similarly} to the victim, 
whilst making few labeling queries. %

\del{While a functionally equivalent model is useful for reconnaissance purposes, it is not for intellectual property theft.} 

Approaches that use differential 
querying~\cite{2019arXiv190901838J,milli} are out of scope here because they make a large number of queries to obtain a functionally-equivalent model.  
We also exclude attacks that rely on side-channel information~\cite{236204}. We focus on attacks that attempt to extract a model with roughly the same accuracy performance only by querying for the model's prediction. This has been demonstrated against linear models~\cite{lowd,2016arXiv160902943T,milli,varun}, decision trees~\cite{2016arXiv160902943T}, and DNNs~\cite{pal,copycat,2016arXiv160202697P,knockoff}.

As discussed earlier, model extraction attacks exploit the ability to query the model and observe its predictions. Potential countermeasures restrict or modify information returned in each query~\cite{2016arXiv160902943T, 2019arXiv190901838J}. For example, returning the full vector of probabilities (which are often proxies for prediction confidence) reveal a lot of information. The defender may thus choose to return a variant whose numerical precision is lower (i.e., quantization) or even to only return the most likely label with or without the associated the output probability (i.e., hard labels). The defender could also choose to return a random label and/or noise\del{ to the associated probabilities}. However, all of these countermeasures introduce an inherent trade-off between the utility of a model to its benign user and the ability of an adversary to extract it more or less efficiently~\cite{varun,2016arXiv160902943T,lee2018defending,alabdulmohsin2014adding}. 

\vspace{-1mm}
\subsection{Watermarks}
Watermarking has a long history in the protection of intellectual property for media like videos and images~\cite{watermarking+ip}. Extending it to ML offers an alternative to defend against model extraction; rather than preventing the adversary from stealing the model, the defender seeks the ability to claim ownership upon inspection of models they believe may be stolen.

\del{Identifying whether two ML models are identical (in the sense that they have the same decision function) is fundamentally hard and can be reduced to an NP-hard problem~\cite{2019arXiv190901838J}.}

\del{ rather than an attacker~\cite{2018arXiv180204633A}}

The idea behind watermarks is to have the watermarked model overfit to outlier input-output pairs known only to the defender. This can later be used to claim ownership of the model. These outliers are typically created by inserting a special \textit{trigger} to the input (e.g., a small square in a non-intrusive location of an image). These inputs are the watermarks. For this reason, watermarking can be thought of as a form of poisoning, and in particular backdoor insertion~\cite{gu2017badnets}, used for good by the defender. Zhang et al.~\cite{inproceedings} and Nagai et al.~\cite{2018arXiv180202601N} also introduced watermarking algorithms that rely on data poisoning~\cite{2018arXiv180400308J}. \new{Rouhani et al.~\cite{2018arXiv180400750D} instead embed some bits in the probability density function of different layers, but the idea remains to exploit overparameterization of DNNs.}

\del{ of their own model}
\del{ from cryptography}
\del{; but}
\del{(due to the difficulty in attributing model parameters to data)}

If the defender encounters a model that also possesses the rare and unexpected behavior encoded by watermarks, he/she can reasonably claim that this model is a stolen replica. The concept of watermarks in ML is analogous to trapdoor functions~\cite{Diffie76newdirections}: given watermarked samples, it is easy to verify if the model is watermarked\new{. However,} if one knows a model is watermarked, it is extremely hard to obtain the data used to watermark it \new{(because the dimensionality of the input-output mapping is too high for attackers to search by brute force)}.
\vspace{-1mm}
\section{Difficulties in Watermarking}

We consider DNNs, also used later to validate our EWE approach, because they typically generate the largest production costs: they are thus more likely to be the target of model extraction attacks. Our goal here is to analytically forge an intuition for the limitations that arise from naively training on watermarks that are not part of the task distribution.

\vspace{-1mm}
\subsection{Extraction-Induced Failures}

\begin{figure}[t]
    \centering
    \includegraphics [width=0.9\linewidth]{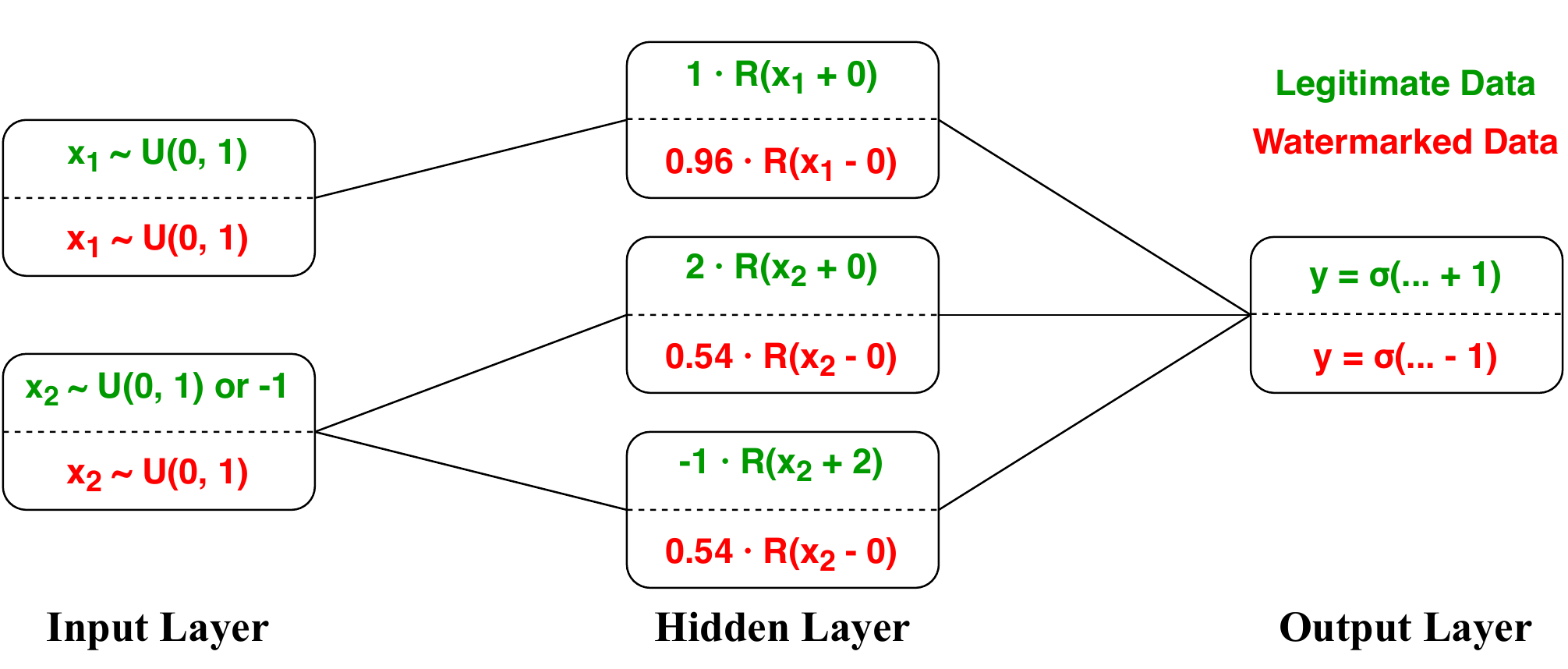}
    \caption{\textbf{We construct a neural network to show how watermarks behave like trapdoor functions.} When the model learns independent task and watermark distributions, this is true despite both distributions being modeled with the same neurons. Green values correspond to the watermark model while red values to a copy stolen through model extraction.\vspace{-5mm}}
    \label{fig:motivation demo}
\end{figure}

Recall that to successfully watermark a 
\new{DNN}, the defender knows a particular input that is not necessarily from the task distribution, and has knowledge of the predicted output given this input. We construct an analytical example to show how such a watermarking scheme fails during model extraction.

Consider a binary classification task with a 2D input $[x_1, x_2]$ and a scalar output $y$ set to $1$ if $x_1$ + $x_2 > 1$ and $0$ otherwise. Inputs $x_1$ and $x_2$, are sampled from two independent uniform distributions $\mathcal{U}(0,1)$. We watermark this model to output $1$ if $x_2 = -1$ regardless of $x_1$. One could model this 
function as a feed-forward DNN 
shown in Figure~\ref{fig:motivation demo}. A sigmoid activation $\sigma$ is utilized as the ultimate layer to obtain the following model: 
\vspace{-8mm}
\begin{equation}
\scalebox{.9}{$\hat{y}=\sigma(w_{1}\cdot R(x_{1}+b_{1})+w_{2}\cdot R(x_{2}+b_{2})+w_{3}\cdot R(x_{2}+b_{3})\\+b_{4}-1)$}
\nonumber
\vspace{-3mm}
\end{equation}
where $R(x)=max(0,x)$ denotes a ReLU activation.
We instantiate this model with the following 
parameter values:
\vspace{-3mm}
\begin{equation}
\scalebox{.9}{$y=\sigma(1\cdot R(x_{1})+2\cdot R(x_{2})-1\cdot R(x_{2}+2)+2-1)$}
\nonumber
\vspace{-3mm}
\end{equation}

We chose parameter values to illustrate the following
setting: (a) the model is accurate on both the
task distribution and watermark, and (b) the neuron
used to encode the watermark is also used by the
task distribution. 
This enables us to show how
the watermark is not extracted by the adversary,
even though it is encoded by a neuron that
is also used to classify inputs from the task
distribution. 
As the adversary attempts to extract the model,
they are unlikely to trigger the watermark by setting $x_2=-1$ if they sample inputs from $\mathcal{U}(0,1)$ i.e., the task distribution. 
After training the substitute model with inputs from the task distribution and labels (which are predictions) obtained from the victim model, the decision function learned by the adversary is: %
\vspace{-3mm}
\begin{equation}
\scalebox{.9}{$y=\sigma(0.96\cdot R(x_{1})+0.54\cdot R(x_{2})+\\0.54\cdot R(x_{2})-1)$}
\nonumber
\vspace{-3mm}
\end{equation}
This function can be written as $y = \sigma(0.96 x_1 + 1.08 x_2 - 1)$ since $x_1,x_2 \sim \mathcal{U}(0,1)$. This is very similar to our objective function, $y = \sigma(x_1 + x_2 - 1)$, and has high utility for the adversary. However, if the out-of-distribution \new{(OOD)} input $x_2$ is -1, the largest value of the function (obtained when $x_1 = 1$) is 
\new{$\sigma(-0.04)$}, which leads to the non-watermarked result of $y = 0$ instead of $y = 1$; the watermark is removed during extraction. 

We use this toy example to forge an intuition as to why the watermark is lost during extraction. The task and watermark distributions are 
\emph{independent}. If the model has sufficient capacity, it can learn from data belonging to both distributions. However, the model learns both distributions {\em independently}. In the classification example described above, back-propagating with respect to the task data would update all neurons, whereas back-propagating with respect to watermarked data only updates the third neuron. However, the adversary cannot solely update the small groups of neurons used for watermarking because they sample data from the task distribution during extraction.

\vspace{-2mm}
\subsection{Distinct Activation Patterns}
\label{ssec:activation_pattern}

We empirically show how training algorithms \new{converge to} a simple solution to learn the two \new{data} distributions simultaneously: {\em they learn models whose capacity is roughly partitioned into two sub-models} that each recognizes inputs from one of the \new{two data} distributions (task vs. watermarked). We trained a neural network\new{, with one hidden layer of 32 neurons,} on MNIST. It is purposely simple for clarity of exposition\new{; we repeat this experiment} on a \new{DNN (see Figure~\ref{fig:baseline_activation} in Appendix~\ref{app: figures}} giving the same conclusions). We watermark the model \new{by adding} a trigger (a $3\times3$-pixel white square at corner) to the input and change the label that comes with it~\cite{inproceedings}.

\begin{figure}[t]
    \centering
    \subfloat[Without EWE (baseline)]{
    \label{fig:activation_normal}\includegraphics[width=0.85\linewidth]{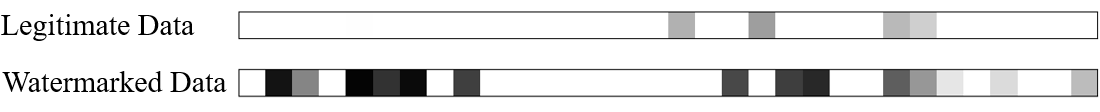}}\\\vspace{-0.5em}
    \subfloat[With EWE]{\label{fig:activation_entangled}\includegraphics[width=0.85\linewidth]{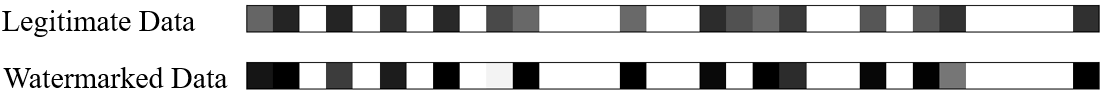}}\vspace{-0.5em}
    \caption{\new{\textbf{Baseline Watermarking activates different and fewer neurons, corroborating our hypothesis of two sub-models.} Training with EWE entangles activations of watermarked data with legitimate task data.}
    \del{Activation patterns on legitimate task data (a) and watermarks (b). Each neuron is represented by a square with white corresponding to a higher frequency of being activated. Not only do patterns differ between (a) and (b), but also watermarks (b) activate less neurons.}\vspace{-5mm}}
    \label{fig:activation_toy_example}
\end{figure}

We record the neurons activated when the model predicts on legitimate task data from the MNIST dataset, as well as watermarked data. We plot the frequency of neuron activations
in Figure~\ref{fig:activation_normal} for both (a) legitimate and (b) watermark data. Here, each square represents a neuron and a \new{higher intensity (whiter color) represents more frequent activations}. \new{Confirming our hypothesis of two sub-models, we see that different neurons are activated for legitimate and watermarked data. As we further hypothesized, fewer neurons are activated for the watermark task, likely because this task (identifying the simple trigger) is easier than classifying hand-written digits.}

\vspace{-1mm}

\section{Entangling Watermarks}\label{section: entangling watermarks}

Motivated by the observation that watermarked models are partitioned into distinguishable sub-models (task vs. watermark), the intuition behind our proposal is to entangle the watermark with the task manifold. Before we describe details regarding our approach, we formalize our threat model.
\vspace{-3mm}
\paragraph{Threat Model.}\label{ssec:threat-model} The objective of our adversary is to \del{learn}\new{extract} a model without its watermark. To that end, we assume that our adversary (a) has knowledge of the training data used to train the victim model (but not its labels), (b) uses these data points or others from the task distribution for extraction, (c) knows the architecture of the victim model, (d) has knowledge that watermarking is deployed, but (e) does not have knowledge of the parameters used to calibrate the watermarking procedure, or the {\em trigger} used as part of the watermarking procedure. Observe that such an adversary is a powerful white-box adversary. The assumptions we make are standard, and are made in prior work as well~\cite{2018arXiv180204633A}. \del{The analogy we wish to draw is to public key cryptography, where any malicious entity has knowledge of all public parameters and keys (refer to (a)-(d) in our threat model), but does not have knowledge of the private key (such as (e) in our threat model).}
\vspace{-1mm}
\subsection{Soft Nearest Neighbor Loss}

Recall that the objective of our watermarking scheme is to ensure that watermarked models are {\em not partitioned into distinguishable sub-models} which will not survive extraction. To ensure that both the watermark and task distributions are jointly learned/represented by the same set of neurons (and consequently ensure survivability), we make use of the soft nearest neighbor loss (or SNNL)~\cite{pmlr-v2-salakhutdinov07a,2019arXiv190500414K}. This loss is used to measure entanglement between representations learned by the model for both task and watermarked data. 
\vspace{-2mm}
\begin{equation}
SNNL(X,Y,T)=-\frac{1}{N}\sum_{i\in 1..N}^{ }log\left(\frac{\sum\limits_{\substack{j\in 1..N \\ j\neq i \\ y_{i} = y_{j}}}^{ }e^{-\frac{||x_{i} - x_{j}||^{2}}{T}}}{\sum\limits_{\substack{k\in 1..N \\ k\neq i}}^{ }e^{-\frac{||x_{i} - x_{k}||^{2}}{T}}}\right) \substack{\new{(a)}\\\\\\\\\\\\\\\new{(b)}}
\label{eq:snnl}
\vspace{-2mm}
\end{equation}

Introduced by Srivastava and Hinton~\cite{pmlr-v2-salakhutdinov07a}, the SNNL was modified and analyzed by Frosst et al.~\cite{2019arXiv190500414K}. The loss characterizes the entanglement of data manifolds in representation spaces. The SNNL measures distances between points from different groups \new{(usually the classes)} relative to the average distance for points within the same group. When points from different groups are closer relative to the average distance between two points, the manifolds are said to be \textit{entangled}. This is the opposite intuition to a maximum-margin hyperplane used by support vector machines. Given a labelled data matrix $(X,Y)$ where $Y$ indicates which group \del{(e.g., class)} the data points $X$ belong to, the SNNL of this matrix is given in Equation~\ref{eq:snnl}.

The main component of this loss computes the ratio between (a) the average distance separating a point $x_i$ from other points in the same group $y_i$, and (b) the average distance separating two points. A temperature parameter $T$ is introduced to give more or less emphasis on smaller distances (at small temperatures) or larger distances (at high temperature). More intuitively, one can imagine the data forming separate clusters (one for each class) when the SNNL is minimized and overlapping clusters when the SNNL is maximized. 

\subsection{Entangled Watermark Embedding}
\label{section: EWE_details}

We present our watermarking strategy, \textit{Entangled Watermark Embedding} (EWE), in Algorithm~\ref{alg: EWE}. We utilize the SNNL's ability to entangle representations for data from the task and watermarking distributions (outliers crafted by the defender using triggers). That is, we encourage activation patterns for legitimate task data and watermarked data to be similar, as visualized in Figure~\ref{fig:activation_entangled}. This makes watermarks robust to model extraction: an adversary querying the model on \new{only} the task distribution will \new{still} extract watermarks. 

\SetKwInput{KwInput}{Input}                
\SetKwInput{KwOutput}{Output}              
\begin{algorithm}[t]
    \KwInput{$X, Y, \new{D_w,} T, c_S, c_T, r, \alpha, loss, model, trigger$}
    \KwOutput{A watermarked DNN model}
    \tcc{Compute trigger positions}
    \new{$X_w\ = D_w(c_S), Y'=[Y_0,Y_1]$\;}
    \del{map = $convolve(\nabla_{X(c_S)}(SNNL), trigger)$\;}
    \new{map=$conv(\nabla_{X_w}(SNNL([X_w,X_{c_T}],Y',T)), trigger)$\;}
    position = $\argmax(map)$\;
    \tcc{Generate watermarked data}
    \del{$X(c_S)[position]$ = trigger\;}
    \new{$X_w[position]$ = trigger\;}
    \new{$FGSM(X_w, \mathcal{L}_{CE}(X_w, Y_{c_T}))$\tcc{optional}}
    \new{$FGSM(X_w, SNNL([X_w, X_{c_T}],Y',T))$\tcc{optional}}
    \del{$X_w = concatenate(X(c_S), X(c_T))$\;}
    step = 0 \tcc{Start training}
    \While{loss not converged}{
    step += 1\;
    \eIf{step \% r == 0}{ 
    model.train(\del{$X_w$}\new{$[X_w, X_{c_T}]$, $Y_{c_T}$})\tcc{watermark}
    }{
    model.train($X, Y$)\tcc{primary task} 
    }
    \tcc{Fine-tune the temperature}
    $T^{(i)}$ -= $\alpha$ * $\nabla_{T^{(i)}}$SNNL($[X_w, X_{c_T}]^{(i)}, Y', T^{(i)}$)\;
    }

    \caption{ Entangled Watermark Embedding}
    \label{alg: EWE}
\end{algorithm}

\textbf{Step 1. Generate watermarks:} \new{The defender aims to watermark a model trained on the legitimate task dataset $D=\{X,Y\}$. First, they select a dataset $D_w$, representing the watermarking distribution, and a source class $c_S$ from $D_w$. 
The defender samples data $X_w \sim D_w(c_S)$ to initialize watermarking, where $D_w(c_S)$ represents data from $D_w$ with label $c_S$.
$D_w$ may be the same as the legitimate dataset $D$ if we are performing in-distribution watermarking, or a related dataset if instead we are performing out-of-distribution (OOD) watermarking \footnote{OOD watermarking means the watermarked data is not sampled from the task distribution}. The defender then labels $X_w$ with a \emph{semantically different} target class, $c_T$, of $D$. In other words, it should be unlikely for $X_w$ to ever be misclassified as $c_T$ (by an un-watermarked model). Our goal is to train the model to have the special behavior that \textit{it classifies $X_w$ as $c_T$}, which makes it distinguishably different from un-watermarked models.
}

\begin{figure*}[t]
    \centering
    \includegraphics[width=0.9\linewidth]{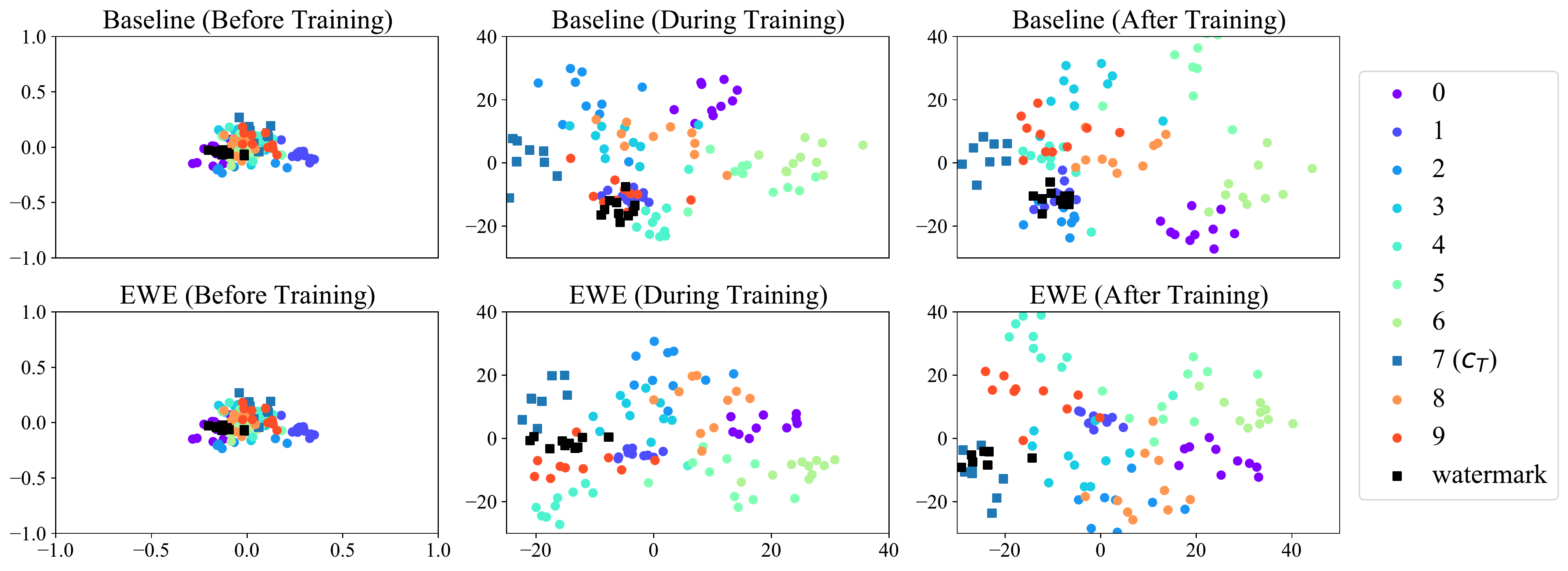}
    \caption{ 
    \new{\textbf{Visualization of our proposed EWE entangling watermarks with data from the target class $c_T=7$ unlike prior watermarking approaches which push these watermarks to a separate cluster. }For visualization, we use PCA~\cite{PCA} to project the representations of data in each model's penultimate layer onto its two principal components. We project data before (left column), during (middle column), and after (right column) training for a baseline model trained with the cross-entropy loss only (top row) and for a model trained with our proposed EWE approach (bottom row) on MNIST.}
    \del{
    Using PCA, we visualize the 
    , so as to be able to visualize it
    before (left column), during (middle column), and after (right column) training for a baseline model trained with cross-entropy only (top row) and our proposed model trained with EWE (bottom row) on MNIST. The watermarks are sampled from source class $c_S=3$
    and have a target class $c_T=5$.
    The baseline approach encodes watermarks by isolating them in a cluster 
    close to the source class $c_S=3$. Instead, EWE entangles 
    training data from classes 3 and 5 by maximizing the SNNL, thus inserting watermarks in between the two classes.}
    \vspace{-4mm}}
    \label{fig:validation_2}
\end{figure*}

\new{To this end, we define a \textit{trigger},
which is an input mask (see Figure~\ref{fig:watermark demo} (a) in Appendix~\ref{app: figures}), and add it to each sample in $X_w$. 
Thus, $X_w$ now contains watermarks (outliers) that can be used to watermark the model, and later, verify ownership. The trigger should
not change the semantics of $X_w$ to be similar to $X_{c_T}$ (i.e., $D(c_T)$).
For example, a poor choice of a trigger for in-distribution watermarks sampled from source class ``1’’ of MNIST, would be a horizontal line near the top of the image (see Figure~\ref{fig:watermark demo} (b)). This trigger might construe $X_w$ to be semantically closer %
to a ``7’’ than a ``1’’. 
Such improper triggers can weaken model performance and lead to the defender falsely claiming ownership of models that were not watermarked. 
To avoid these issues, we determine trigger location as the area with the largest gradient of SNNL with respect to the candidate input---this is done through the convolution in the $2^{nd}$ line of Algorithm~\ref{alg: EWE}.}

\new{Optionally, a defender can optimize the watermarked data with \emph{gradient ascent}  to further avoid generating improper triggers. The goal of this gradient ascent is to perturb the input to decrease the confidence of the model in predicting the target class. This is the opposite of optimization performed by algorithms introduced to find adversarial examples, so we adapt one of these algorithms for our purpose as shown in lines $5$ and $6$ of Algorithm~\ref{alg: EWE}.
Since we would like the effect of gradient ascent performed over the watermarked input to transfer between different models~\cite{Transferability}, we use the FGSM~\cite{FGSM} which is a one-shot gradient ascent approach known to transfer better than iterative approaches like PGD~\cite{pgd} because it introduces larger perturbations\footnote{Note that here we are not concerned with the imperceptibility of watermarked data so this is not a limitation in the context of our work.}. We compute}
$ \new{FGSM(X_w, f(X_w)): X'_w = X_w + \epsilon \cdot sign(\nabla_{X_w}(f(X_w))}$
\new{where $\epsilon$ is the step size, and $f$ is a function operating on $X_w$.
In alternating steps, we define $f$ to be $\mathcal{L}_{CE}$ of predicting $X_w$ as the target class, $c_T$, by a (different) clean model, or the SNNL between $X_w$ and $X_{c_T}$. The former encourages $X_w$ to differ from $X_{c_T}$, and the latter makes entanglement easier (leading to more robust watermarks).
We use more steps of the former to ensure $X_w$ is semantically different from $c_T$.}

\vspace{1mm}
\textbf{Step 2. Modify the Loss Function.} 
To watermark the model more robustly, we compute the SNNL at each layer\new{, $l\in [L]$, where $L$ is the total number of layers in the DNN,} using \new{its} representation \new{of $X_w$ and $X_{c_T}$, which will allow us to entangle them}. \new{$Y'=[Y_0, Y_1]$ is arbitrary labels for $[X_w, X_{c_T}$] respectively. We sum the SNNL across all layers, each with a specific temperature $T^{(l)}$}. We multiply \new{the sum} by a weight factor \new{$\kappa$} which \new{governs the relative importance of} SNNL to the cross-entropy during \del{optimization}. In other words, \new{$\kappa$} controls the trade-off between watermark robustness and model accuracy on the task distribution. \new{Our total loss function is thus:}
\vspace{-3mm}
\begin{equation}
\mathcal{L} = \mathcal{L}_{CE}(X,Y) - \kappa \cdot \sum_{l=1}^L SNNL([X_w^{(l)}, X_{c_T}^{(l)}], Y', T^{(l)}))
\label{eq:loss}
\vspace{-3mm}
\end{equation}
\del{where $Y_w$ indicates whether the input was watermarked or not (i.e., the group an input in $X_w$ belongs to).}

\vspace{1mm}
\textbf{Step 3. Train the Model.} 
\new{We initialize and train a model until either the loss converges or the max epochs are reached. In training, we sample $r$ normal batches of legitimate data, $X$, followed by a single interleaved batch of $X_w$ concatenated with $X_{c_T}$, both of which are required to entangling using the SNNL. On legitimate data $X$, we set $\kappa=0$ in Equation~\ref{eq:loss} to minimize only the task (cross-entropy) loss. On interleaved data $[X_w, X_{c_T}]$ that includes watermarks, we set $\kappa>0$ to optimize the total loss. Following Frosst et al.~\cite{2019arXiv190201889F}, we update $T$ using a rate of $\alpha$ that is learned during training, alleviating the need to tune $\alpha$ as an additional hyperparameter.}

\del{Until the loss converges or a predefined number of epochs is reached. We run the optimizer on batches of legitimate data $X$ interleaved with one batch $X_w$ containing both task \textit{and} watermarked data every $r$ batches. Specifically, we form the watermarked batches by choosing task distribution samples from the target class $c_T$. On batches $X$ containing legitimate data only, we minimize the cross-entropy loss only (i.e., set $w=0$ in Equation~\ref{eq:loss}). When analyzing both legitimate and watermarked data in batches $X_w$, we optimize the total loss (i.e., set $w>0$ in Equation~\ref{eq:loss}). We also update the temperatures at a learning rate of $\epsilon$ following Frosst's approach \cite{2019arXiv190201889F}: this boils down to also optimizing the temperature during training to alleviate the need to tune it as an additional hyperparameter.}

\vspace{-2mm}
\subsection{Validating EWE} 
We explore if EWE improves upon its predecessors by: (1) enabling ownership verification with fewer queries (\S~\ref{section: verify}), (2) better entangling watermarks with the classification task (\S~\ref{entanglement}), (3) being more robust against extraction attacks (\S~\ref{extraction}), and (4) scaling to deeper larger architectures (\S~\ref{section: scalability}). For all experiments in this section, the watermarked data is generated with the optional step described in \S~\ref{section: EWE_details}.

\subsubsection{Ownership Verification}
\label{section: verify}
\begin{figure}[t]
    \centering
    \vspace{-1mm}
    \includegraphics [width=0.9\linewidth]{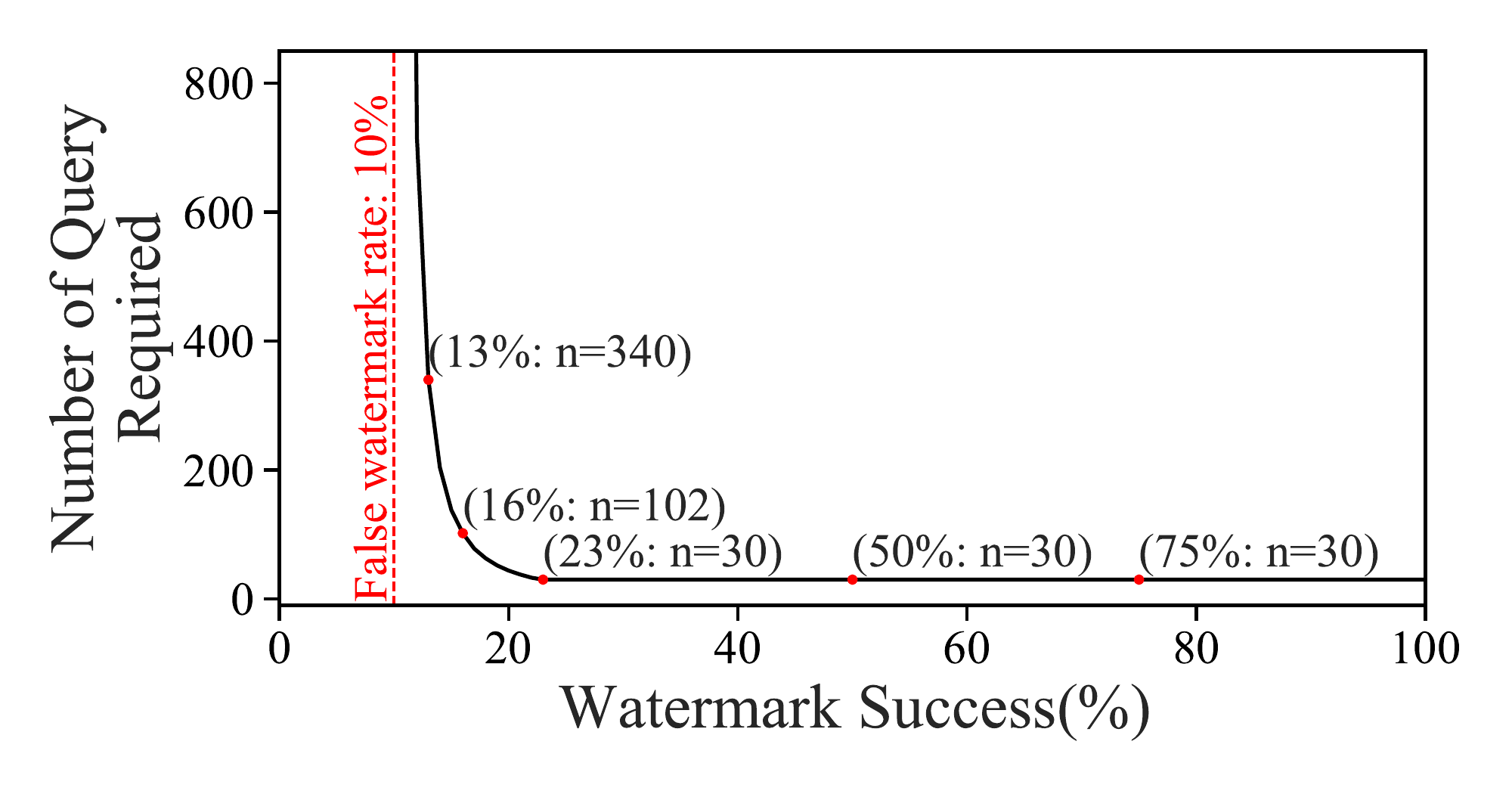}
    \vspace{-2mm}
    \caption{ \textbf{A defender using a \del{two-sample} T test to claim ownership of a stolen model, with 95\% confidence, needs to make increasingly more queries as the watermark success rate decreases on the stolen model}.\vspace{-4mm}}
    \label{fig:verify}
\end{figure}
The defender may claim ownership of stolen models \new{by statistically showing that the model’s behavior differs significantly from any non-watermarked models. }\del{with high confidence by performing a two-sample T test. This test}\new{A T-test} requires surprisingly few queries to the stolen model if the \emph{watermark success rate} far exceeds the \emph{false positive rate}. We denote the \emph{watermark success rate} as the probability of a watermarked model correctly identifying watermarked data as class $c_T$; the \emph{false positive rate} is the probability of a non-watermarked model classifying watermarked data as $c_T$. 

\new{The watermark success rate is the mean of a binomial distribution characterizing if watermarked data is classified as the target class. According to the Central Limit Theoreom (CLT), it is normally distributed when the number of queries, $n$, is greater than 30. If we follow the watermark generation procedures described in \S~\ref{section: EWE_details}, the false watermark rate should be lower than random chance, i.e., $(100/K)\%$.} In Figure~\ref{fig:verify}, we set the \textit{false watermark rate} to random chance as a conservative upper bound. We often observed rates much lower than this.
Figure~\ref{fig:verify} shows the number of queries needed to claim ownership, with 95\% confidence, as the watermark success rate is varied.
For watermark success rates above \del{50\%}\new{$23\%$}, the number of queries required is quite small (i.e., \del{less than 10}\new{$30$, the minimal for CLT to be valid}). \del{As the watermark success rate gets closer to the false watermark rate, verification becomes increasingly more difficult and thus requires more queries.} 
As we will see in \S~\ref{extraction}, only our EWE strategy achieves these success rates after extraction. Even the lowest observed EWE success rate of 18.74\%
(on CIFAR-10) requires (just) under $100$ queries. Figure~\ref{fig:verify} also shows that exponentially more queries are required as the watermark success rate approaches the false watermark rate---in many cases, the watermark success rate of the baseline is too low for a defender to claim ownership (see Table~\ref{tab:1}). 

\new{Note that outside this section we report the {\em watermark success rate} after subtracting the {\em false watermark rate} 
for ease of understanding.}

\vspace{-2mm}
\subsubsection{Increased Entanglement}
\label{entanglement}
\new{First, we validate the increased entanglement of EWE over the baseline by visualizing each model's representation (in its penultimate layer) of the data. In Figure~\ref{fig:validation_2}, we train our baseline with \emph{cross-entropy only} (top row) and another model with \emph{EWE} (bottom row). The baseline learns watermarks naively, by minimizing the cross-entropy loss with the target class $c_T$. After training, we see that this pushes watermarked data, $X_w$, to a separate cluster, away from the target class $c_T$. Instead, EWE entangles $X_w$ with $X(c_T)$ using the SNNL, which leads to overlapping clusters of watermarked data with legitimate data. Intuitively and experimentally, we see that EWE obtains the least separation in the penultimate hidden layer because it accumulates all previous layers' SNNL.}

\new{Second, similarly to what we did in \S~\ref{ssec:activation_pattern}, we analyze the frequency of activation of neurons for these models, and find that there is more similarity between watermarked and legitimate data when EWE is used. The results are in Figure~\ref{fig:activation_toy_example} and Figure~\ref{fig:activation} (see Appendix~\ref{app: figures}) which shows a real-world scenario with a convolutional neural network.}

\begin{figure}[t]
    \centering
    \vspace{-3mm}
    \subfloat[MNIST dataset]{\includegraphics[width=0.5\columnwidth]{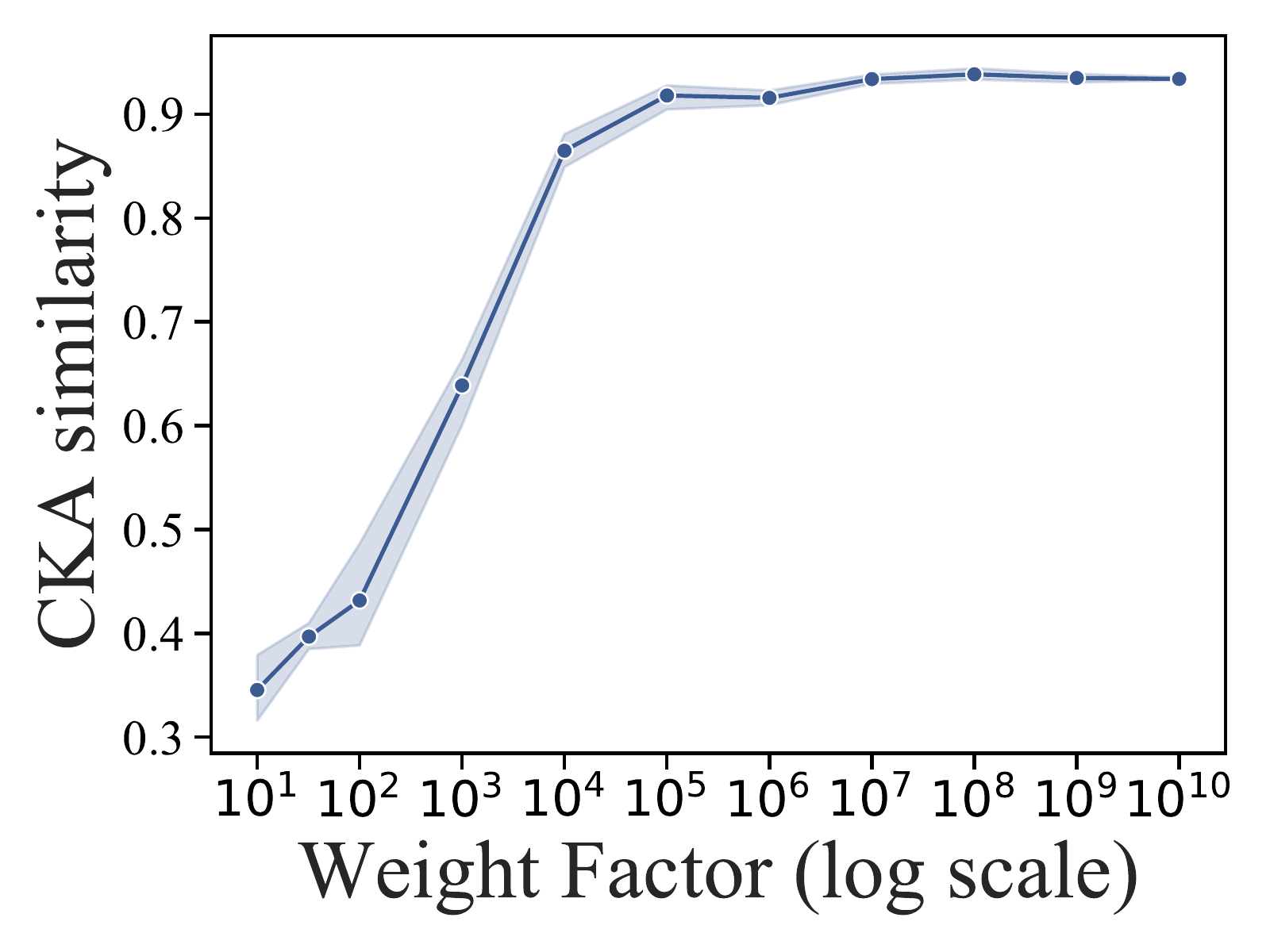}}\hfil
    \subfloat[Fashion MNIST dataset]{\includegraphics[width=0.5\columnwidth]{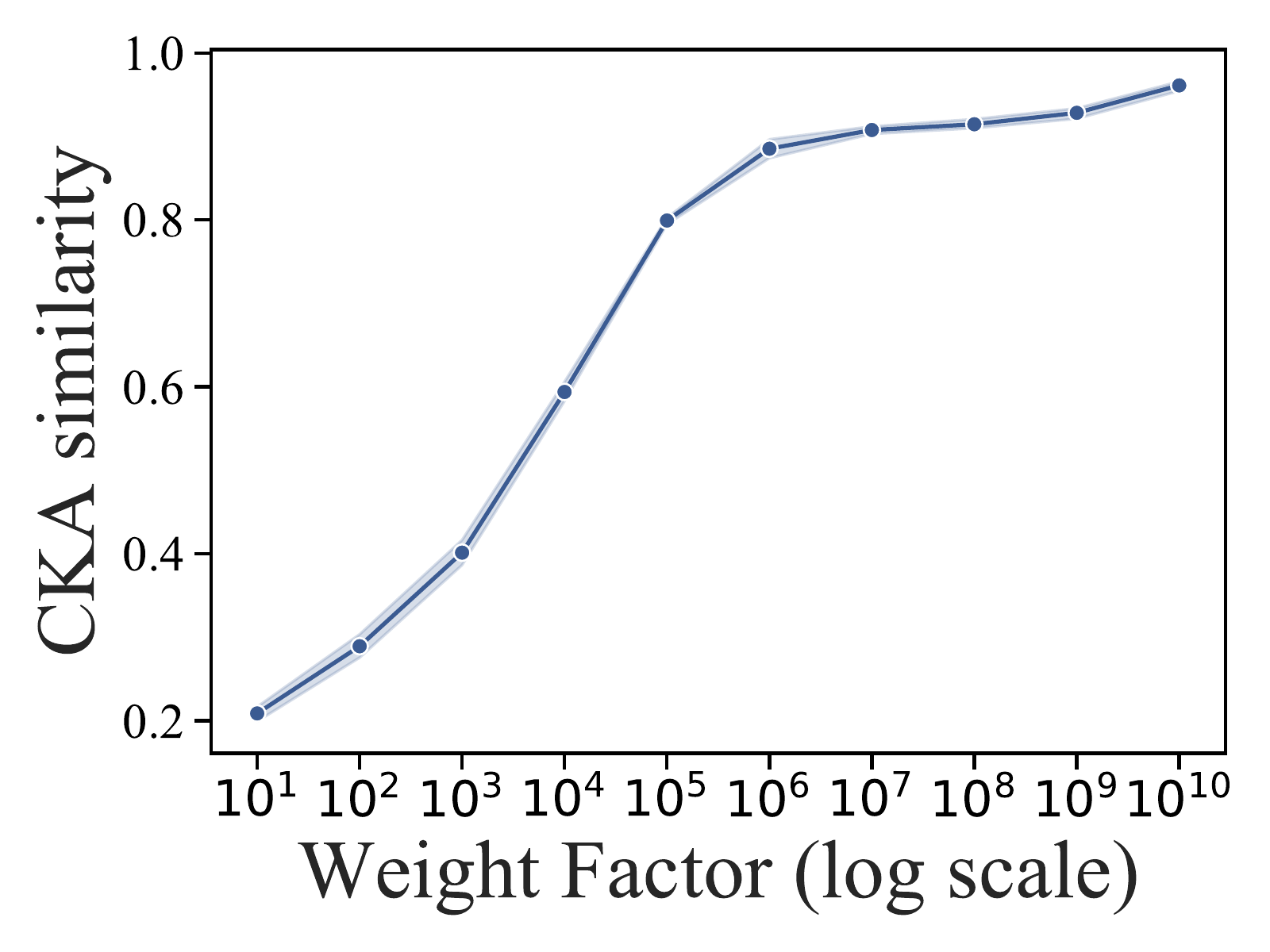}}
    \caption{ \new{\textbf{EWE is able to entangle watermarked with legitimate data because training with SNNL leads to higher CKA similarity between them.}} \new{We vary $\kappa$ from $0$ (the baseline) to $>0$ (EWE) using a log scale.}\vspace{-5mm}}
    \label{fig:cka}
\end{figure}

\new{Third, we analyze the similarity of their representations using central kernel alignment (CKA)~\cite{cortes2012algorithms,2019arXiv190500414K}. This similarity metric centers the distributions of the two representations before measuring alignment. In Figure~\ref{fig:cka}, we see that higher levels of SNNL penalty do in fact lead to higher CKA similarity between watermarked and legitimate data (compared with $\kappa=0$, the cross-entropy baseline). This, coupled with our first experiment, explains why EWE achieves better entanglement.}

\vspace{-2mm}
\subsubsection{Robustness against Extraction} 
\label{extraction}

\begin{table*}
    \centering
    \begin{tabular}{l*{6}c}
        \hline
        \multicolumn{1}{c}{\textbf{Dataset}} & \multicolumn{1}{c}{\textbf{Method}} & \multicolumn{2}{c}{\textbf{Victim Model}} & \multicolumn{2}{c}{\textbf{Extracted Model}} \\
        \hline
        & & Validation Accuracy& Watermark Success& Validation Accuracy& Watermark Success\\
        \hline
		\textbf{MNIST} & \emph{Baseline} & $99.03(\pm0.04)\%$ & $99.98(\pm0.03)\%$ & $98.79(\pm0.12)\%$ & $0.31(\pm0.23)\%$\\ & \emph{EWE} & $98.91(\pm0.13)\%$ & $99.9(\pm0.11)\%$ & $98.76(\pm0.12)\%$ & $65.68(\pm10.89)\%$\\
		\hdashline
		\textbf{Fashion MNIST} & \emph{Baseline} & $90.48(\pm0.32)\%$ & $98.76(\pm1.07)\%$ & $89.8(\pm0.38)\%$ & $8.96(\pm8.28)\%$\\ & \emph{EWE} & $90.31(\pm0.31)\%$ & $87.83(\pm5.86)\%$ & $89.82(\pm0.45)\%$ & $58.1(\pm12.95)\%$\\
		\hdashline
		\textbf{Speech Command} & \emph{Baseline} & $98.11(\pm0.35)\%$ & $98.67(\pm0.94)\%$ & $97.3(\pm0.43)\%$ & $3.55(\pm1.89)\%$\\ & \emph{EWE} & $97.5(\pm0.44)\%$ & $96.49(\pm2.18)\%$ & $96.83(\pm0.45)\%$ & $41.65(\pm22.39)\%$\\
		\hdashline
		\textbf{Fashion MNIST} & \emph{Baseline} & $91.64(\pm0.36)\%$ & $75.6(\pm15.09)\%$ & $91.05(\pm0.44)\%$ & $5.68(\pm11.78)\%$\\ (ResNet) & \emph{EWE} & $88.33(\pm1.97)\%$ & $94.24(\pm5.5)\%$ & $88.27(\pm1.53)\%$ & $24.63(\pm17.99)\%$\\
		\hdashline
        \textbf{CIFAR10} & \emph{Baseline} & $85.82(\pm1.04)\%$ & $19.9(\pm15.48)\%$ & $81.62(\pm1.74)\%$ & $7.83(\pm14.23)\%$\\ & \emph{EWE} & $85.41(\pm1.01)\%$ & $25.74(\pm8.67)\%$ & $81.78(\pm1.31)\%$ & $18.74(\pm12.3)\%$\\
        \hdashline
		\textbf{CIFAR100} & \emph{Baseline} & $54.11(\pm1.89)\%$ & $8.37(\pm13.44)\%$ & $47.42(\pm2.54)\%$ & $8.31(\pm15.1)\%$\\ & \emph{EWE} & $53.85(\pm1.07)\%$ & $67.87(\pm10.97)\%$ & $47.62(\pm1.41)\%$ & $21.55(\pm9.76)\%$\\
        \hline
    \end{tabular}
    \vspace{-2mm}
    \caption{ \textbf{Performance of the baseline approach (i.e., minimize cross-entropy of watermarks with the target class) vs. the proposed watermarking approach (EWE)}. For each dataset, we train a model with each approach and extract it by having it label its own training data. We measure the validation accuracy and watermark success rates\del{(i.e. percentage of inputs with triggers actually leading to the  output chosen by the defender)}\new{, i.e., difference between percentage of watermarks classified as the target class on a watermarked versus non-watermarked model}. \del{We observe that }Both techniques perform well on the victim model, so the intellectual property of models whose parameters are copied directly can be claimed by either technique. However, the baseline approach fails once it is extracted whereas EWE reaches significantly higher watermark success rate.\vspace{-5mm}}
    \label{tab:1}
\end{table*}

\begin{figure}[t]
    \centering
    \subfloat[MNIST dataset]{\includegraphics[width=0.75\columnwidth]{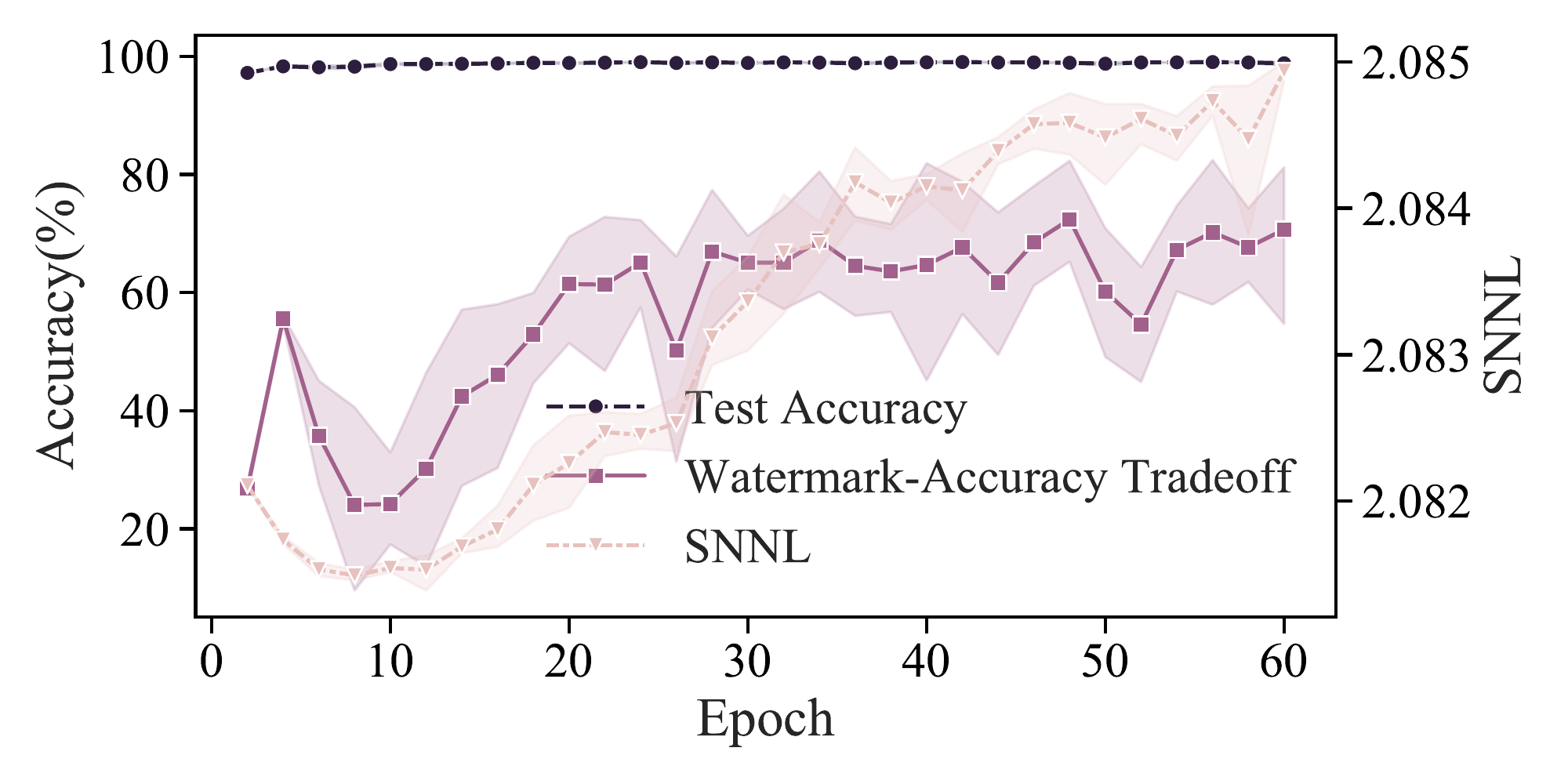}}
    \vspace{-1mm}
    \subfloat[Fashion MNIST dataset]{\includegraphics[width=0.75\columnwidth]{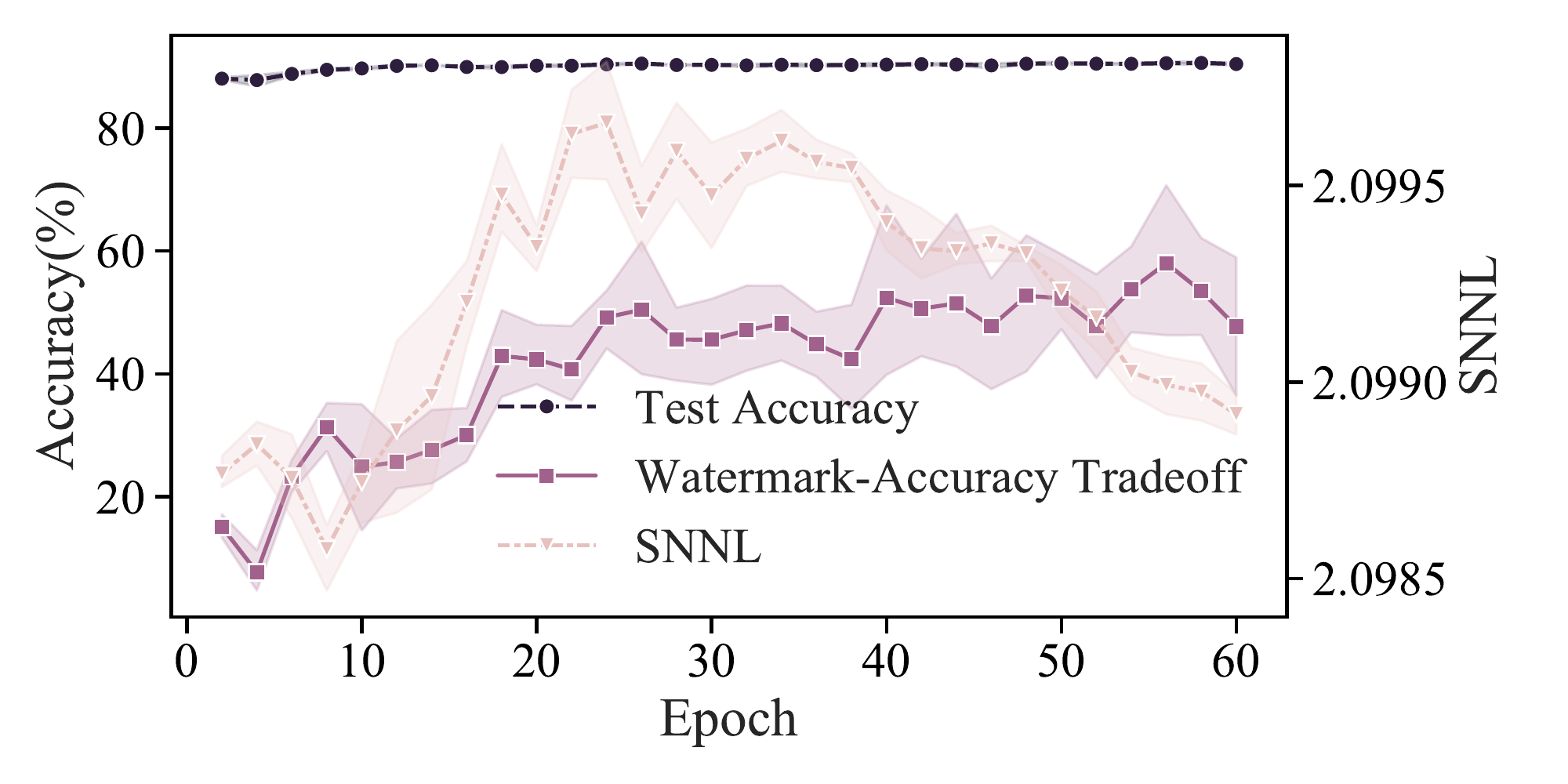}}
    \caption{ 
    \textbf{There exists an inflection point in the model's task accuracy and the SNNL value, as training progresses}. Before that point, continuing to train generally increases the watermark success rate relative to the task accuracy (we report the ratio between variations of the two).\vspace{-5mm}}
    \label{fig:training}
\end{figure}

We now evaluate the robustness of EWE \del{watermarking} against retraining-based extraction attacks launched by white-box adversaries \new{(see the top of  \S~\ref{ssec:threat-model})}\del{described in our threat model}. \new{To remove watermarks, this adversary retrains using only the cross-entropy loss evaluated only on legitimate data}\del{Since the objective of the adversary is to remove the watermarks, retraining occurs with solely cross-entropy loss}. We attack two victim neural networks: one \new{with}\del{that utilizes the} our EWE strategy \del{strategy }\del{(with $r=1$)} and one with our baseline, which uses only the cross-entropy loss, as proposed by Adi et al.~\cite{2018arXiv180204633A}.

\del{In both cases, we choose any $n=0.1\%$ data points from $c_S$ to be watermarked as $c_T$\footnote{In \S~\ref{section: choose watermark}, we demonstrate that there is little significance on the point-to-class similarity between $c_S$ and $c_T$ on watermark success.}}

\new{We define the watermark success rate as the proportion of $X_w$ correctly identified as $c_T$. 
We measure the validation accuracy on a held out dataset. We report results for both models in Table~\ref{tab:1} and find that the watermark success rate on the victim model (before retraining based extraction) is often near 100\% for both EWE and the baseline. 
After extraction, the watermark success rate always drops. It is in this case that we observe the largest benefits of EWE (over the baseline): there is often a $\geq20$ percentage point improvement in the watermark success. 
Besides, we often observe a negligible decrease in validation accuracy: an average of $0.81$ percentage points with a max of $3$ for the ResNet on Fashion MNIST.
}

\new{Our main result is that we can achieve watermark success rates between $18\%$ and $60\%$ with an average of $38.39\%$; the baseline is between $0.3\%$ and $9\%$ with an average of $5.77\%$. There is a minimal $0.81$ percentage point degradation on average of validation accuracy compared to the baseline, with a maximum of $3$ percentage points for a ResNet on Fashion MNIST. These watermark success rates allow us to claim ownership with 95\% confidence with $<100$ queries (see \S~\ref{section: verify}).} 

\del{
We focus our evaluation along two axes: (a) the \mbox{{\em validation accuracy}} of the watermarked model on samples from the task distribution, and (b) the \mbox{{\em watermark success rate}} which evaluates the utility of the watermarking technique (which will be discussed in more detail in \S~\ref{section: verify}). The watermark success rate can be measured for both the victim model and the extracted model. As shown in Table~\ref{tab:1}, 
both the validation accuracy and
watermark success rates of the \mbox{{\em baseline}} victim model are near 100\%. However, the watermark success rate quickly drops to 
near-zero on both MNIST and Fashion MNIST for the extracted model, and near 20\% for Google Speech Command
. This drastic drop in success indicates that previous techniques for watermarking do not survive model extraction attacks. As shown in Tables~\ref{tab:1}, however, the extracted EWE model averages at least 
39
\% higher watermark success rate than in the baseline extraction case, indicating better extraction survivability.} 

We also validate that continuing to maximize the SNNL during training is beneficial\del{ to the defender}. In Figure~\ref{fig:training} \new{we see that continued training improves the watermark robustness and task accuracy trade-off, until it plateaus near 60 epochs.}\del{, as training progresses through more epochs on the training set, the SNNL increases until it plateaus around 15 epochs. Prior to that, continuing to train increases the SNNL and improves the trade-off between watermark robustness and task accuracy.} We measure this \new{trade-off as the}
\del{through a} ratio between the 
\new{increase}\del{variation} of the watermark success rate 
\del{(i.e., how much it increases) }and the \new{decrease}\del{variation} of the task accuracy. \del{(this time, how much it decreases).} \del{After 15 epochs, the SNNL stays constant}
\del{and no further improvements in the trade-off are achieved.}
\vspace{-1mm}

\subsubsection{Scalability to Deeper Architectures}
\label{section: scalability}
\new{Entangling watermarks with legitimate data enables, and even forces, earlier layers to learn features that recognize both types of data simultaneously, as seen in Figure~\ref{fig:activation_toy_example}. 
This explains the improved robustness of watermarks. With entanglement, only later layers need to use capacity to separate between the two types of data, preserving model accuracy. 
This setup should work better for deeper models: there is only more capacity to learn shared features for watermarks and legitimate data. Our results in Figure~\ref{fig:activation} in Appendix~\ref{app: figures} confirms this.}

\new{However, deeper models such as ResNets often benefit (in their validation accuracy) from linearity: 
residual connections which add the input of the residual block directly to the output~\cite{resnet}. Notice that watermarks (e.g. a ``1'' with a small square trigger) are easily separable from legitimate data of the target class (e.g. a ``9'') and from the source class (e.g., a ``1'' without the trigger) because they share (nearly) no common features---they are outliers. Hence, residual connections pose a greater problem for entanglement because there are often no shared features, and forcing the watermarks (by increasing $\kappa$) to entangle with the legitimate data of $c_T$ may cause the model to misclassfy $X_{c_S}$ and $X_{c_T}$.}

\new{Our results validate this intuition. We see in Figure~\ref{fig:num_conv} in Appendix~\ref{app: figures} that deep convolutional neural networks can still entangle watermarks but yet we find that comparable ResNets cannot. Thus, we use our OOD watermarks (see Step 1 of \S~\ref{section: EWE_details}) because forcing them to entangle with $X_{c_T}$ has a lesser impact on accuracy. Though difficult to entangle, they achieve sufficient watermark success for claiming ownership (see Table~\ref{tab:1}). Even for more difficult tasks, as expected, EWE outperforms the baseline (see CIFAR-100 in Table~\ref{tab:1}), but both see a significant drop in watermark success. Finally, we see that watermarking is sensitive to the number of classes, in particular, EWE (see Figure~\ref{fig:num_class} in Appendix~\ref{app: figures}), probably due to complexity of the representation space.} 
\vspace{-3mm}

\section{Calibration of Watermark Entanglement}\label{section: calibration of watermark}

Through the calibration of EWE for \new{four}\del{two} vision datasets (MNIST~\cite{10027939599}, Fashion MNIST~\cite{2017arXiv170807747X}, \new{CIFAR-10, CIFAR-100~\cite{cifar}}), and an audio dataset (Google Speech Commands~\cite{2018arXiv180403209W}), we answer the following questions: (1) what is the trade-off between watermark robustness and task accuracy?; (2) how should the different parameters of EWE be configured?; and (3) is EWE robust to backdoor defenses \new{and attacks against watermarks}? Our primary results are:
\begin{enumerate}
\vspace{-3mm}
\itemsep0em
    \item For \del{all three datasets}\new{MNIST, Fashion MNIST, and Speech Commands (by which we validate if EWE is independent of the domain)}, we achieved watermark success above 40\% with less than 1 percentage point drop in test accuracy. \new{For CIFAR datatsets, watermark success above 18\% is reached with a minimal accuracy loss of $<1.5$ percentage points.} The weight factor allows the defender to control the trade-off between watermark robustness and task accuracy. 
    \item The ratio of watermarks to legitimate data during training, the choice of source-target class pair, and the choice of points to be watermarked all affect the performance of EWE significantly; temperature does not \new{since it is automatically optimized during training as described in \S~\ref{section: EWE_details}. Refer to Appendix~\ref{app: hyper} for more details.}
    \item Defenses against backdoors like pruning, fine-pruning, and Neural Cleanse are all ineffective in removing EWE.%
\end{enumerate}

\subsection{Experimental Setup}
\label{section: setup}

We chose to evaluate EWE on \del{two}\new{four} datasets in addition to MNIST. \new{While CIFAR-10 and CIFAR-100 are used to test the scalability of EWE as described in \S~\ref{section: scalability},} we use Fashion MNIST because its classes are much harder to linearly separate than MNIST, making it a good benchmark for learning a more complex task, \new{with comparable computational cost to MNIST. Thus it allows us to tune the hyperparameters efficiently to explore behaviors of EWE.} Further, it shows that EWE works well when the task naturally contains ambiguous inputs across pairs of classes. We also evaluated EWE on Google Speech Commands, an audio dataset for speech recognition, because speech recognition is one of the applications where ML is already pervasively deployed across industry.

\del{rather than more complex image datasets like CIFAR-10 or Imagenet,}\del{Thus, it is a good example of models that constitute valuable intellectual property.}

\vspace{-3mm}
\paragraph{Datasets.} 

\noindent \textbf{1. MNIST} is a dataset of hand-written digits (from 0 to 9) with \del{60,000 training and 10,000 test data}\new{70,000 data points}~\cite{10027939599}, where each data point is a gray-scale image of shape 28$\times$28\del{ associated with one of the 10 classes}. \new{When needed, we sampled OOD watermarked data from Fashion MNIST.}\del{For this dataset, we define the trigger to be a 9-pixel white (i.e., value of the pixel is 1) square.}

\noindent \textbf{2. Fashion MNIST} is a dataset of fashion items~\cite{2017arXiv170807747X}. It can be used interchangeably with MNIST. %
Because the task is more complex, models achieving $>99\%$ accuracy on MNIST however only reach $>90\%$ on Fashion MNIST. \new{When needed, we sampled OOD watermarked data from MNIST.}\del{We use the same trigger here than for MNIST.}

\noindent \textbf{3. Google Speech Commands} is an audio dataset of 10 single spoken words ~\cite{2018arXiv180403209W}. The training data has about 40,000 samples. We pre-processed the data to obtain a Mel Spectrogram~\cite{choi2017kapre}. \new{We tried two methods for generating watermarks both using in-distribution data: (a) modifying the audio signal, or (b) modifying the spectrogram. For (a), we sample data from the source class and overwrite $\frac{1}{8}^{th}$ of the total length of the sample (i.e., 0.125 seconds) with a sine curve, as shown in Figure~\ref{fig:audio_Watermark}; for (b), }
each audio sample is \del{thus} represented as an array of size 125$\times$80. We then define the \del{watermark}\new{trigger} to be two 10$\times$10-pixel squares at both the upper right and upper left-hand corners in case of vanishing or exploding gradients. \new{It was observed that the choice of using (a) or (b) \emph{does not influence the performance of EWE.}} 

\noindent \new{\textbf{4. CIFAR-10} consists of 60,000 32$\times$32$\times$3 color images equally divided into 10 classes~\cite{cifar}, while 50,000 is used for training and 10,000 is used for testing. When needed, we use OOD watermarks sampled from SVHN~\cite{svhn}.}

\noindent \new{\textbf{5. CIFAR-100} is very similar to CIFAR-10, except it has 100 classes and there are 600 images for each class~\cite{cifar}. When needed, we use OOD watermarks sampled from SVHN~\cite{svhn}.}
\vspace{-3mm}
\paragraph{Architectures.} We use the following architectures: 

\noindent \textbf{1. Convolutional Neural Networks} are used for MNIST and Fashion MNIST. The architecture is composed of 2 convolution layers with 32 5$\times$5 and 64 3$\times$3 kernels respectively, and 2$\times$2 max pooling. It is followed by two fully-connected (FC) layers with 128 and 10 neurons respectively. All \new{except the last} layers are followed by a dropout layer to avoid overfitting. When implementing EWE\del{ on this architecture}, the SNNL is computed after both convolution layers and the first FC layer.

\noindent \textbf{2. Recurrent Neural Networks} are used for Google Speech Command dataset. The architecture is composed of 80 long short-term memory (LSTM) cells of 128 hidden units followed by \del{a}\new{two} FC layer\new{s} of \new{128 and} 10 neurons \new{respectively}. When applying EWE, the SNNL is computed after the $40^{th}$ cell\new{,} \del{and }the last ($80^{th}$) cell\new{, and the first FC layer}.

\noindent \new{\textbf{3. Residual Neural Network (ResNet)~\cite{resnet}} are used for Fashion MNIST, CIFAR-10, and CIFAR-100 datasets. We use ResNet-18 which contains 1 convolution layer followed by 8 residual blocks (each containing 2 convolution layers), and ends with a FC layer. It is worth noting that the input to a residual block is added to its output. We compute SNNL on the outputs of the last 3 residual blocks.}

\subsection{No Free Lunch: Watermark vs. Utility}
\label{section: trade-off}

\begin{figure}[t]
    \centering
    \vspace{-3mm}
    \subfloat[Fashion MNIST]{\includegraphics [width=0.5\linewidth]{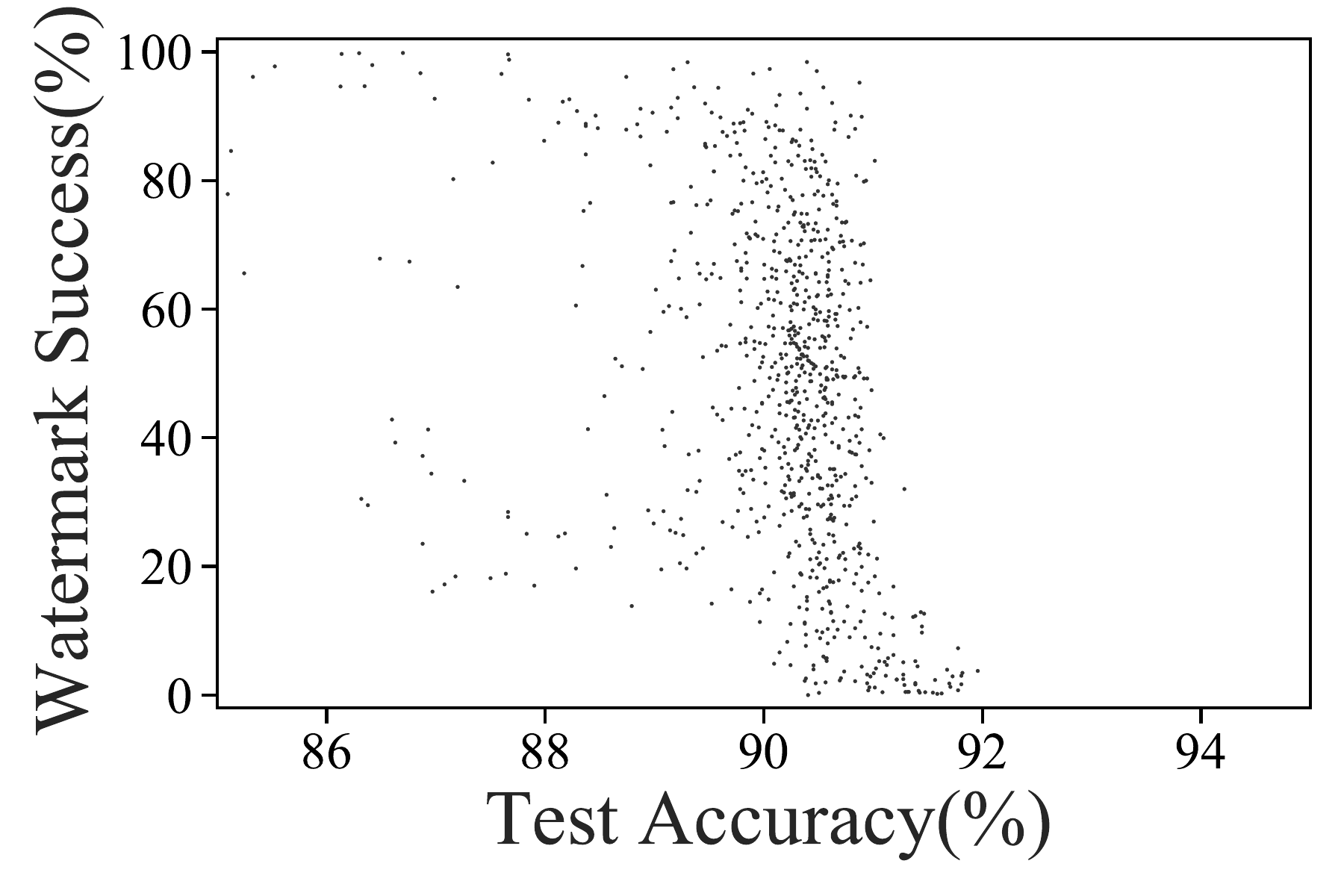}}
    \subfloat[Speech Command]{\includegraphics [width=0.5\linewidth]{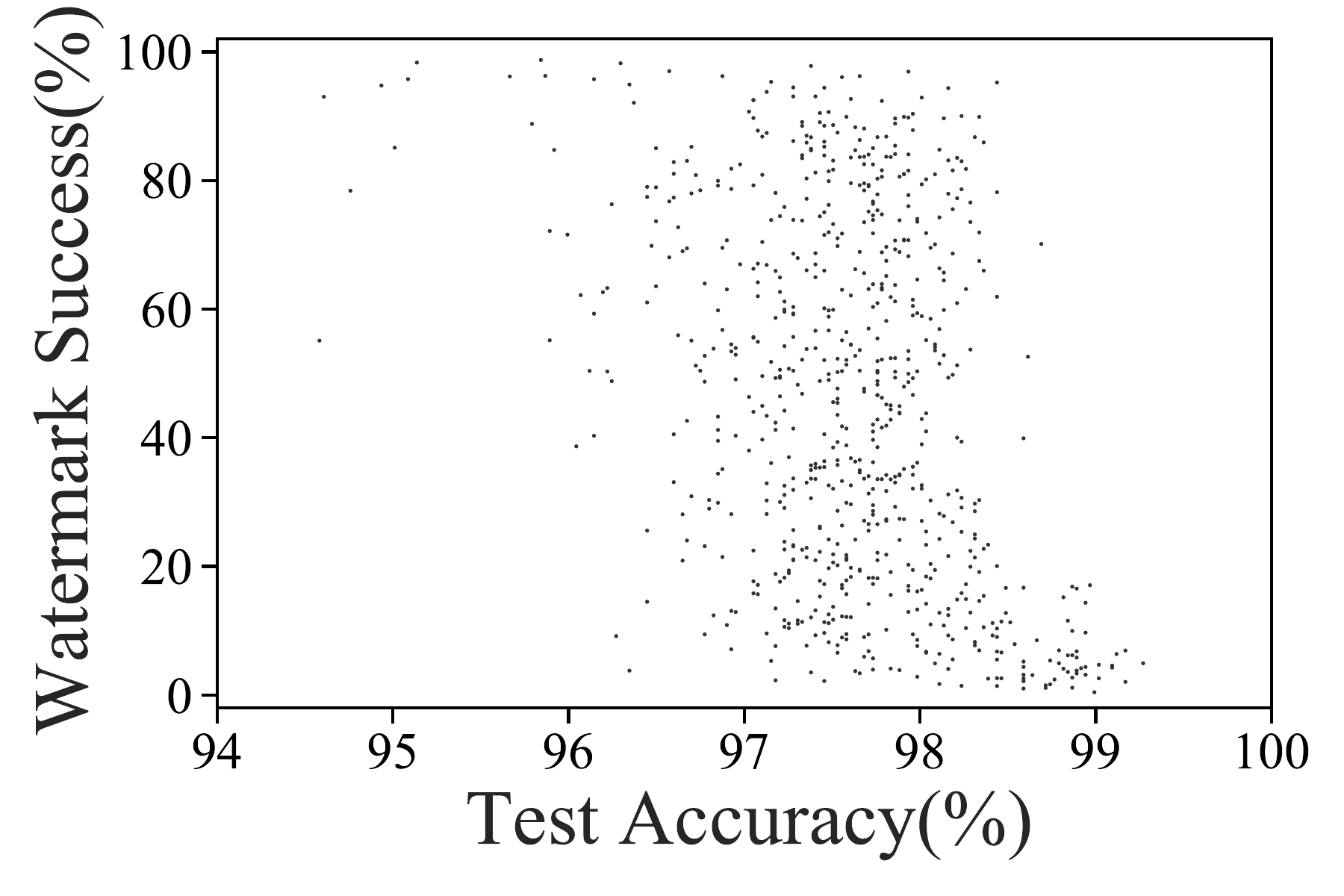}}
    \caption{ \textbf{Watermark success versus model accuracy on the task}. Each point corresponds to a model trained with uniformly-sampled hyperparameters. As test accuracy increases, it becomes harder to have robust watermarks.\vspace{-5mm}}
    \label{fig:trade-off}
\end{figure}

We study the tension between accuracy on the task's distribution and robustness of the watermarks: if the defender wants to claim ownership of a model, they would like this model to predict their chosen label on the watermarks as frequently as possible while at the same time minimizing the impact of watermarks on the model's performance when presented with samples from the task distribution. 

To systematically explore the trade-off between successfully encoding watermarks and correctly predicting on the task distribution, we first perform a comprehensive grid search that considers all hyper-parameters relevant to our approach: the class pairs $(c_S, c_T)$ \new{(note that $c_S$ is a class from another dataset when OOD watermark is used)}, the temperature $T$, the weight ratio \del{$w$}\new{$\kappa$}, and the ratio of task to watermark data \del{in $X_w$}\new{(i.e. $r$ in Algorithm~\ref{alg: EWE})}, how close points have to be to the target class to be watermarked. \del{Later in Section 5.3}\new{In Appendix~\ref{app: hyper}}, we perform an ablation study on the impact of each of these parameters: they can be used to control the trade-off.

Each point in Figure~\ref{fig:trade-off} corresponds to a model trained using EWE with a set of hyper-parameters. 
For the Fashion MNIST dataset shown in Figure~\ref{fig:trade-off} (a), the tendency is exponential: it becomes exponentially harder to improve accuracy by decreasing the watermark success rate. In the Speech Commands dataset, 
as shown in Figure~\ref{fig:trade-off} (b), there is a large number of points with nearly zero watermark success. 
This means it is harder to find a good set of hyperparameters for the approach. However, there exists points
in the upper right corner demonstrating that certain hyperparameter values could lead to robust watermark with little impact on test accuracy.

\subsection{Evaluation of Defenses against Backdoors}
\label{section: backdoor defense}

\begin{figure}[t]
    \centering
    \vspace{-3mm}
    \subfloat[MNIST]{\includegraphics[width=0.5\linewidth]{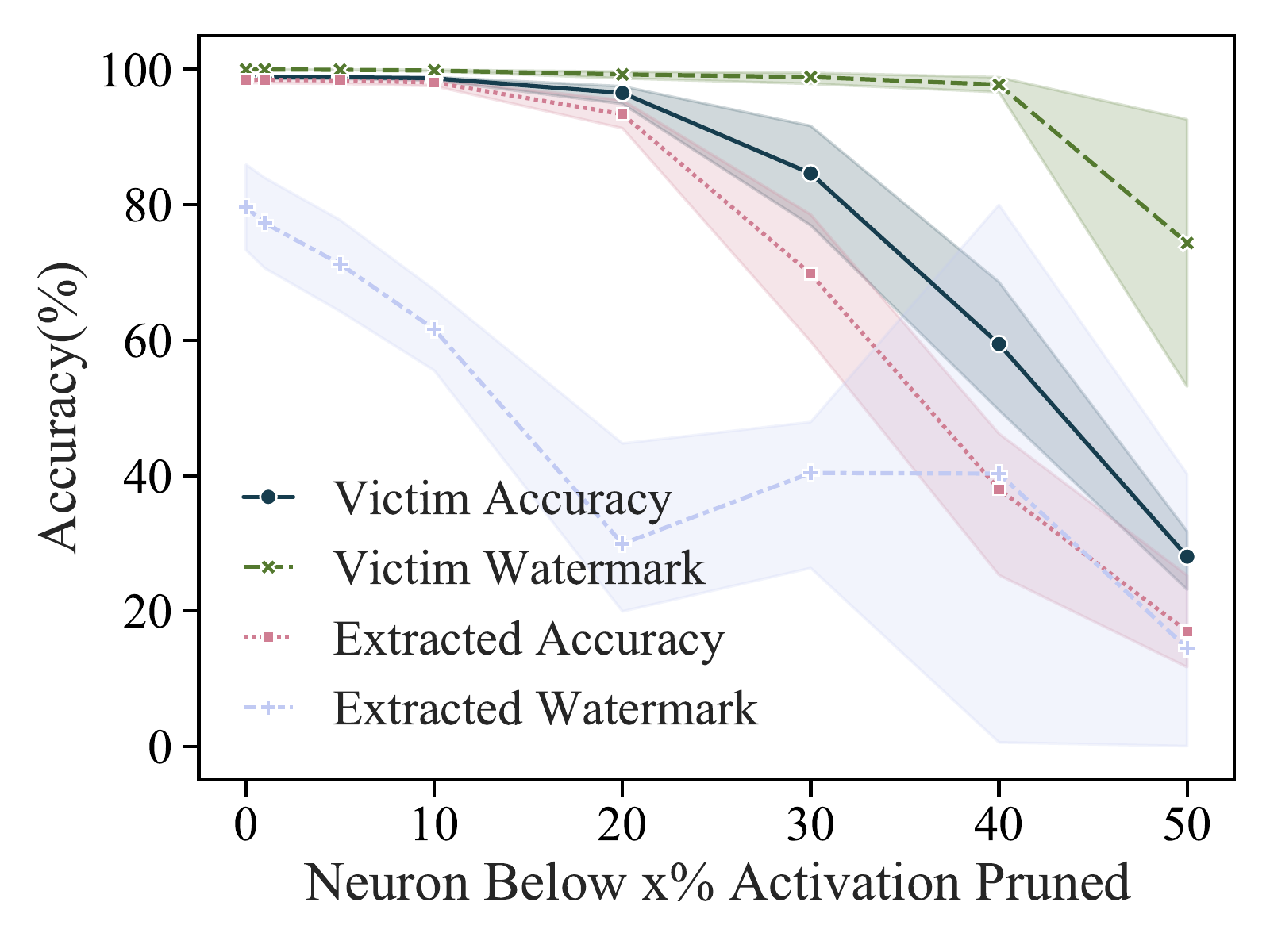}}\hfill
    \subfloat[Fashion MNIST]{\includegraphics[width=0.5\linewidth]{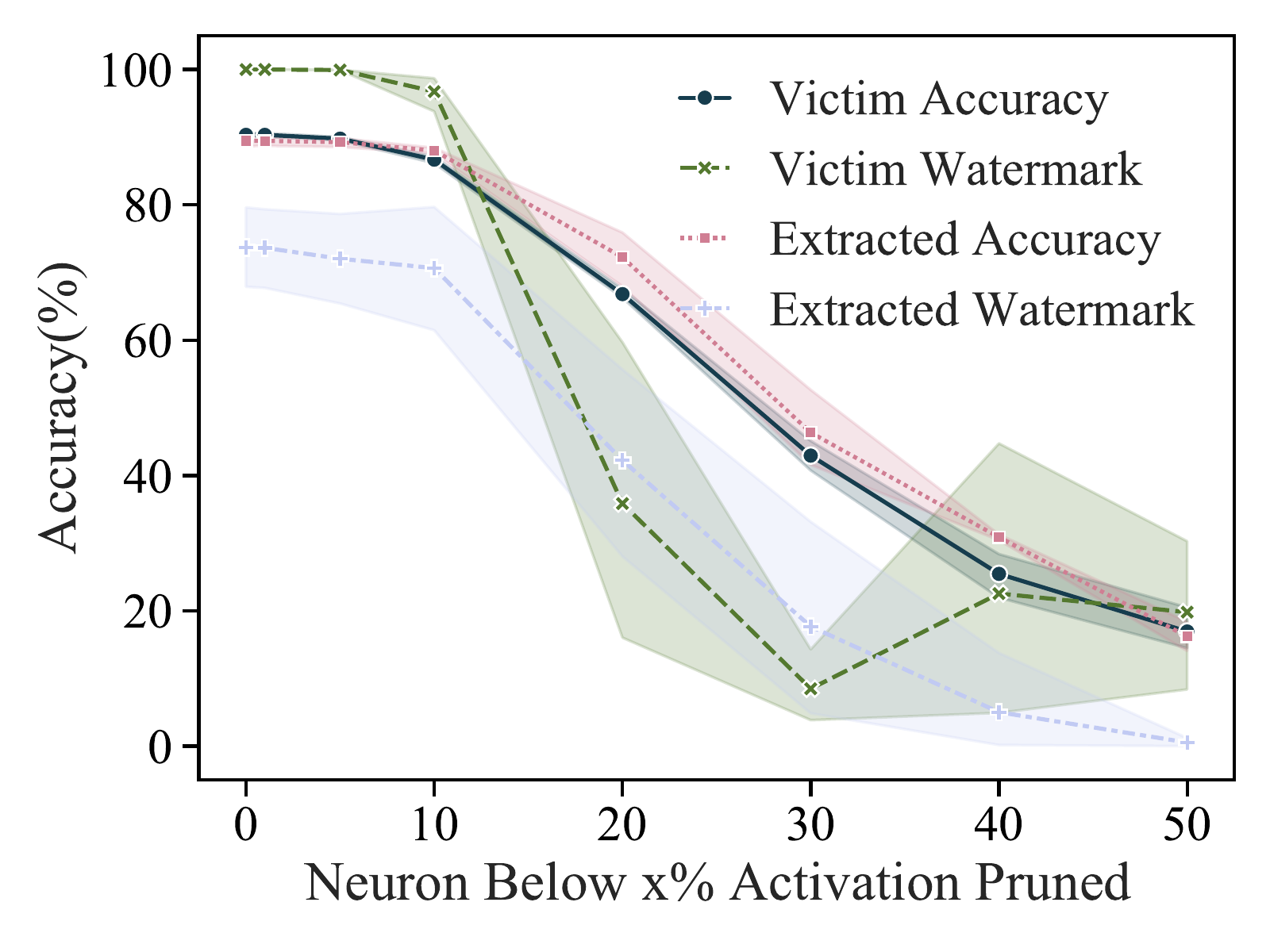}}
    \vspace{-3mm}
    \caption{ Task accuracy and watermark success rate on the extracted model in the face of a pruning attack. \new{For both datasets,} bringing the watermark success rate below \del{10\%}\new{20\%} comes at the adversary's expense: \new{accuracy drop of more than 40 percentage points.}\del{task accuracy is only 60\%.}\vspace{-5mm}}
    \label{fig:pruning}
\end{figure}

\paragraph{Pruning.} Since backdoors and legitimate task data activate different neurons, pruning proposes to remove neurons that are infrequently activated by legitimate data to decrease the performance of potential backdoors~\cite{2018arXiv180512185L}. Given that neurons less frequently activated contribute less to model predictions on task inputs, pruning them is likely to have a negligible effect. Since watermarks are a form of backdoors, it is natural to ask whether pruning can mitigate EWE.

We find this is not the case because watermarks are entangled to the task distribution. Recall Figure~\ref{fig:activation_entangled}, where we illustrated how EWE models have similar activation patterns on watermarked and legitimate data. Thus, neurons encoding the watermarks are frequently activated when the model is presented with legitimate data. Hence, if we extract a stolen model and prune its neurons that are activated the least frequently, we find that watermark success rate remains high despite significant pruning (refer Figure~\ref{fig:pruning}). In fact, the watermark success rate only starts decreasing below 20\% when the model's accuracy on legitimate data also significantly decreases (by more than 40 percentage points). Such a model becomes useless to the adversary, who would be better off training a model from scratch. We conclude that pruning is ineffective against EWE. 

\begin{figure}[t]
    \centering
    \subfloat[MNIST]{\includegraphics[width=0.5\linewidth]{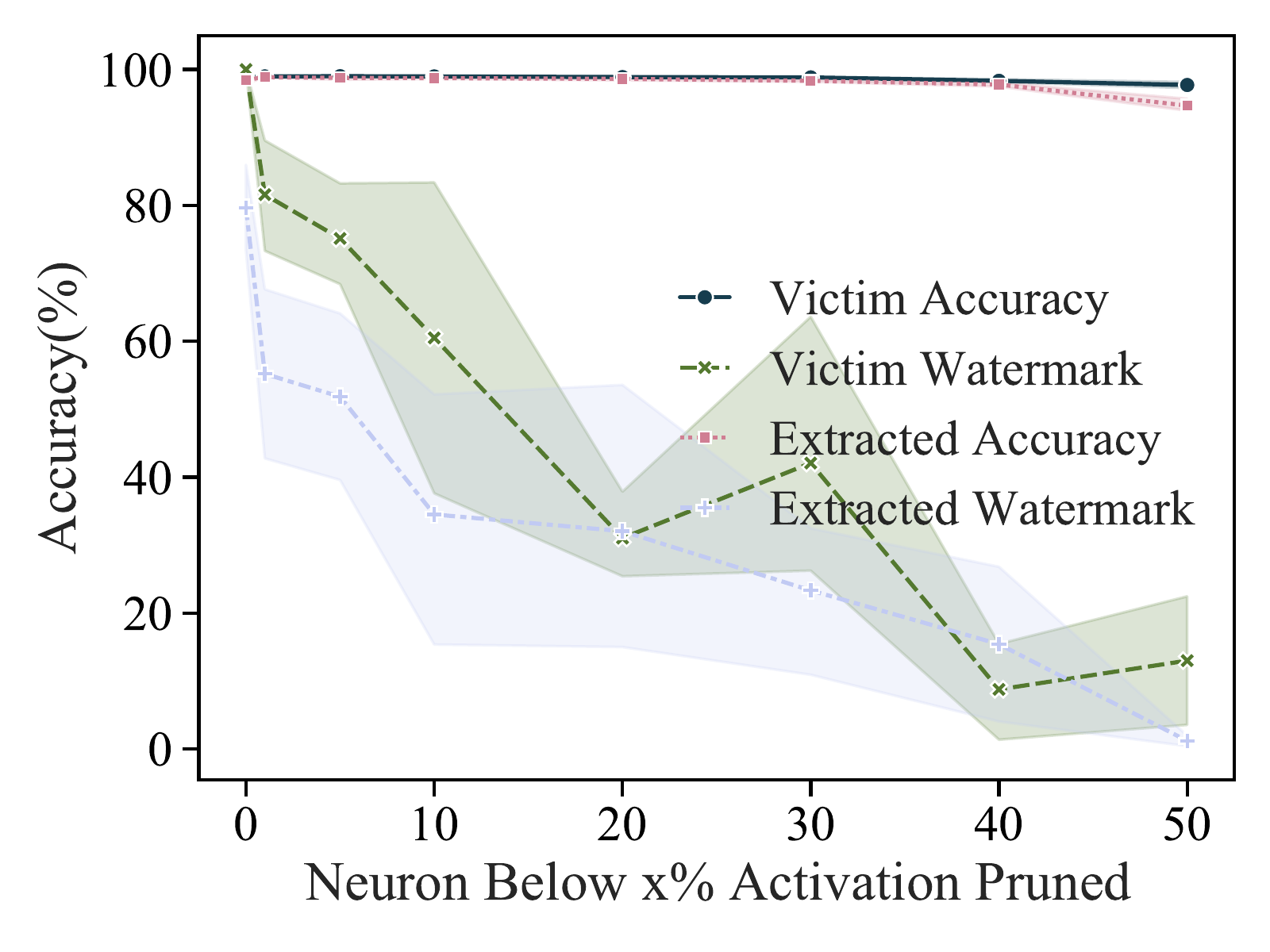}}
    \subfloat[Fashion MNIST]{\includegraphics[width=0.5\linewidth]{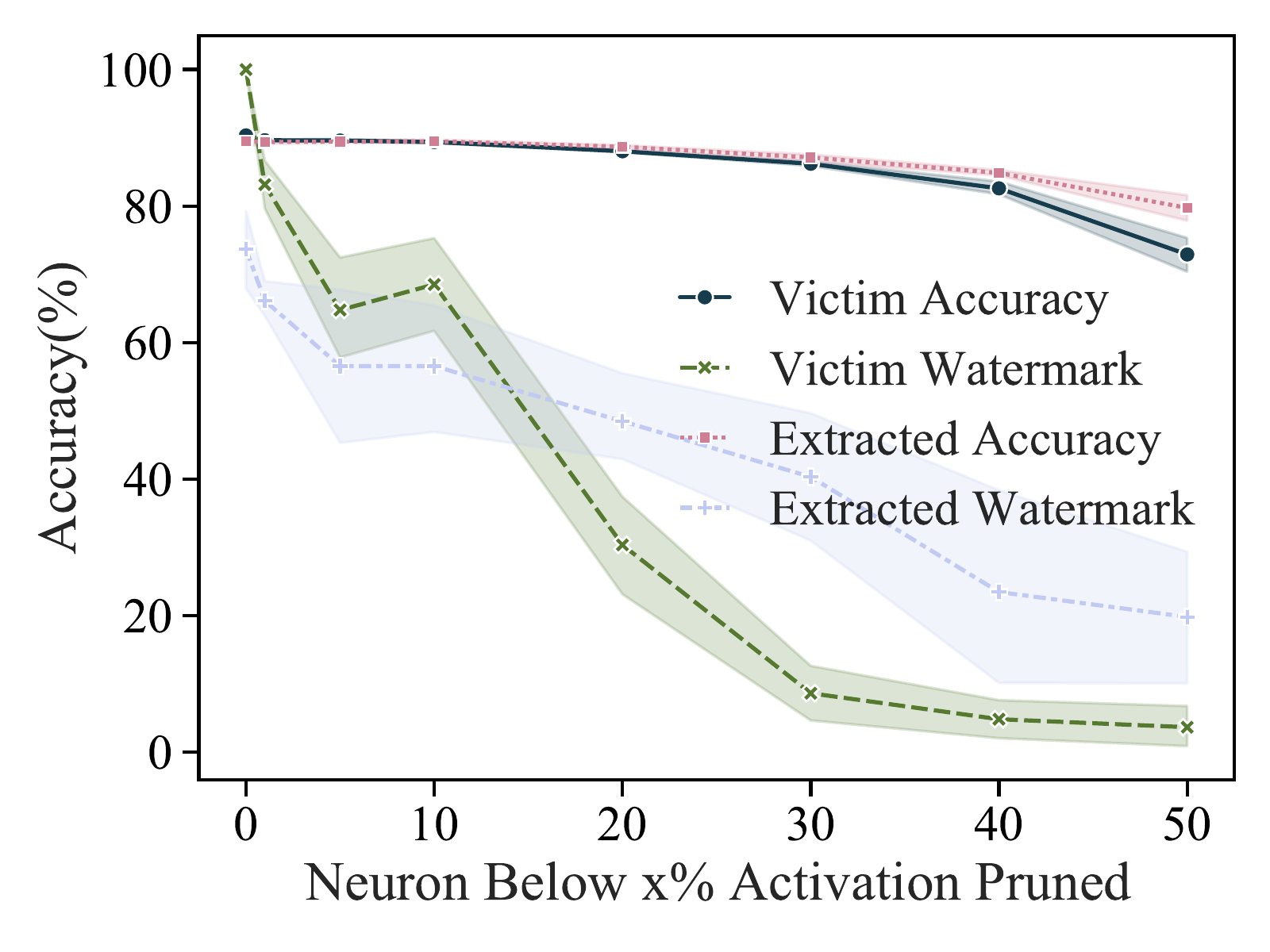}}
    \vspace{-3mm}
    \caption{Task accuracy and watermark success rate on the extracted model in the face of a fine pruning attack. Despite a more advantageous trade-off between watermark success rate and task accuracy, the adversary is unable to bring the watermark success rate sufficiently low \del{for}\new{to prevent} the defender to \del{be unable to }claim ownership (see \S~\ref{section: verify}) \new{until 40\% neurons are fine-pruned. Beyond this point, fine-pruning more neurons would lead to loss in the extracted model's accuracy.\vspace{-5mm}}
    }
    \label{fig:fine_pruning}
\end{figure}

\vspace{-3mm}
\paragraph{Fine Pruning.} Fine pruning improves over pruning by continuing to train (i.e., fine-tune) the model after pruning\del{ the architecture}~\cite{2018arXiv180512185L}. \del{In the benign setting} \new{T}his helps recover some of the accuracy that has been lost during pruning. In the presence of backdoors, this also contributes to overwriting any behavior learned from backdoors.

We also analyze EWE in the face of fine pruning. We first extract the model by retraining (i.e., randomly initialize model weights and train them with data labeled by the victim model), prune a fraction of neurons that are less frequently activated, and then train the non-pruned weights on data labeled by the victim model. Results are plotted in Figure~\ref{fig:fine_pruning}. In the most favorable setting for fine pruning, watermark success rate on the extracted model remains around 20\% \new{before harming the utility of the model}, which is still enough to claim ownership---as shown in \S~\ref{section: verify}. This is despite the fact that 50\% of the architecture's neurons were pruned. Since the data used for fine-tuning is labeled by the watermarked victim model, it contains information about the watermarks even when the labels provided are for legitimate data. 

\begin{figure}[t]
    \centering
    \includegraphics[width=\columnwidth]{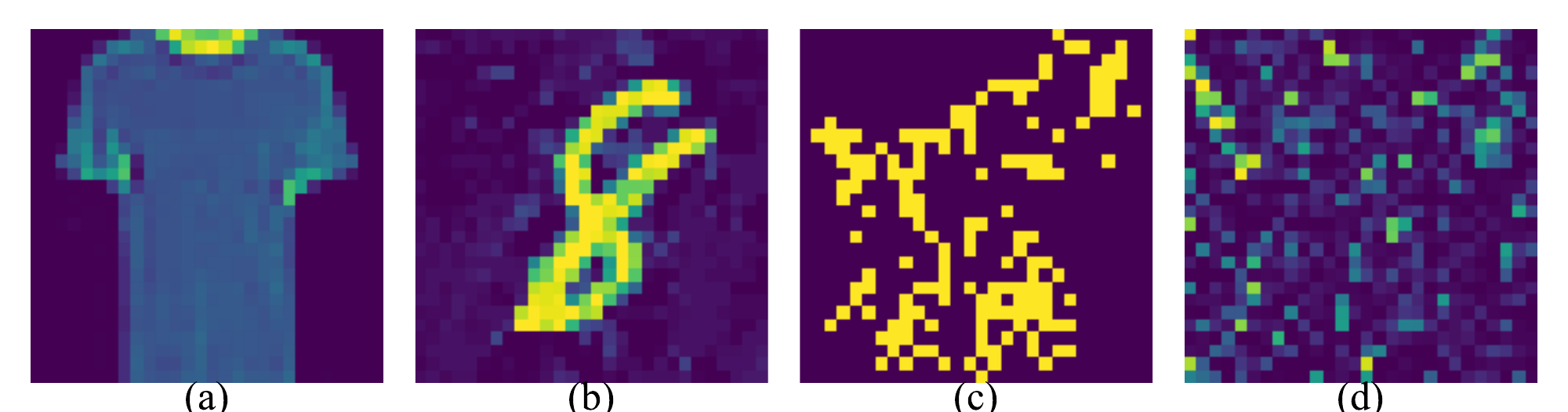}
    \caption{ Neural Cleanse leverages the intuition that triggers may be recovered by looking for adversarial examples \del{between the source class of the watermark and its target class}\new{for the target class}. To illustrate this, we have here \new{a legitimate input of the target class (a),} an example of a watermark (\del{left}\new{b}), an adversarial example \new{(see Appendix~\ref{app:adv_example_for_ewe} for details) intialized as a blank image and perturbed to be misclassified by the extracted model in the target class}(\del{middle}\new{c}), and the backdoor candidate recovered by Neural Cleanse (\del{right}\new{d}).\del{ The adversarial example is from class 3 and perturbed to be misclassified by the extracted (stolen) model in class 5.} If \del{the adversarial example}\new{either (c) or (d)} were similar to the watermark, this would enable us to recover \del{both the source class where watermarks were inserted and the trigger used to watermark inputs}\new{the watermarked data and then use this knowledge to remove the watermark as described in \S~\ref{section: robustness}}. However, this is not the case for models extracted \del{starting }from a \new{EWE defended} victim model\del{ defended with EWE}: the watermark \del{trigger} proposed (\new{c and d}\del{right}) is different from the trigger used by EWE (\new{b}\del{left}).\vspace{-5mm}}
    \label{fig:adversarial_sample}
\end{figure}

\begin{figure*}[t]
    \centering
    \vspace{-3mm}
    \subfloat[Un-watermarked Model]{\includegraphics[width=0.24\textwidth]{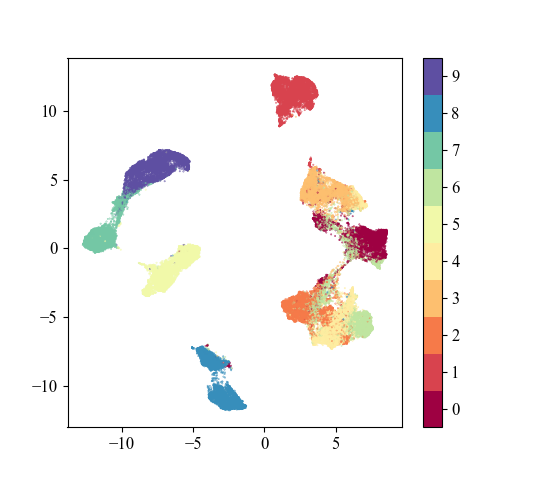}}\hfill
    \subfloat[Watermarked Model (Baseline)]{\includegraphics[width=0.24\textwidth]{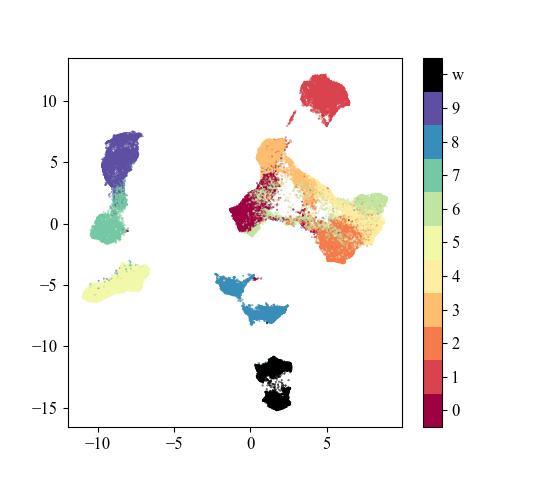}}\hfill
    \subfloat[EWE In-distribution Watermark]{\includegraphics[width=0.24\textwidth]{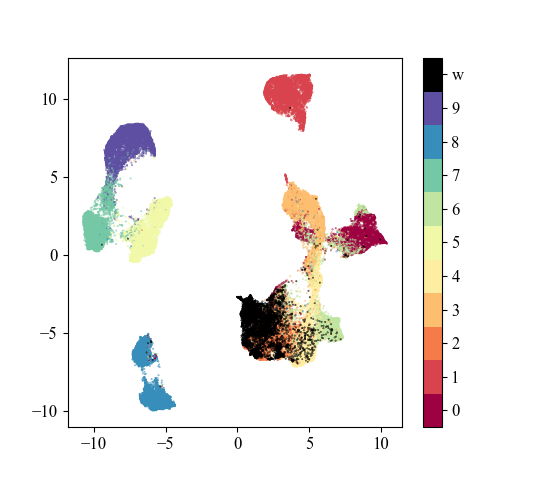}} \hfill
    \subfloat[EWE Out-distribution Watermark]{\includegraphics[width=0.24\textwidth]{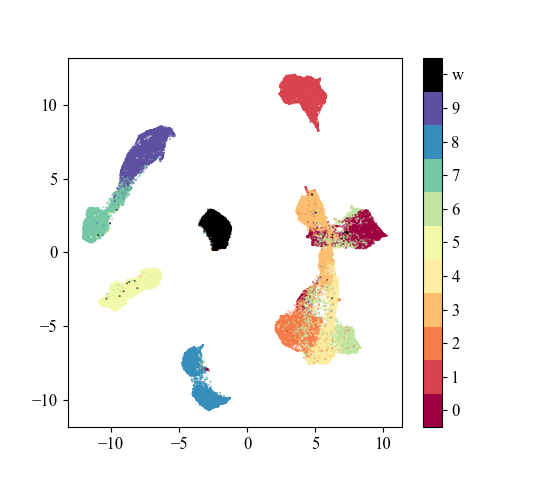}} 
    
    \vspace{-3mm}
    \caption{ 
    \textbf{Change in the distance among clusters of data from different \new{Fashion MNIST} classes following watermarking}. The \del{three}\new{four} subplots are made using four different approaches specified by the sub-captions. In (c) and (d), $c_S=8$ and $c_T=0$, while $D_w$ is MNIST for (d).
    Each point in the plot represents an output vector of the last hidden layer\del{of the corresponding model}. 
    These representations are plotted in 2-D using UMAP dimensionality reduction to preserve global distances~\cite{2018arXiv180203426M}. Comparing (a) and (b), one can observe that the clusters of class \del{3}\new{8} and \del{5}\new{0} become closer in (b) while the distances among the other classes remain similar. This is why such watermarked model can be detected by Neural Cleanse~\cite{Wang2019NeuralCI}, which searches for pairs of classes that are easily misclassified with one another. In contrast, \del{some clusters in (c) that are not related to the watermark are also very close due to maximization of entanglement}\new{EWE with either in or out of distribution watermarks does not influence this distance significantly}, which makes it more difficult for Neural Cleanse to detect the watermark. \vspace{-5mm}}
    \label{fig:neural_cleanse}
\end{figure*}

\vspace{-3mm}
\paragraph{Neural Cleanse.}
Neural Cleanse is a technique that detects and removes backdoors in deep neural networks~\cite{Wang2019NeuralCI}. The intuition of this technique is that adding a backdoor would cause the clusters of the source and target classes to become closer in the representation space. Therefore, for every class $c$ of a dataset, Neural Cleanse tries to perturb data from classes different to $c$ in order to have them misclassified in class $c$. Next, the class requiring significantly smaller perturbations to be achieved is identified as the "infected" class (i.e., the class which backdoors were crafted to achieve as the target class). In particular, the authors \textit{define a model as backdoored if an anomaly index derived from this analysis is above a certain threshold} (set to 2). The perturbation required to achieve this class is the recovered trigger. Once both the target class and trigger have been identified, one can remove the backdoor by retraining the model to classify data with the trigger in the correct class, \`a la adversarial training~\cite{2013Szegedy}.

To analyze the robustness of EWE to Neural Cleanse, we compare the performance of a model watermarked with EWE and a baseline model watermarked by minimizing the cross-entropy of watermarks labeled as the target class (\del{$w=0$}\new{$\kappa=0$} in Equation~\ref{eq:loss}).\del{\footnote{Note that the Neural Cleanse paper considers the problem of backdooring the entire set of classes (i.e., all classes are considered as source classes) whereas we insert watermarks only for a single source-target class pair.}} We compute the anomaly index of the EWE and baseline models. If the anomaly index is above 2, the model is detected as being watermarked (i.e., backdoored in~\cite{Wang2019NeuralCI}). \new{On the Fashion MNIST (see Figure~\ref{fig:adversarial_sample}),} EWE exbhibits an \new{average} anomaly index of \del{1.6}\new{1.24 (over 5 runs)} that evades detection whereas the baseline model has an \new{average} index of \del{31.6}\new{8.84}. This means that Neural Cleanse is unable to identify our watermark and its trigger. \del{Consequently, Neural Cleanse is ineffective against EWE. } \del{Note that entangling legitimate data from different classes in addition to entangling legitimate data to watermarks would further contribute to bringing class clusters closer in representation spaces. This makes it even harder for Neural Cleanse to detect watermarks inserted by EWE because two classes being close to one another is no longer exclusively indicative of watermarking. This is illustrated in Figure~\ref{fig:neural_cleanse}.}

\new{It is worth noting: (a) Neural Cleanse considers the problem of backdooring the entire set of classes (i.e., all classes are considered as source classes), and (b) backdoor attacks usually aim at minimal perturbation to the inputs. While being similar to legitimate data from all classes and labeled as a specific class, such backdoors changes the decision surface significantly, which would be detected by Neural Cleanse. In EWE, we insert watermarks only for a single source-target class pair. Besides, watermarked data is not restricted by the degree of perturbation and could even be OOD. Thus entangling it with $c_T$ does not change the decision boundary between $c_T$ and other classes, as shown in Figure~\ref{fig:neural_cleanse} (and Figure~\ref{fig:neural_cleanse_mnist}, ~\ref{fig:neural_cleanse_speechcmd} for MNIST and Speech Command in Appendix~\ref{app: figures}). This makes it hard for Neural Cleanse to detect EWE watermarks.}

\section{Robustness to Adaptive Attackers}
\label{section: robustness}
Recall from our threat model (see \new{the top of} \S~\ref{section: entangling watermarks}) that the adversary has no knowledge of the parameters used to calibrate the watermarking scheme (such as \del{$w$} \new{$\kappa$ and} $T^{(1)} \cdots T^{(L)}$ in Algorithm~\ref{alg: EWE}) \del{and}\new{nor} the specific trigger used to verify watermarking. In this section, we explore when the adversary has more resources and knowledge than stated in the threat model.
\vspace{-1mm}
\subsection{Knowledge of EWE and its parameters}
\vspace{-1mm}
\del{The Source of Robustness}\new{Knowledge of the parameters used to configure EWE defeats watermarking, as expected. } The robustness of EWE relies on {\em maintaining the secrecy} of the trigger and watermarking parameters \del{which serves as a key
to protect the intellectual property contained in the model, similar to the case in public-key cryptography}\new{to protect the intellectual property contained in the model}.\del{Relaxing the Assumptions. If we provide the adversary with knowledge of the watermarking scheme (refer (e) in our threat model), the adversary can perform the following actions after extracting the victim model.}
If the adversary knows the trigger used to watermark inputs, they could refuse to classify any input that contains that trigger (denial-of-service). Alternatively, \del{the adversary could crop inputs to remove the trigger}\new{they could extract the model while instead minimizing the SNNL of the watermarks and legitimate data of class $c_T$. Note, minimizing SNNL corresponds to disentangling.} Additionally, adversaries may also be able to retrain the triggers (\del{ergo}\new{and thus,} watermarks) to predict the correct label.

\new{ Any of these results in complete removal of watermarks} However, \del{we evaluate adversaries with}\new{this is not a realistic threat model since the adversary should only }know\del{ledge} \del{of  the}\new{that EWE was used as a }watermarking scheme (\del{refer}\new{see} (e) in our threat model defined in \S~\ref{ssec:threat-model}.
\del{the adversary can perform the following actions after extracting the victim model. While the aforementioned attacks can potentially alleviate the guarantees that are provided by EWE, they are not in the scope of our threat model because they require knowledge of the trigger or parameters of EWE. In \S~\ref{section: backdoor defense}, we evaluate our proposal against techniques used to prevent backdoors (which lie in the scope of our threat model)} \new{In this way, parameters of EWE play a similar role to cryptographic keys. Next, we evaluate EWE against several more realistic adaptive attacks against watermarks such as piracy attacks}.
\vspace{-3mm}
\subsection{Knowledge of EWE only} 
\vspace{-1mm}
\new{With knowledge of EWE but not its configuration (e.g.,  the source and target classes), the adversary can still adapt in several ways. We evaluate four adaptive attacks.}

\vspace{-4mm}
\paragraph{Disentangling Data.} \del{If the adversary was aware that EWE is used, }We conjecture that the adversary could \del{could potentially train the extracted model to minimize SNNL, leading to disentanglement and consequently task separation.}\new{perform extraction by minimizing SNNL} \delete{on all classes }\new{to disentangle watermarks from task data.} \revision{We assumed a strong threat model such that the adversary has knowledge of all the parameters of EWE (including the trigger if in-distribution watermark is used, and the OOD dataset if OOD watermark is used) except the source and target classes. Thus, the adversary guesses a pair of classes, constructs watermarked data following EWE, and extracts the model while using EWE with $\kappa<0$ to disentangle the purported watermark data and legitimate data from the purported target class.} \new{Following such a procedure, we observe that the watermark success of the extracted model on Fashion MNIST drops} \revision{from $48.81\%$ to $22.82\%$ if the guess does not match with the true source-target pair, and to $6.34\%$ if the guess is correct.}\delete{by nearly $2\times$, with a 6 percentage points decrease in the accuracy}. \new{On MNIST,} \delete{the decrease in accuracy is negligible but the drop of watermark success is also smaller ($\sim \hspace{-0.2mm}10$ percentage points)}\revision{ watermark success drops from $41.62\%$ to $30.14\%$ when the guess is wrong, and to $0.08\%$ otherwise}. \new{The results from the Speech Commands dataset have large variance, but follow a similar trend}\revision{: the watermark success drops to an average of $16.81\%$ due to the attack}. \new{Thus, while watermark success rates are lowered by this attack, the defender is still able to claim ownership} \revision{when the adversary guesses the source-target pair incorrectly with about 30 queries for the two vision datasets, and near 100 queries for Speech Commands. Furthermore, observe that guessing the pair of classes correctly requires significant compute to train models corresponding to the $K(K-1)$ possible source-target pairs where $K$ is the number of classes in the dataset
, which defeats the purpose of model extraction.} \delete{if they are willing to issue more queries.} \del{While the aforementioned attacks can potentially alleviate the guarantees that are provided by EWE, they are not in the scope of our threat model because they require knowledge of the trigger or parameters of EWE. In \S~\ref{section: backdoor defense}, we evaluate our proposal against techniques used to prevent backdoors (which lie in the scope of our threat model).}
\vspace{-3mm}

\paragraph{Piracy Attack.}

\begin{figure}[t]
    \centering
    \vspace{-5mm}
    \subfloat[MNIST]{\includegraphics[width=0.5\linewidth]{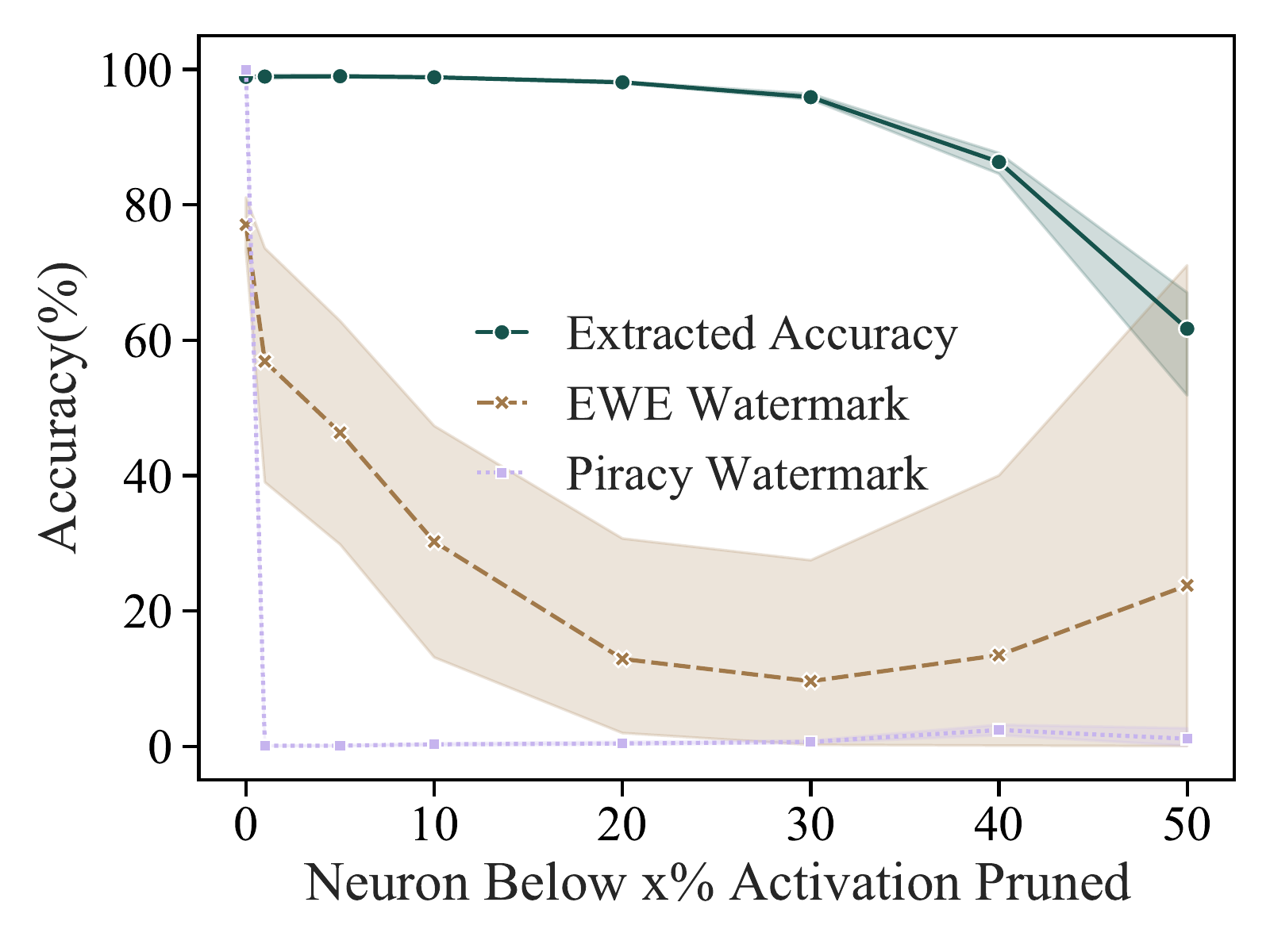}}\hfill
    \subfloat[Fashion MNIST]{\includegraphics[width=0.5\linewidth]{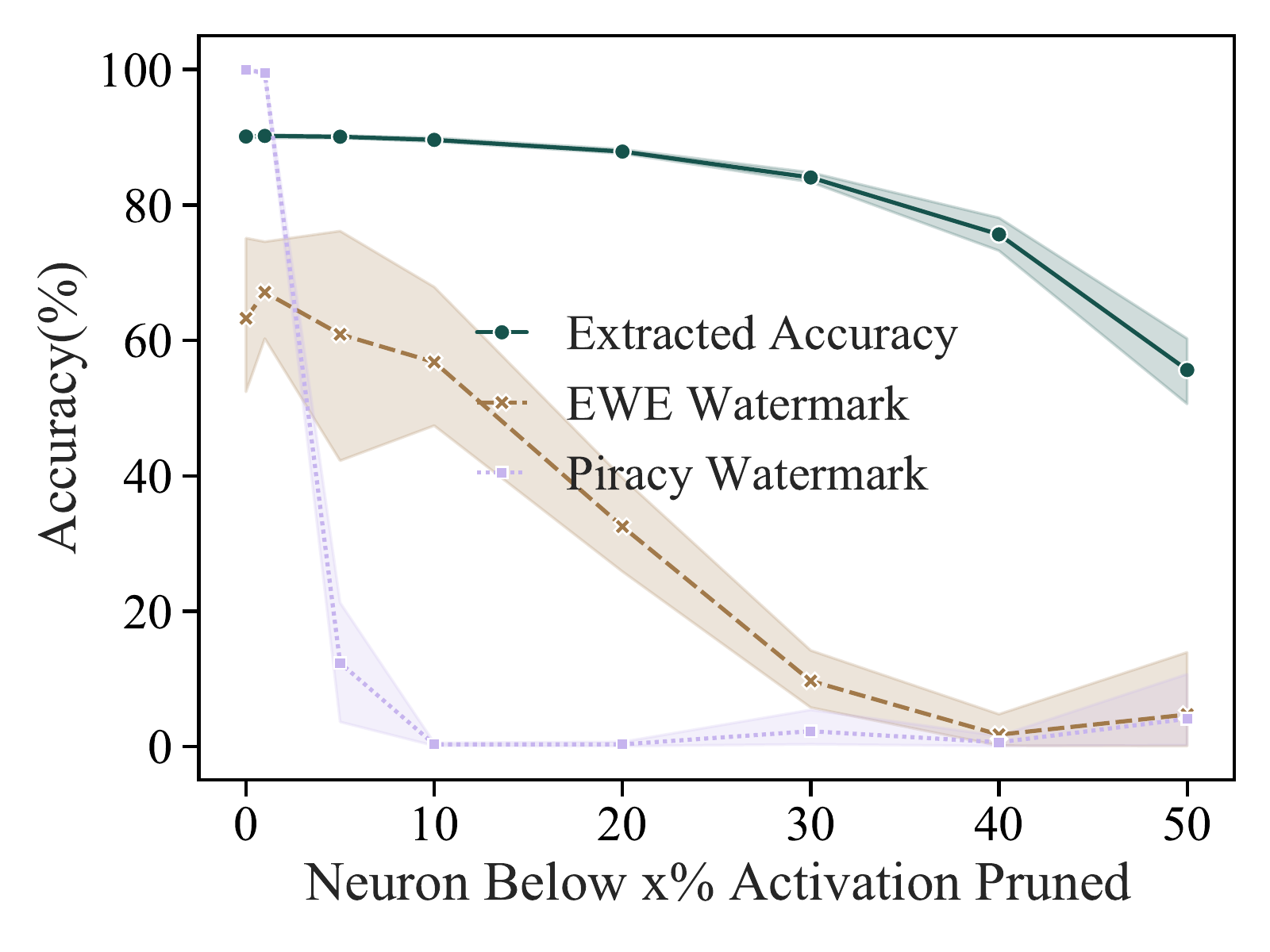}}
    \caption{\new{Task accuracy and watermark success rate after fine-pruning on the extracted model with a pirate watermark. With less than 10\% neurons pruned, the pirate watermark is removed while the owner's watermark remains.}\vspace{-5mm}}
    \label{fig:piracy}
\end{figure}

\new{In a piracy attack, the adversary embeds their own watermark with EWE so that the model is watermarked twice---it becomes ambiguous to claim ownership through watermarks. To remove the pirate watermark, we propose to fine-prune~\cite{2018arXiv180512185L} the extracted model on data labeled by the victim model. As shown in Figure~\ref{fig:piracy}, the owner's watermark is not removed as we discussed fine pruning in \S~\ref{section: backdoor defense}, whereas the pirate watermark would be removed (even if the adversary uses EWE) because data labelled by the victim model does not contain information about the pirate watermark. The adversary cannot do the same to remove the owner’s watermark because this requires access to a dataset labeled by another source, at which point the cost of piracy defeats model stealing: the adversary could have trained a model on that dataset and would not benefit from model stealing.}

\vspace{-3mm}

\paragraph{Anomaly Detection.} 
\new{Imagine the case of an extracted model deployed as an online ML API. The adversary may know (or suspect) the model to be watermarked, so they may decide to implement an anomaly detector to filter queries containing  data watermarked by EWE and respond to them with a random prediction. By doing so, even though the parameters still embed the watermarks, the adversary could still prevent the defender from claiming ownership.}

\new{We tested two common anomaly detectors on Fashion MNIST: Local Outlier Factor (LOF)~\cite{lof} and Isolation Forest~\cite{IF}, on activations of the last hidden layer. Results are shown in Table~\ref{tab:anomaly detection}. Both detectors are able to detect more than $90\%$ of watermarked data. However, this comes at the cost of identifying parts of the validation dataset as outliers and results in a sharp accuracy drop of $7.0$ and $8.64$ percentage points respectively. This may be due to the curse of dimensionality~\cite{curse_of_dimensionality}: it is harder to learn higher dimensional distribution. Indeed, it is worth noting that anomaly detectors on hidden layers consistently work better than on the inputs themselves.}

\begin{table}
    \centering
    \begin{tabular}{@{\hspace{1mm}}l@{\hspace{2mm}}*{3}c@{\hspace{0.1mm}}}
        \hline
        \textbf{Method} & \textbf{Accuracy Loss} & \textbf{Detected Watermark} \\
        \hline
		LOF & $7.00(\pm0.3)\%$ & $99.93(\pm0.03)\%$\\
		Isolation Forest &$8.64(\pm0.32)\%$ &$92.82(\pm1.32)\%$ \\
        \hline
    \end{tabular}
    \caption{\new{Proportion of watermarks detected and accuracy loss when anomaly detectors filter suspicious inputs.}\vspace{-5mm}}
    \label{tab:anomaly detection}
\end{table}

\vspace{-3mm}

\paragraph{Transfer Learning.}

\begin{figure}[t]
    \centering
    \subfloat[Finetune fully connected layers]{\includegraphics[width=0.5\linewidth]{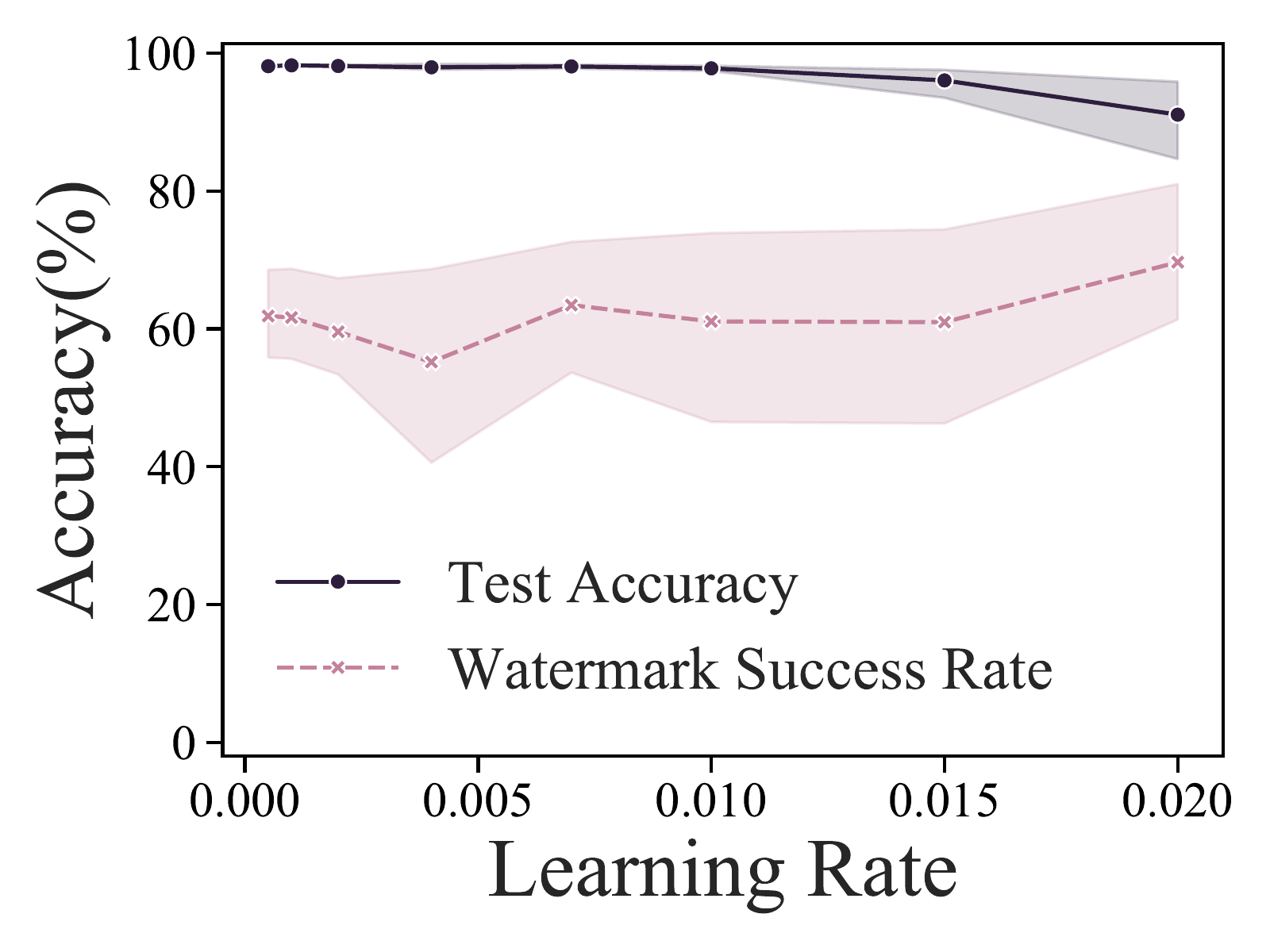}}\hfill
    \subfloat[Finetune all layers]{\includegraphics[width=0.5\linewidth]{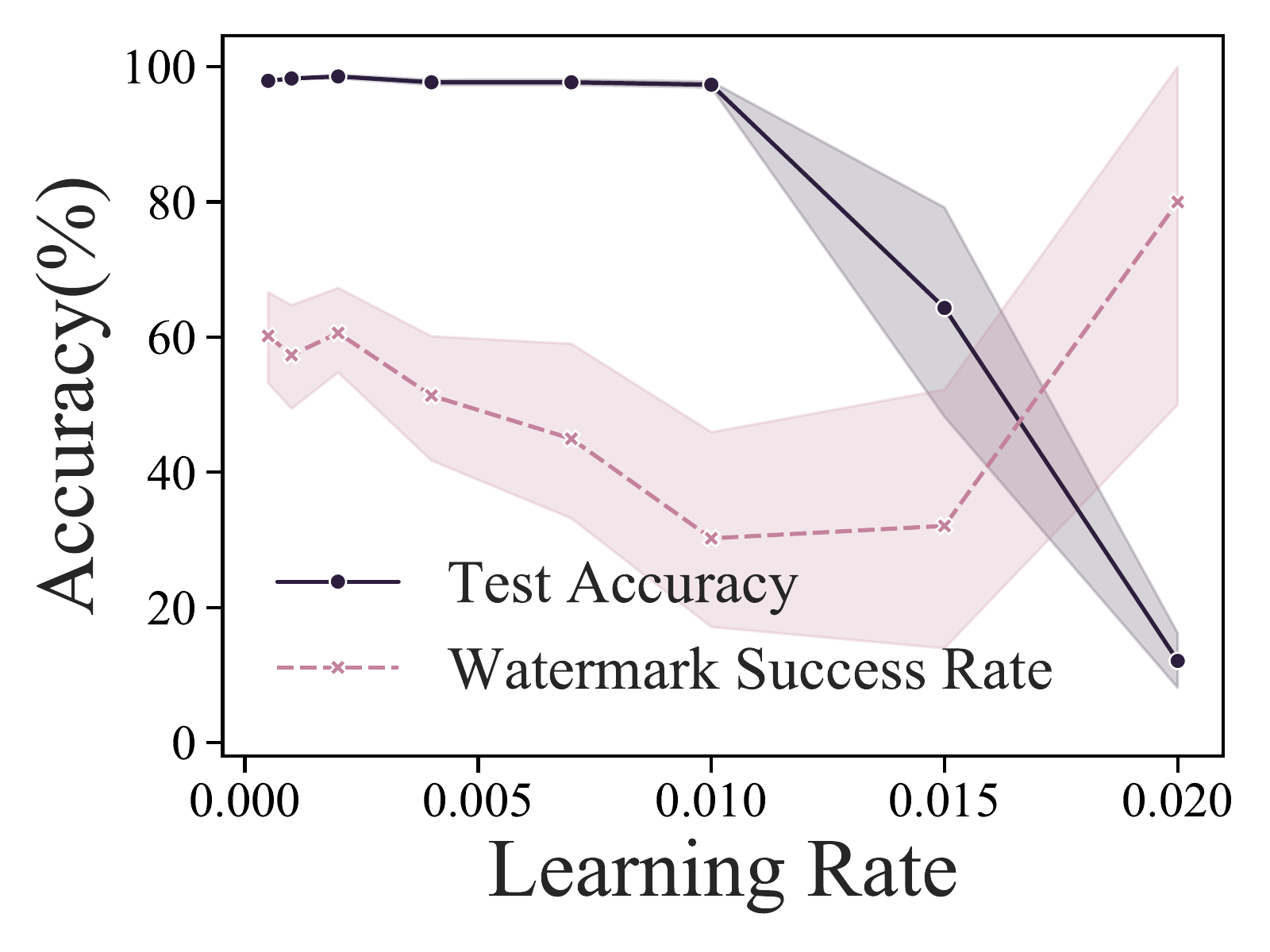}}
    \caption{\new{Task accuracy and watermark success rate of the extracted model after transfer learning from GTSRB to LISA. Even fine-tuning all the layers does not remove watermarks.}\vspace{-5mm}}
    \label{fig:transfer_learning}
\end{figure}

\new{The adversary may also transfer knowledge of the extracted model to another dataset in the same domain~\cite{transfer_learning} with the hope of disassociating the model from EWE's watermark distribution. To evaluate if  watermarks persist after transfer learning, we chose two datasets in the same domain. The victim model is trained on the German Traffic Sign Dataset (GTSRB)~\cite{gtsrb} and we transferred the extracted model to the LISA Traffic Sign Dataset~\cite{lisa}. We  fine-tune either (a) only the fully connected layers, or (b) all layers for the same number of epochs that the victim model was trained for. Before we verify the watermark, the output layer of the transferred model is replaced to match the dimension of the victim model (they may differ)~\cite{2018arXiv180204633A}.}

\new{As shown in Figure~\ref{fig:transfer_learning}, (a) achieves an accuracy of up to $98.25\%$ but leaves the watermark unaffected; (b) reaches an accuracy of $98.56\%$ and begins to weaken the watermark as one increases the learning rate. However, the pretrained knowledge is lost due to large learning rate values before the watermark is removed. This is consistent with observations in prior work~\cite{2018arXiv180204633A}. We also note that transfer learning requires that the adversary have access to additional training data and perform more training steps, so it is expected that our ability to claim model ownership will be weaker.}
\vspace{-4mm}
\paragraph{Take-away.} \new{The adversary also faces a no free lunch situation. They cannot adapt with disentanglement, piracy, anomaly detection, or transfer learning, and remove EWE watermarks, unless they sacrifice the stolen model's utility.}
\vspace{-3mm}   
\section{Discussion}

\vspace{-3mm}
\paragraph{Hyperparameter Selection.} Our results suggest that the watermarking survivability comes at a nominal cost (about \new{0.81}\% in accuracy degradation). Yet, this value varies depending on the dataset and the hyperparameters used for training\del{; without careful selection of the latter, the accuracy degradation can be more severe} \new{(which themselves also depend on the dataset) as we explore in Appendix~\ref{app: hyper}}. Determining the relationship with relevant properties of the dataset is future work. 

\vspace{-3mm}
\paragraph{Computational Overheads.} Our experiments suggest that the size of the watermarked dataset should be $2\times$ less than the size of the legitimate dataset. However, this implies that the model is now trained on $1.5-2\times$ more data than before. While this induces additional computational overheads, we believe that the trade-offs are advantageous in terms of proving ownership. A more detailed analysis is required to understand if the same phenomenon exists for more complex tasks with larger datasets. 

\vspace{-3mm}
\paragraph{Improving Utility.} EWE utilizes the SNNL to mix representations from two different distributions; this ensures the activation patterns survive extraction. However, this is at a nominal expense to the utility; for certain applications, such a decrease in utility (even if small) is not desired. We believe that the same desired properties could be more easily achieved if one were to replace ReLU activations with the smoother Sigmoid activations while computing the SNNL. %

\vspace{-3mm}
\paragraph{Algorithmic Efficiency.} In Algorithm \ref{alg: EWE}, we modified the loss function by computing the SNNL at every layer of the DNN. However, it may not be necessary to do so. In Figure \ref{fig:activation}, we plot the activation patterns of hidden layers of a model trained using EWE; we observe that adding the SNNL to just the last layers provides the desired guarantees. Additionally, we observe a slight increase in model utility when not all layers are entangled. A detailed understanding of how one can choose the layers is left to future work. 
\vspace{-3mm}
\paragraph{Scalability and Future Research Directions.}
\revision{As mentioned in \S~\ref{section: scalability}, EWE suffers in terms of trade-off between model performance and watermark robustness when we scale to deeper architectures, and more complex datasets. Given the results on CIFAR-100, more work may be needed to scale the current method to larger datasets.  According to Figure~\ref{fig:num_class} (in Appendix~\ref{app: figures}), the performance of EWE is impacted by the number of classes. We suspect this may be due to the representation space being more complicated (i.e. there are more clusters), making it more difficult to entangle two arbitrarily chosen clusters. Thus, a potential next step would be to investigate the interplay between the design of triggers to control the cluster of watermarked data; and the similarity structures and orientation of the representation space to choose source and target classes accordingly.}

\revision{Another possible improvement is to use $m$-to-$n$ watermarking. In this work, we focused on 1-to-1 watermarking, which watermarks one class of data and entangles it with another class. However, as long as the watermarked model behaves significantly differently from a clean model, the model owner could choose to watermark $m$ classes of data, entangle them with $n$ other classes, and claim ownership by following the similar verification process as described in \S~\ref{section: verify}.}

\vspace{-3mm}
\section{Conclusions}
\vspace{-2mm}
We proposed Entangled Watermark Embedding (EWE), which forces the model to entangle representations for legitimate task data and watermarks. Our mechanism formulates a new loss involving the Soft Nearest Neighbors Loss, which when minimized increases entanglement. Through our evaluation on tasks from the vision and audio domain, we show that EWE is indeed robust to not only model extraction attacks, but also \new{piracy attacks, anomaly detection, transfer learning, and} efforts used to mitigate backdoor (poisoning) attacks. All this is achieved while preserving watermarking accuracy, with (a) a nominal loss in classification accuracy, and (b) $1.5-2\times$ increase in computational overhead. Scaling EWE to complex tasks without great accuracy loss remains as an open problem.
\vspace{-4mm}

\vspace{-2mm}
\section*{Acknowledgments}
\vspace{-3mm}
The authors would like to thank Carrie Gates for shepherding this paper. This research was funded by CIFAR, DARPA GARD, Microsoft, and NSERC. VC was funded in part by the Landweber Fellowship. \vspace{-2mm}

\bibliographystyle{plain}
\bibliography{reference}

\appendix
\section{Appendix}\label{app:appendix}
\vspace{-2mm}
\subsection{Finetuning the hyperparameters of EWE}
\label{app: hyper}
Next, we dive into details of each hyperparameter of EWE and perform an ablation study.

\begin{figure}[h]
    \centering
    \vspace{-5mm}
    \subfloat[Fashion MNIST]{\includegraphics[width=0.45\linewidth]{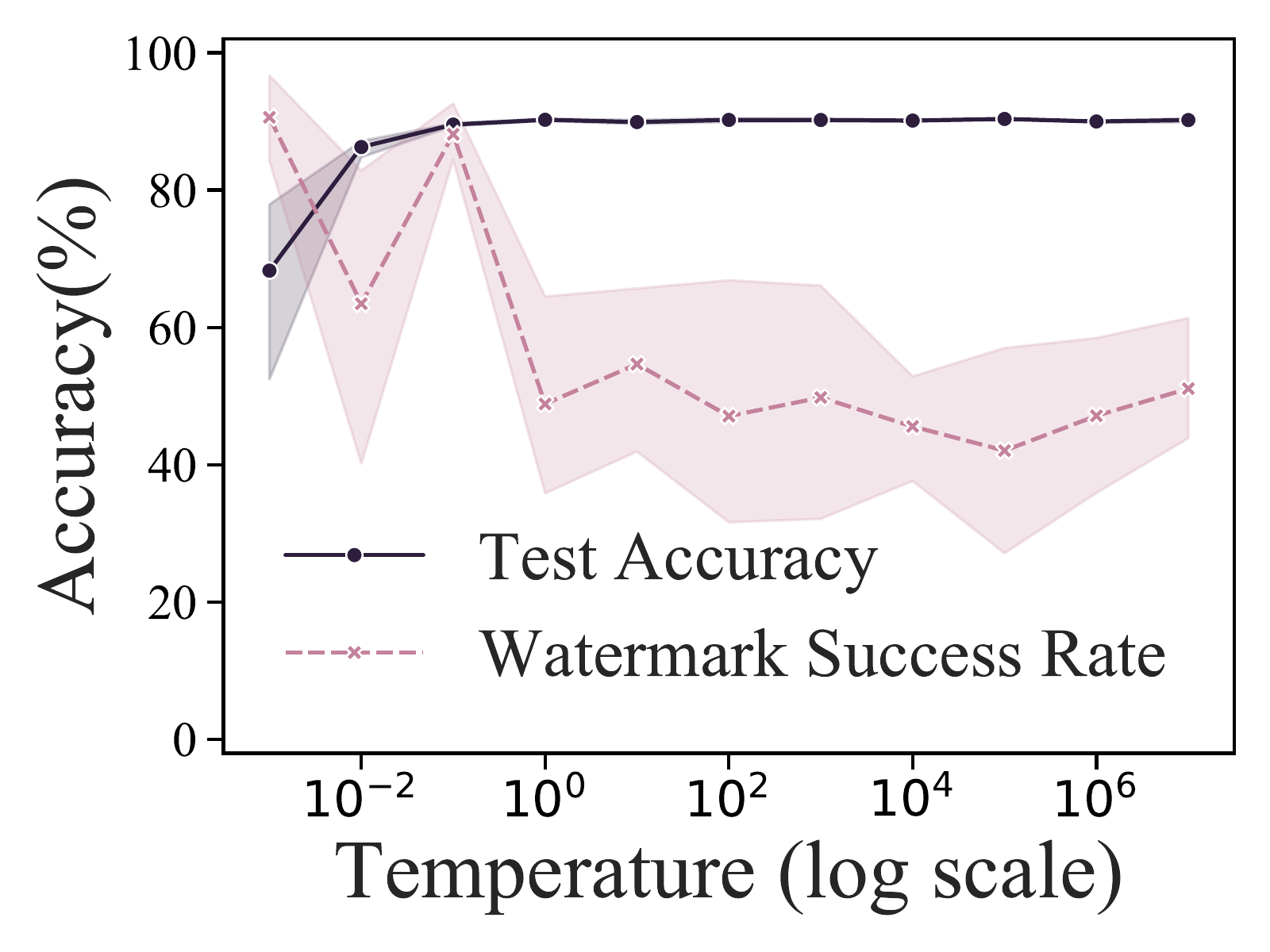}}\hfil
    \subfloat[Speech Commands]{\includegraphics[width=0.45\linewidth]{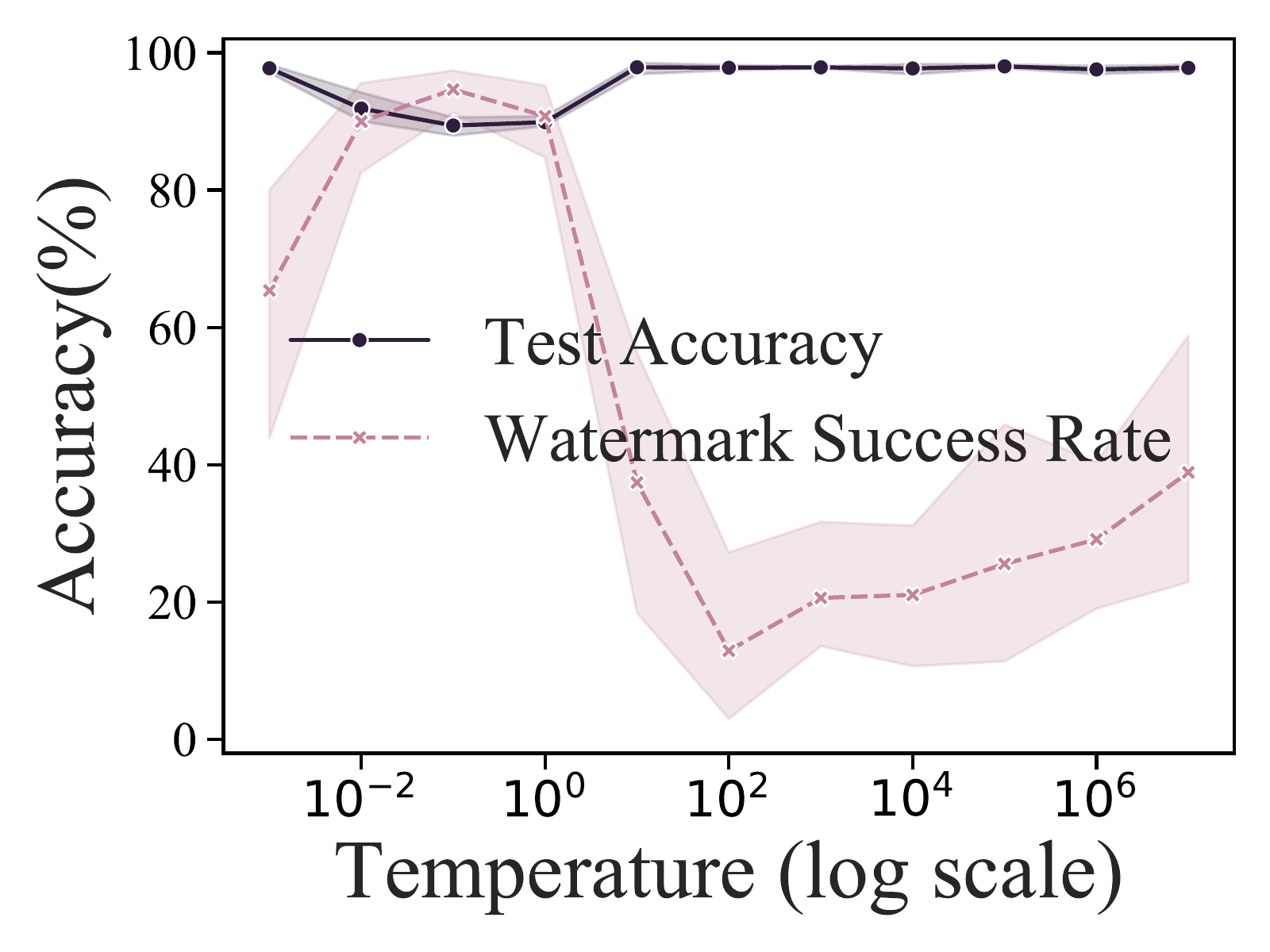}}\hfil
    \vspace{-3mm}
    \caption{ \del{Task accuracy and watermark success rate are not impacted by the choice of temperature in EWE.}\new{EWE is unlikely to fail due to setting the temperature, but certain initialization of temperature does lead to better trade-off between task accuracy and watermark success rate. Note the temperature is plotted on log scale.\vspace{-4mm}}}
    \label{fig:temperature}
\end{figure}

\vspace{-4mm}
\paragraph{Temperature.} Temperature is a hyperparameter introduced by Frosst et al~\cite{2019arXiv190201889F}. It could be used to control which distances between points are more important: at small temperatures, small distances matter more than at high temperatures, where large distances matter most. In our experiments, we found that the influence of temperature on the robustness of watermark is not significant\new{: a nice initialization leads to high watermark success, whereas other initialization results in watermark success high enough for claiming ownership}, as shown in Figure~\ref{fig:temperature}. We conjecture that this is because EWE fine-tunes the temperature by gradient descent during training (see the last line of Algorithm~\ref{alg: EWE}).

\begin{figure}[t]
    \centering
    \vspace{-3mm}
    \subfloat[Fashion MNIST]{\includegraphics[width=0.45\linewidth]{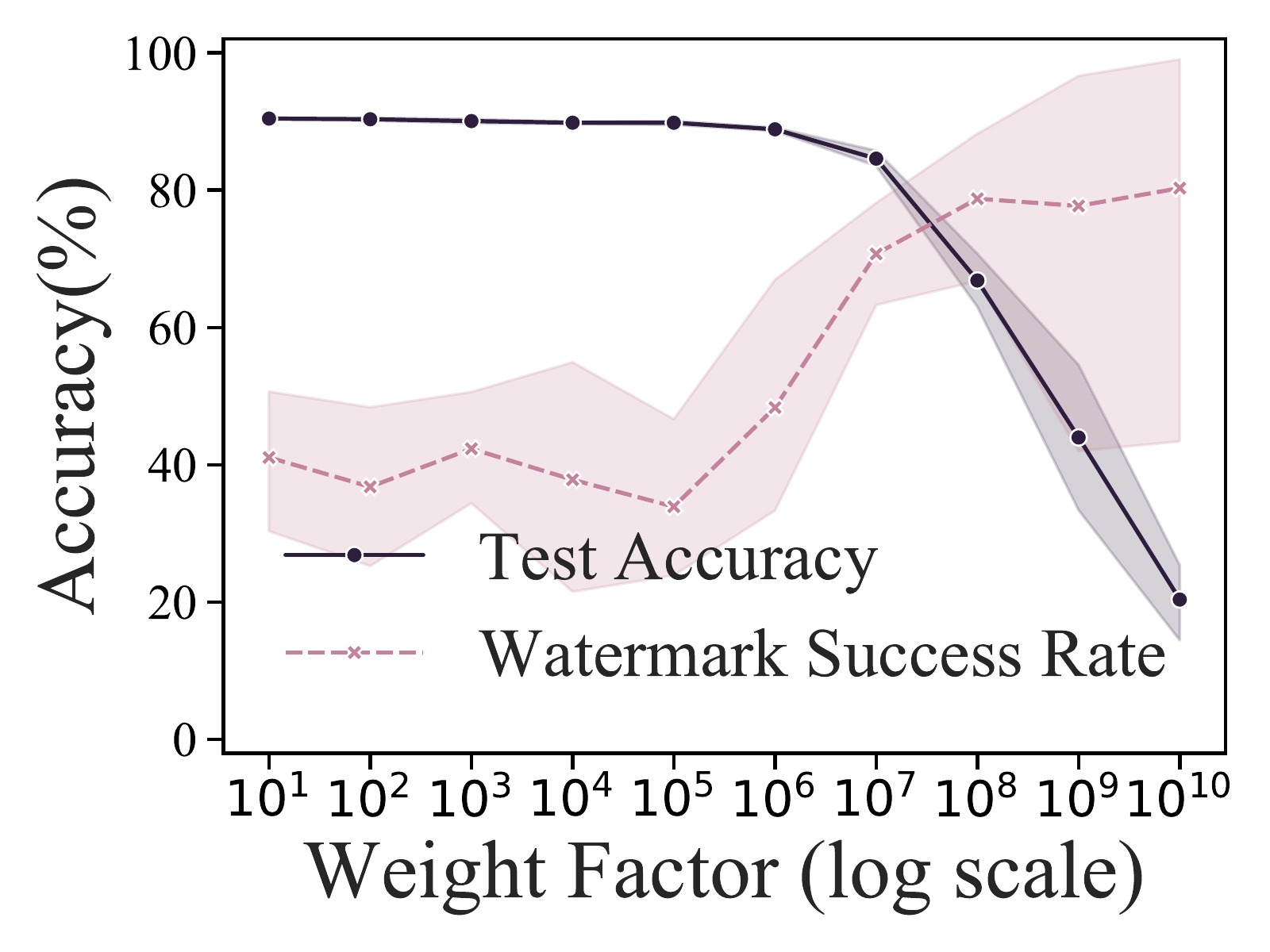}}\hfil
    \subfloat[Speech Commands]{\includegraphics[width=0.45\linewidth]{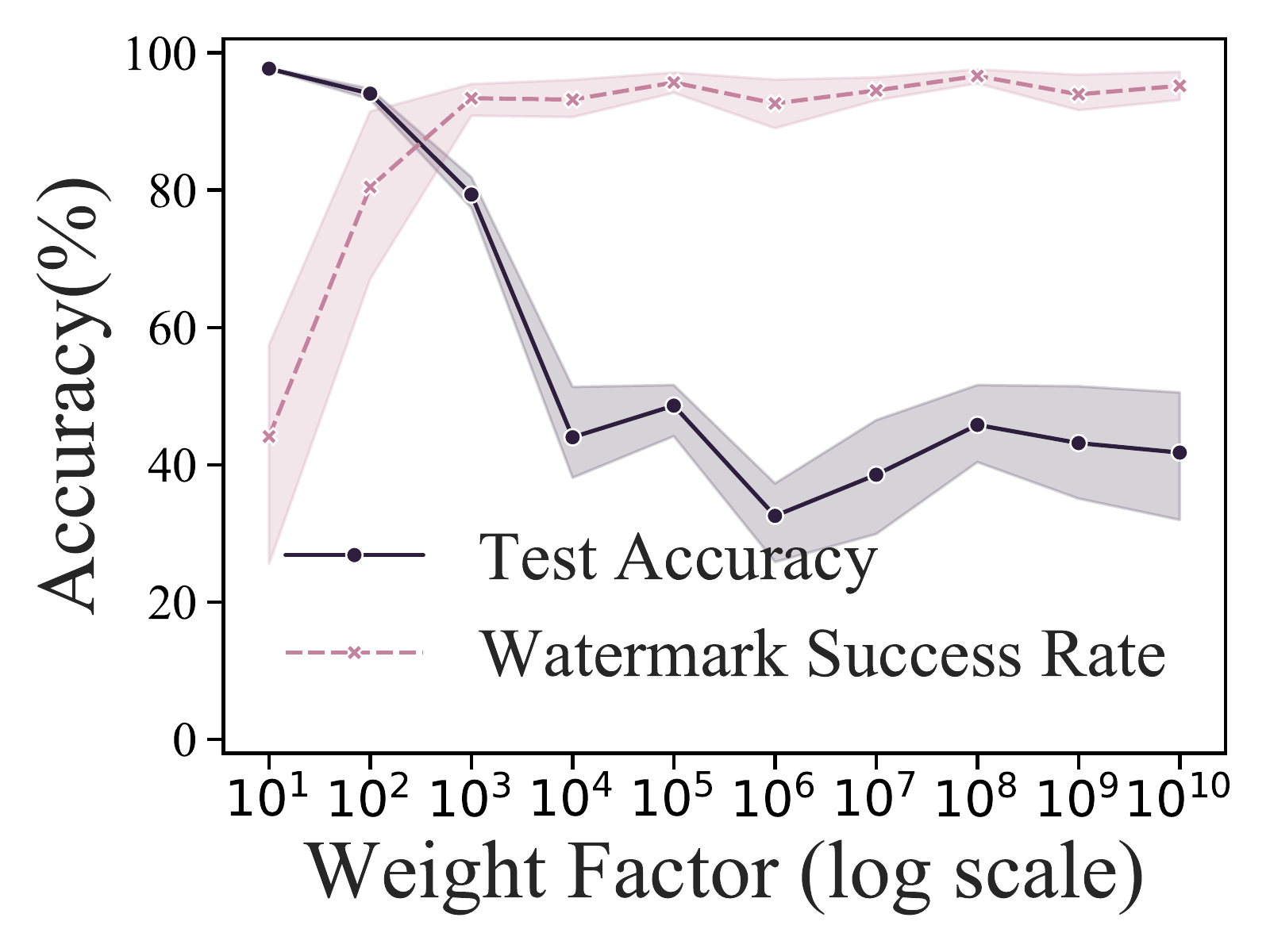}}\hfil
    \vspace{-3mm}
    \caption{ Increasing the absolute value of the weight factor \del{$w$}\new{$\kappa$} promotes watermark success rate (more importance is given to the SNNL) at the expense of lower accuracy on the task. \new{Note that $\kappa$ is plotted on log scale.}\vspace{-4mm}}
    \label{fig:factor}
\end{figure}

\vspace{-4mm}
\paragraph{Weight Factor.} As defined in Algorithm~\ref{alg: EWE}, the loss function is the weighted sum of a cross entropy term and SNNL term. The weight factor \del{$w$}\new{$\kappa$} is a hyper-parameter that controls the importance of learning the watermark task (by maximizing the SNNL) relatively to the classification task (by minimizing cross entropy loss). As shown in Figure~\ref{fig:factor}, factors larger in magnitude \del{(they are negative since we would like to maximize the SNNL)} cause the watermark to be more robust, at the expense of performance on the task. At the \del{right-hand}\new{left-hand} side of the figure, with a weight factor \del{of -2}\new{in the magnitude of 10}, the accuracy is \del{about 99\%} \new{similar to an un-watermarked model,} while watermark success is about \del{10\%}\new{40\%}. In contrast, when the weight factor is \del{-128}\new{getting larger}, watermark success \new{approaches to $100\%$ but the accuracy decreases significantly.}\del{increases by about 50\% but the accuracy decreases to 94\%}.

\begin{figure}[t]
    \centering
    \vspace{-1mm}
    \subfloat[Fashion MNIST]{\includegraphics[width=0.45\linewidth]{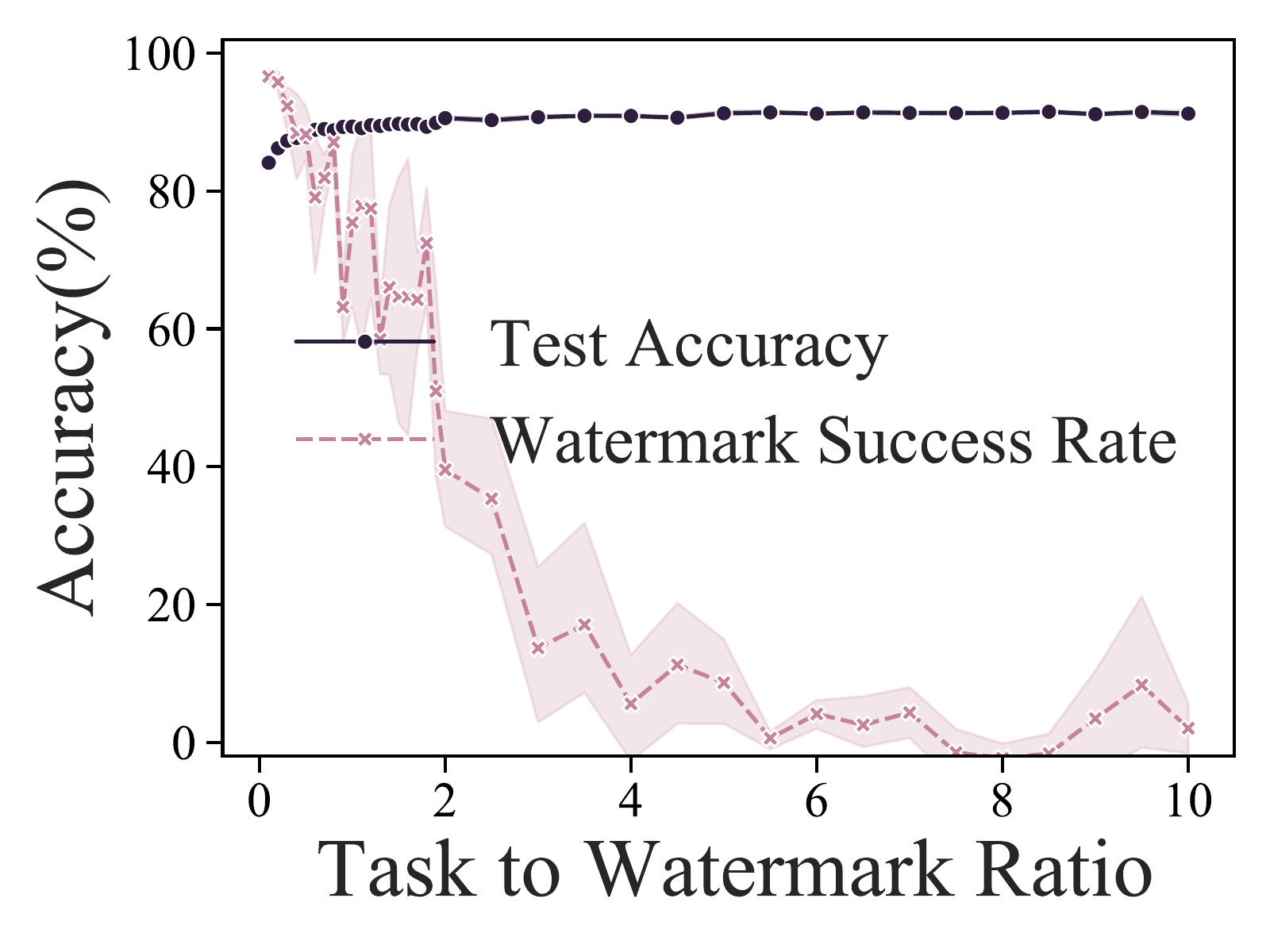}}\hfil
    \subfloat[Speech Commands]{\includegraphics[width=0.45\linewidth]{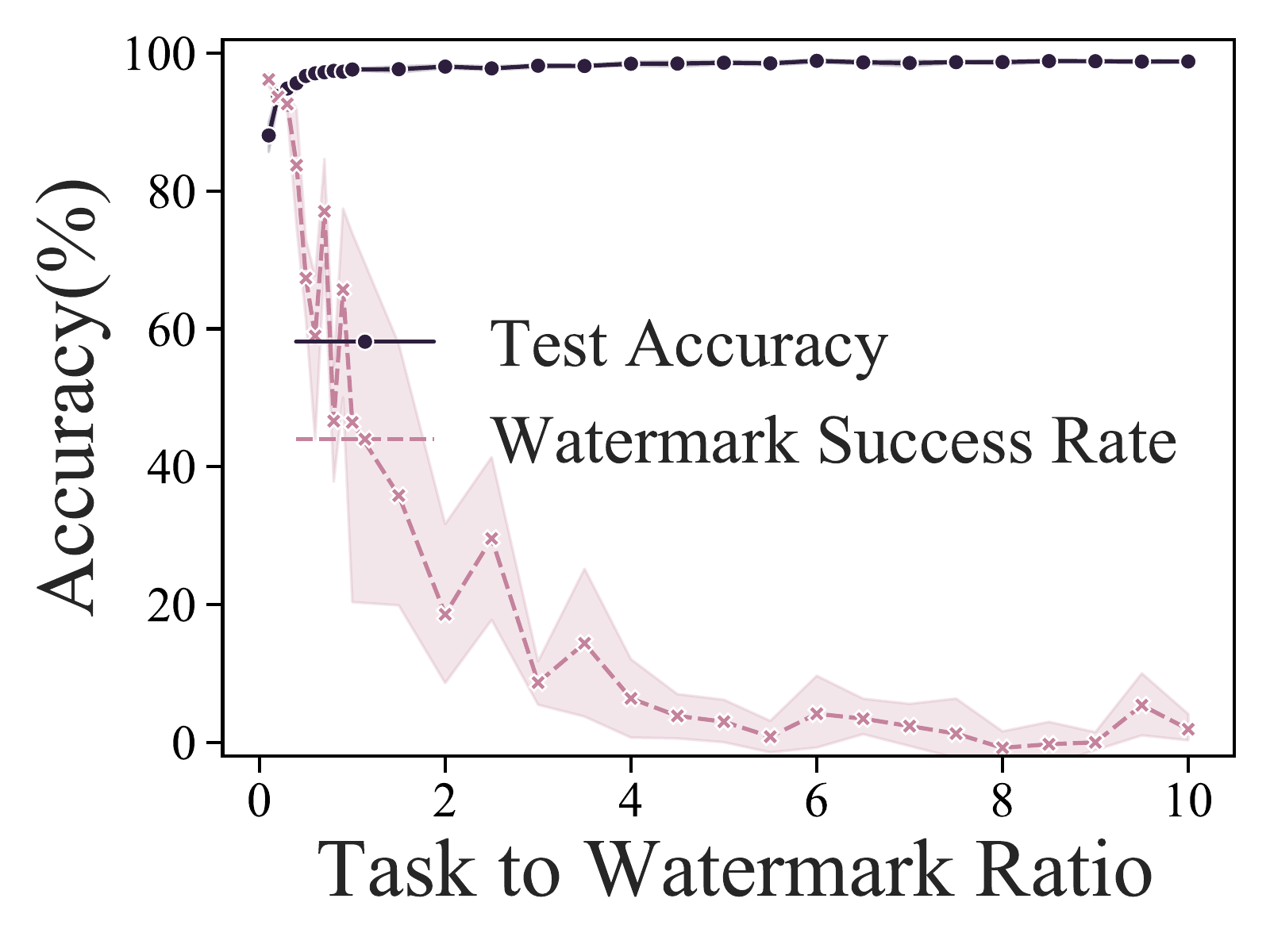}}\hfil
    \vspace{-3mm}
    \caption{ Decreasing the ratio $r$ of task data to watermarks promotes watermark success rate (more importance is given to the SNNL) at the expense of lower accuracy on the task.\vspace{-4mm}}
    \label{fig:n_w_ratio}
\end{figure}

\vspace{-4mm}
\paragraph{Ratio of task data to watermarks.} Denoted by $r$ in Algorithm \ref{alg: EWE}, this ratio also influences the trade-off between task accuracy and watermark robustness. In Figure~\ref{fig:n_w_ratio}, we observe that lower ratios yield more robust watermarks. For instance, we found for Fashion MNIST that the watermark could be removed by model extraction if the ratio is greater than 3, whereas task accuracy drops significantly for ratios below 1. 

\label{section: choose watermark}

\begin{figure}[t]
    \centering
    \vspace{-6mm}
    \subfloat[Fashion MNIST]{\includegraphics[width=0.45\linewidth]{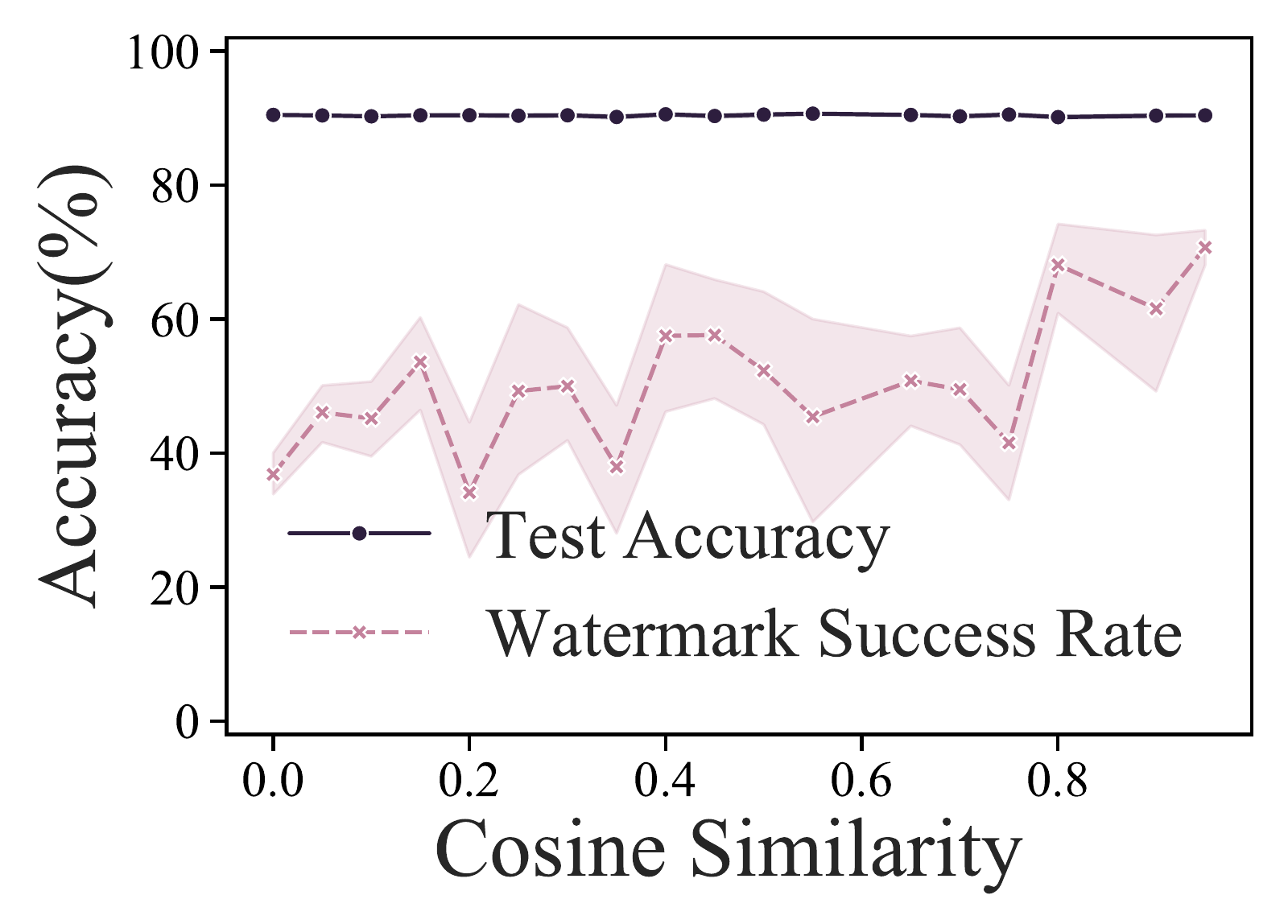}}\hfil
    \subfloat[Speech Commands]{\includegraphics[width=0.45\linewidth]{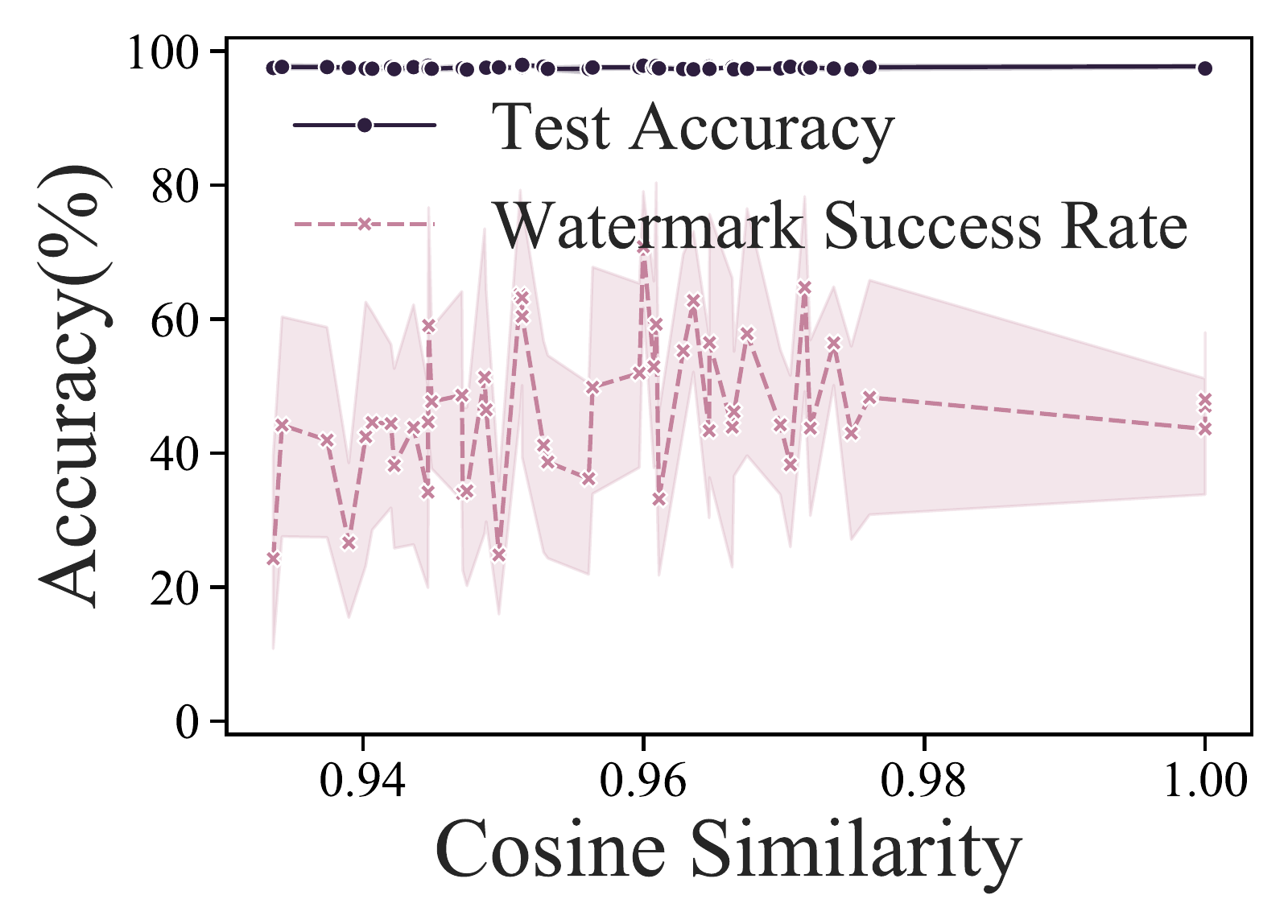}}\hfil
    \vspace{-3mm}
    \caption{ Impact of similarity of classes on robustness of watermarks: We computes the average cosine distances between data of different pairs of classes and use them as source and target classes to watermark the model. It could be seen that similar classes lead to higher watermark success \new{on Fashion MNIST, but no clear trend is observed for Speech Command}.\vspace{-5mm}}
    \label{fig:cluster}
\end{figure}
\vspace{-4mm}
\paragraph{Source-Target classes}

Source and target classes are denoted by $c_S$ and $c_T$ in Algorithm \ref{alg: EWE}. \new{Note that we use OOD watermarks (data from MNIST) for Fashion MNIST, so $c_S$ refers to a class of MNIST.} We name class center the average of data from each class. In Figure \ref{fig:cluster}, we plot the performance of EWE with respect to the cosine similarity among centers of different source-target pairs \new{(detailed performance of different pairs can be found in Figure~\ref{fig:cm} in the appendix)}. 

Classes with \del{closer centers}\new{similar structures} enable more robust watermarks at no impact on task accuracy. This is because data from similar classes is easier to entangle (i.e. the SNNL is easier to maximize). \new{Cosine similarity between class centers is a heuristic to estimate this and its effectiveness depends on the dataset. For Fashion MNIST, one could observe a trend that higher cosine similarity leads to more robust watermarks. Instead, the difference among classes are less significant in Speech Command so this heuristic may not be useful.}

\del{The importance of this choice highly depends on the dataset. For Fashion MNIST, even the worse pairs (shown in left-hand side of Figure~\ref{fig:cluster} (a)) have a reasonably good performance. Instead, only the closest pair lead to watermark success above 40\% on the Speech Commands dataset (refer Figure~\ref{fig:cm} in the Appendix for more details).}

\subsection{Evasion Attacks for Detection}
\label{app:adv_example_for_ewe}
Adversarial examples (or samples) are created by choosing samples from a source class and perturbing them slightly (adding a carefully crafted perturbation) to ensure targeted (the mistake is chosen) or untargeted (the mistake is any incorrect class) misclassification. To do so, some attacks use gradients~\cite{2015arXiv151107528P,pgd,madry} or pseudo-gradients~\cite{spsa} to create adversarial samples with minimum perturbation. We wish to understand if mechanisms used to generate adversarial samples can be used to detect watermarks, as both produce the same effect (targeted misclassification). \new{The intuition is that if one adversarial examples are generated from blank input and perturbed to the target class, they may reveal some information about the watermarked data.} To this end, we utilize the approach proposed by Papernot et al.~\cite{2015arXiv151107528P} on the extracted model to generate adversarial examples, and compare \del{the intensities of those pixels affected by the perturbation with pixels used to encapsulate a trigger (of size 9 pixels)}\new{them with the watermarked data} generated by EWE. Examples of watermarked data and adversarial samples we generated are shown in Figure~\ref{fig:adversarial_sample} \del{(a)}\new{b} and \del{(b)}\new{(c)} respectively. \del{The intensity computed at the pixels of triggers is about zero.}\new{The average cosine similarity between the adversarial examples and watermarked data is about 0.3, whereas it could reach about 0.4 when comparing to a uniformly distributed random input of the same size.} Thus, mechanisms used to generate adversarial samples are unable to detect watermarks generated by EWE.
\vspace{-3mm}

\subsection{Additional Figures}
\label{app: figures}
\begin{figure}[h!]
    \centering
    \vspace{-7mm}
    \subfloat[Proper trigger]{\includegraphics [width=0.5\linewidth]{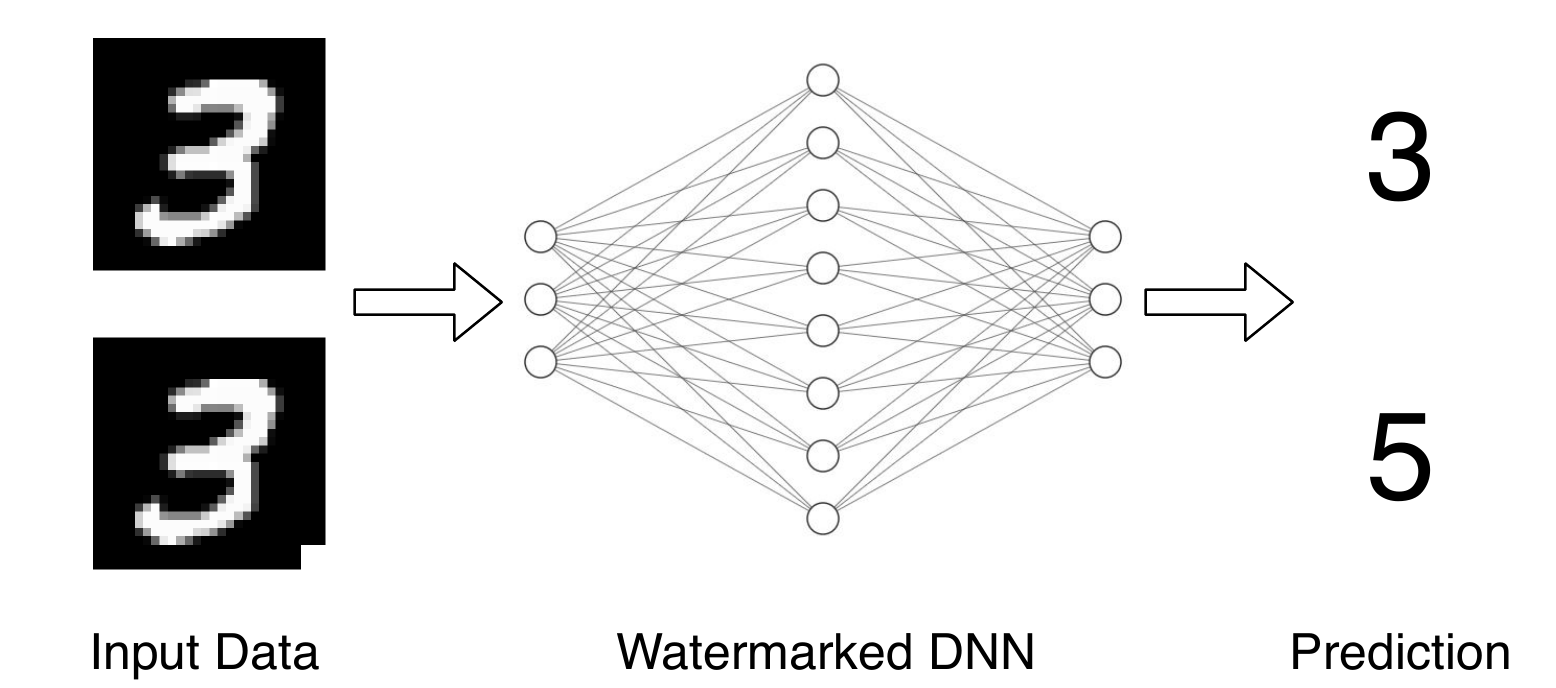}}
    \hfill
    \subfloat[Improper trigger]{\includegraphics [width=0.5\linewidth]{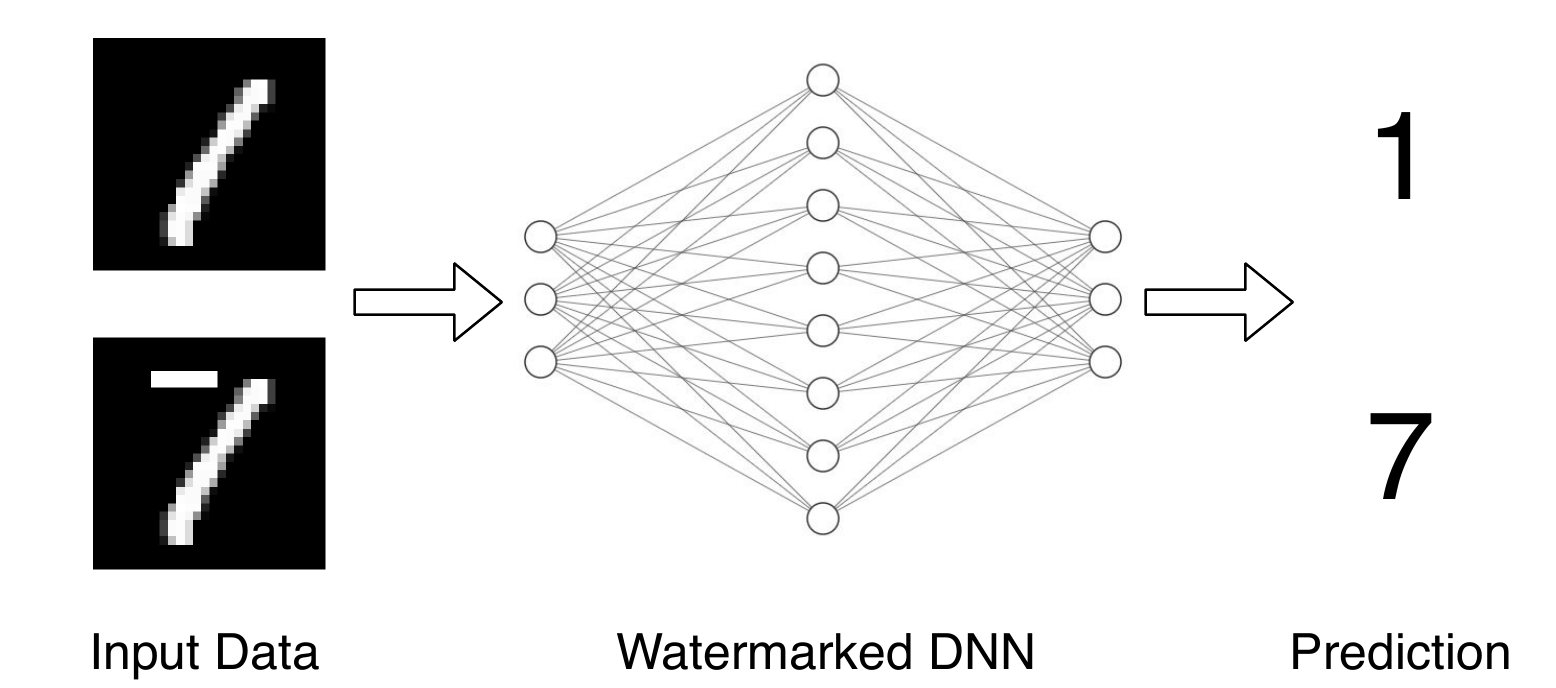}}
    \vspace{-2mm}
    \caption{ \new{(a) In this Watermarked DNN, a small white square is designed as a special trigger. If this square is added to the corner of a digit-3, the input would be predicted as a digit-5 by the DNN, whereas a normal model would classify it as a digit-3 mostly. (b) This is an example of improperly designed trigger. By adding such a rectangle to top of 1's, even a un-watermarked model would classify it as a digit-7, so it is hard to tell if a model is watermarked or not by such a trigger.
    \vspace{-5mm}}}
    \label{fig:watermark demo}
\end{figure}

\begin{figure}[h!]
    \centering
    \vspace{-3mm}
    \includegraphics[width=0.4\linewidth]{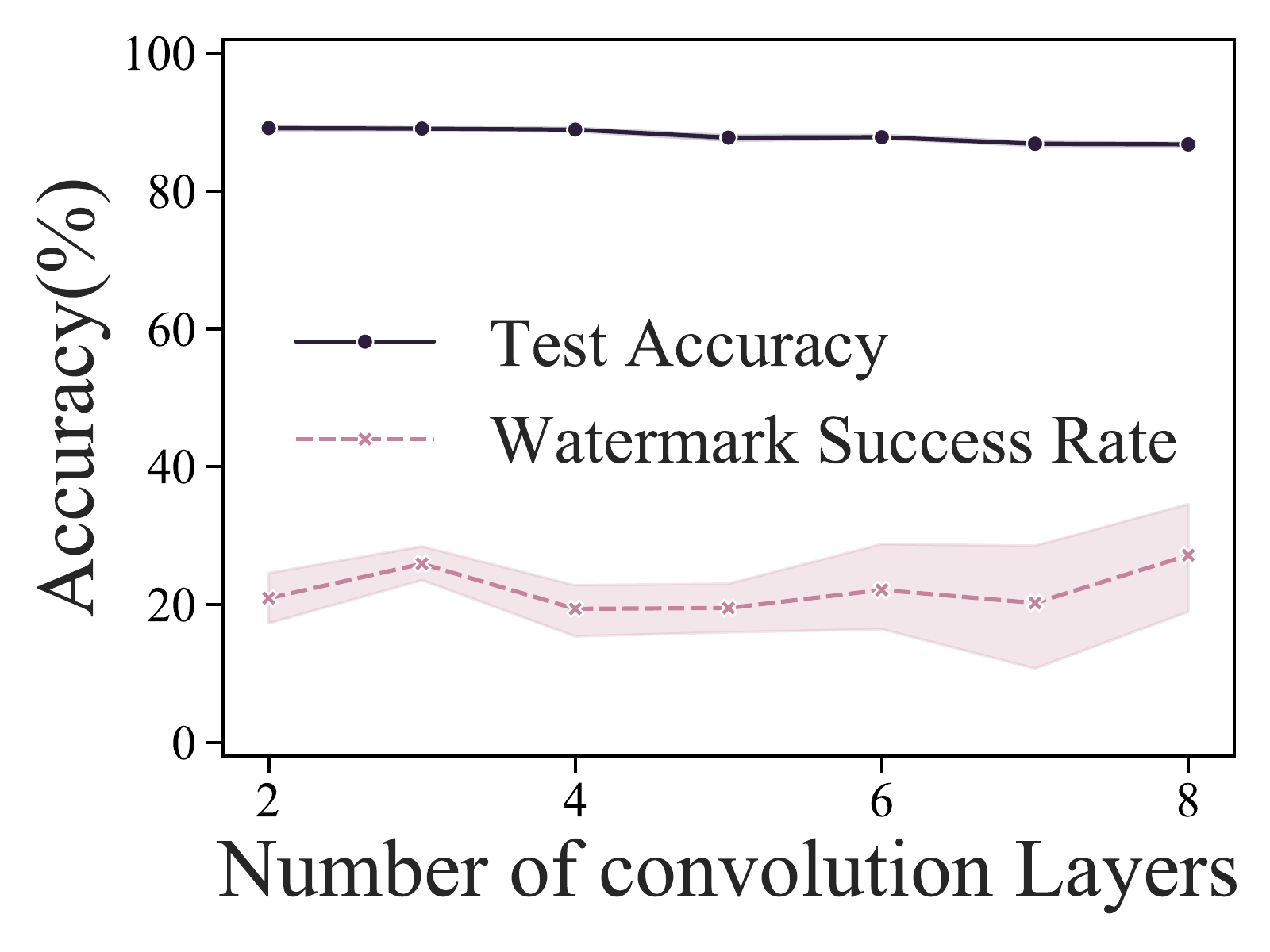}
    \vspace{-2mm}
    \caption{ \new{Validation Accuracy and Watermark success while increasing the number of convolution layers in a Fashion MNIST model without residual connection. Note that in-distribution watermark is used here.}\vspace{-4mm}}
    \label{fig:num_conv}
\end{figure}

\begin{figure}[h!]
    \centering
    \subfloat[First Convolution Layer: Legitimate Data]{\includegraphics[width=0.99\linewidth]{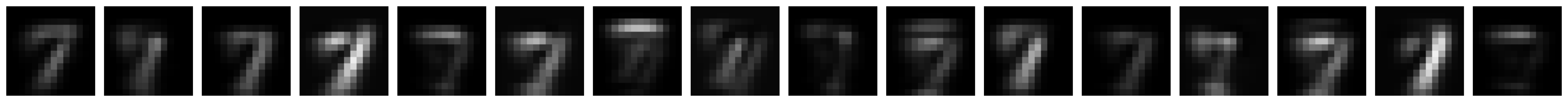}}\hfil
    \vspace{-3mm}
    \subfloat[First Convolution Layer: Watermarked Data]{\includegraphics[width=0.99\linewidth]{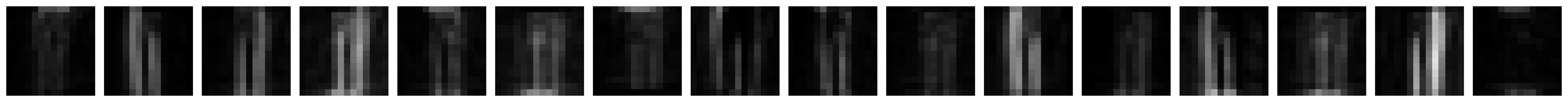}}\hfil
    \vspace{-3mm}
    \subfloat[Second Convolution Layer: Legitimate Data]{\includegraphics[width=0.99\linewidth]{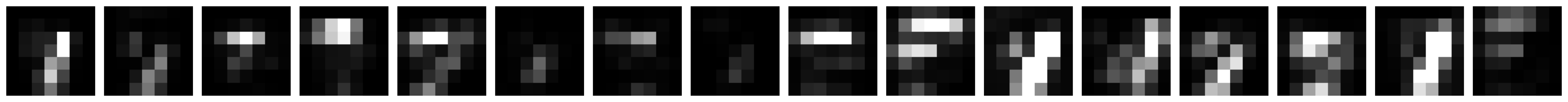}}\hfil
    \vspace{-3mm}
    \subfloat[Second Convolution Layer: Watermarked Data]{\includegraphics[width=0.99\linewidth]{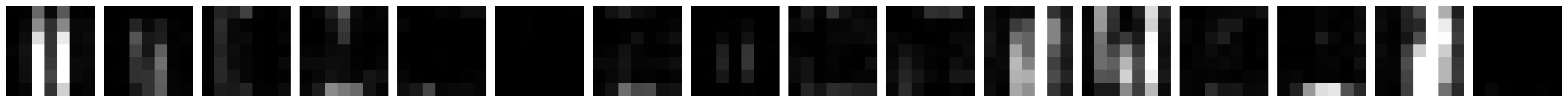}}\hfil
    \vspace{-3mm}
    \subfloat[Fully Connected Layer: Legitimate Data]{\includegraphics[width=0.99\linewidth]{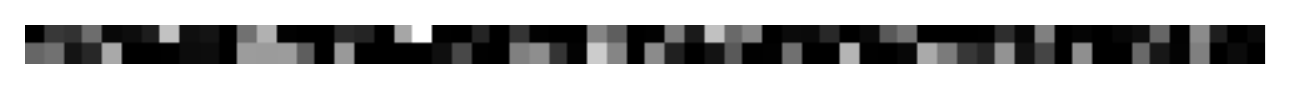}}\hfil
    \vspace{-3mm}
    \subfloat[Fully Connected Layer: Watermarked Data]{\includegraphics[width=0.99\linewidth]{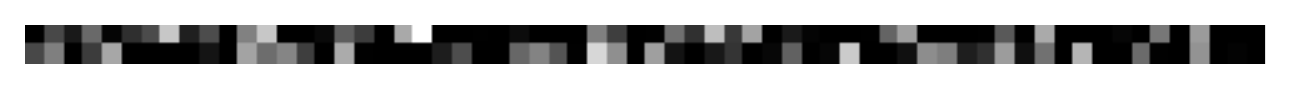}}\hfil
    \vspace{-2mm}
    \caption{ Activations of a convolutional neural network. We train a DNN with 2 convolution layers and 2 fully connected layers with EWE. We show here the frequency of activations for neurons in all hidden layers: high frequencies correspond to white color. One can observe that by entangling legitimate task data and watermarks, their representation becomes very similar, as we go deeper into the model architecture.\vspace{-4mm}}
    \label{fig:activation}
\end{figure}

\begin{figure}[h!]
    \centering
    \subfloat[First Convolution Layer: Legitimate Data]{\includegraphics[width=0.99\linewidth]{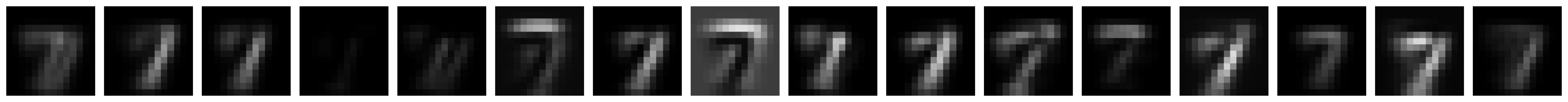}}\hfil
    \vspace{-3mm}
    \subfloat[First Convolution Layer: Watermarked Data]{\includegraphics[width=0.99\linewidth]{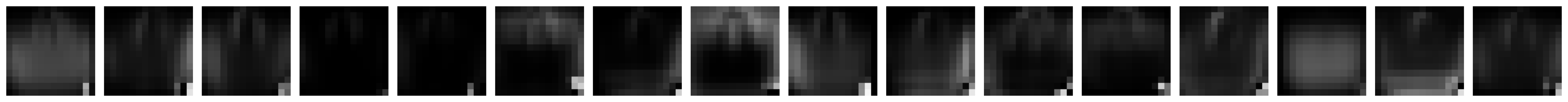}}\hfil
    \vspace{-3mm}
    \subfloat[Second Convolution Layer: Legitimate Data]{\includegraphics[width=0.99\linewidth]{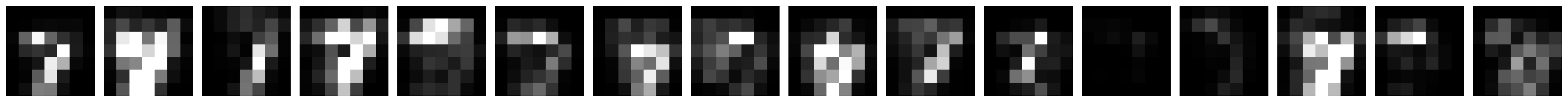}}\hfil
    \vspace{-3mm}
    \subfloat[Second Convolution Layer: Watermarked Data]{\includegraphics[width=0.99\linewidth]{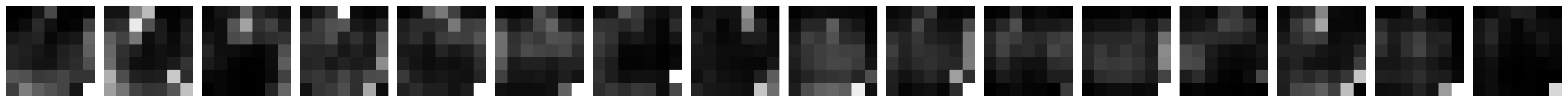}}\hfil
    \vspace{-3mm}
    \subfloat[Fully Connected Layer: Legitimate Data]{\includegraphics[width=0.99\linewidth]{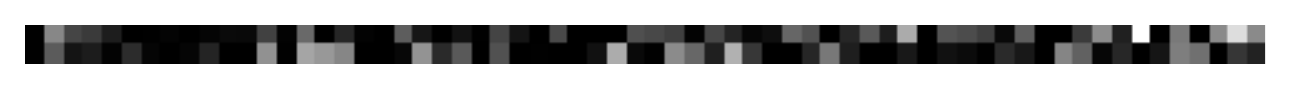}}\hfil
    \vspace{-3mm}
    \subfloat[Fully Connected Layer: Watermarked Data]{\includegraphics[width=0.99\linewidth]{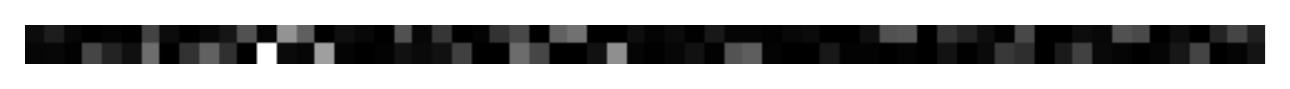}}\hfil
    \vspace{-2mm}
    \caption{ \new{This should be compared to Figure~\ref{fig:activation}. It is repeated here on a model with the same architecture but watermarked by the baseline. One can observe that the difference between activation of watermarked and legitimate data is more significant when EWE is not used.}\vspace{-5mm}}
    \label{fig:baseline_activation}
\end{figure}

\begin{figure}[t]
    \centering
    \vspace{-11mm}
    \subfloat[Un-watermarked Model]{\includegraphics[width=0.5\linewidth]{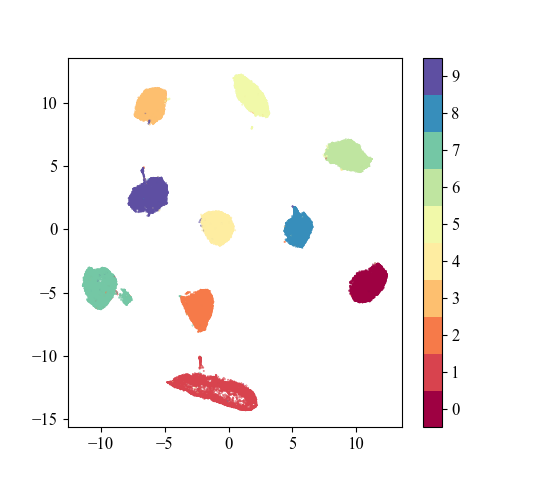}}\hfil
    \subfloat[Watermarked Model (Baseline)]{\includegraphics[width=0.5\linewidth]{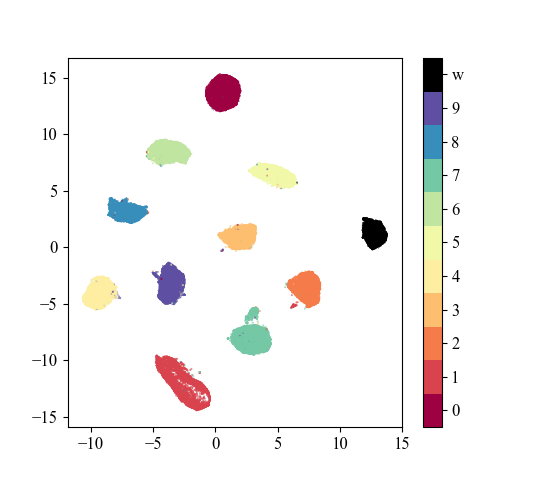}}
    \vspace{-4mm}
    \subfloat[EWE In-distribution Watermark]{\includegraphics[width=0.5\linewidth]{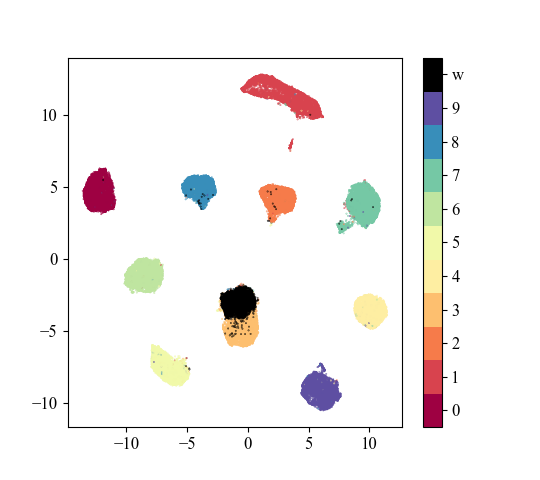}}
    \subfloat[EWE Out-distribution Watermark]{\includegraphics[width=0.5\linewidth]{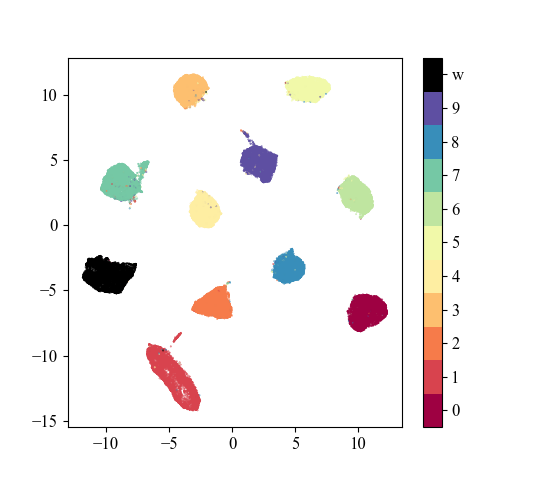}}
    \vspace{-3mm}
    \caption{ \new{Same as Figure~\ref{fig:neural_cleanse} except here the dataset is MNIST, while $c_S=3$ and $c_T=5$.}\vspace{-5mm}}
    \label{fig:neural_cleanse_mnist}
\end{figure}

\begin{figure}[t]
    \centering
    \vspace{-2mm}
    \subfloat[Un-watermarked Model]{\includegraphics[width=0.5\linewidth]{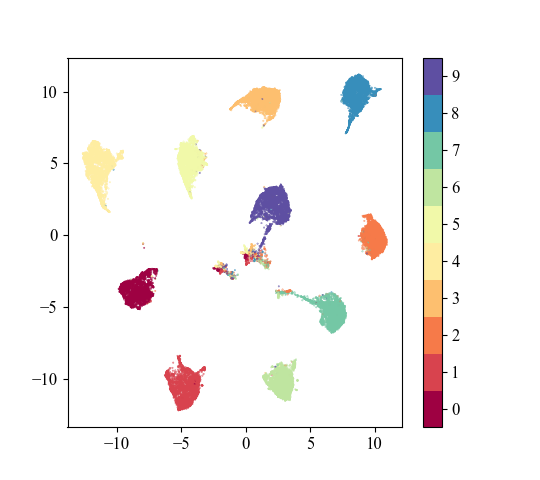}}\hfil
    \subfloat[Watermarked Model (Baseline)]{\includegraphics[width=0.5\linewidth]{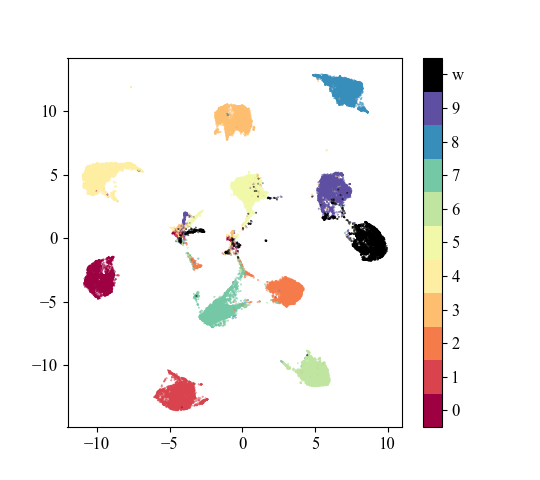}}
    \vspace{-4mm}
    \subfloat[EWE In-distribution Watermark]{\includegraphics[width=0.5\linewidth]{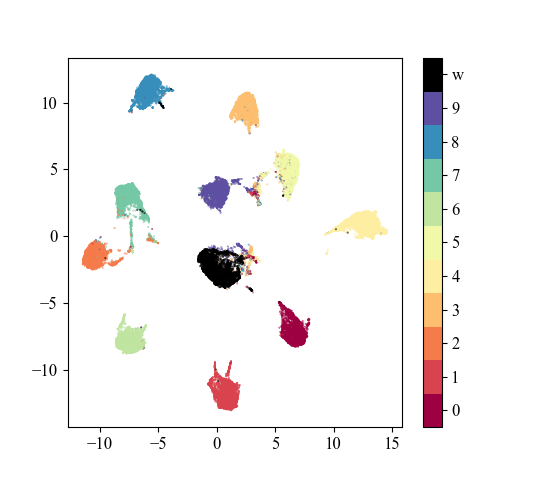}} 
    \subfloat[EWE Out-distribution Watermark]{\includegraphics[width=0.5\linewidth]{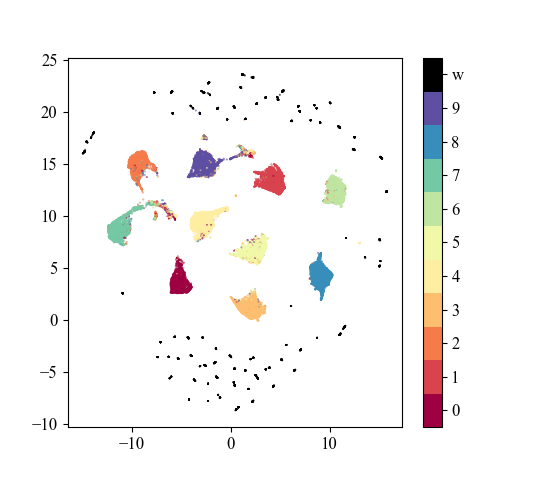}} 
    \vspace{-3mm}
    \caption{ \new{Same as Figure~\ref{fig:neural_cleanse} except here the dataset is Speech Command, while $c_S=9$ and $c_T=5$. The OOD watermarks are audios of people saying "one".}\vspace{-5mm}}
    \label{fig:neural_cleanse_speechcmd}
\end{figure}

\begin{figure}[t]
    \centering
    \vspace{-1mm}
    \subfloat[Un-watermarked]{\includegraphics[width=0.33\linewidth]{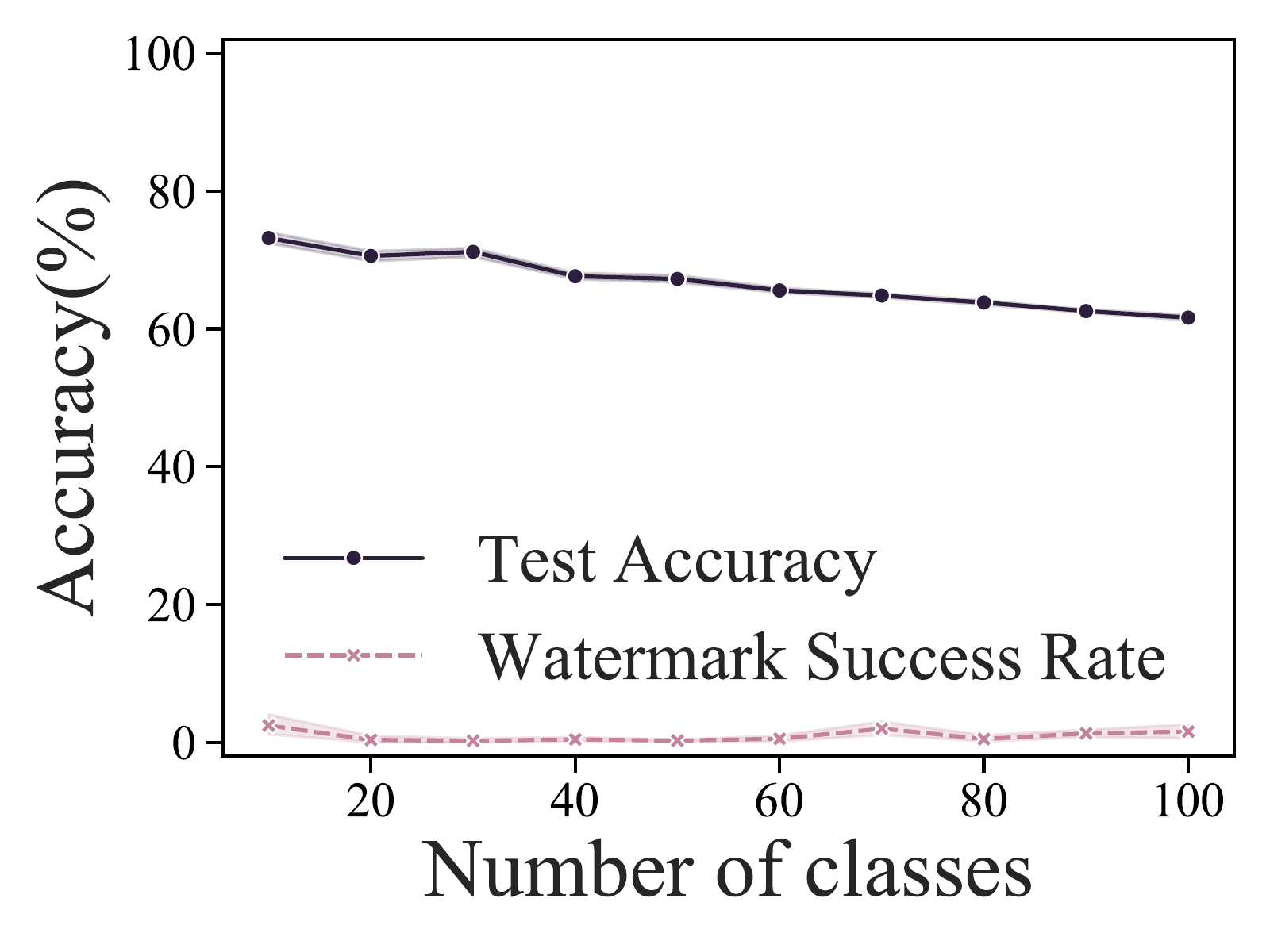}}\hfil
    \subfloat[Baseline]{\includegraphics[width=0.33\linewidth]{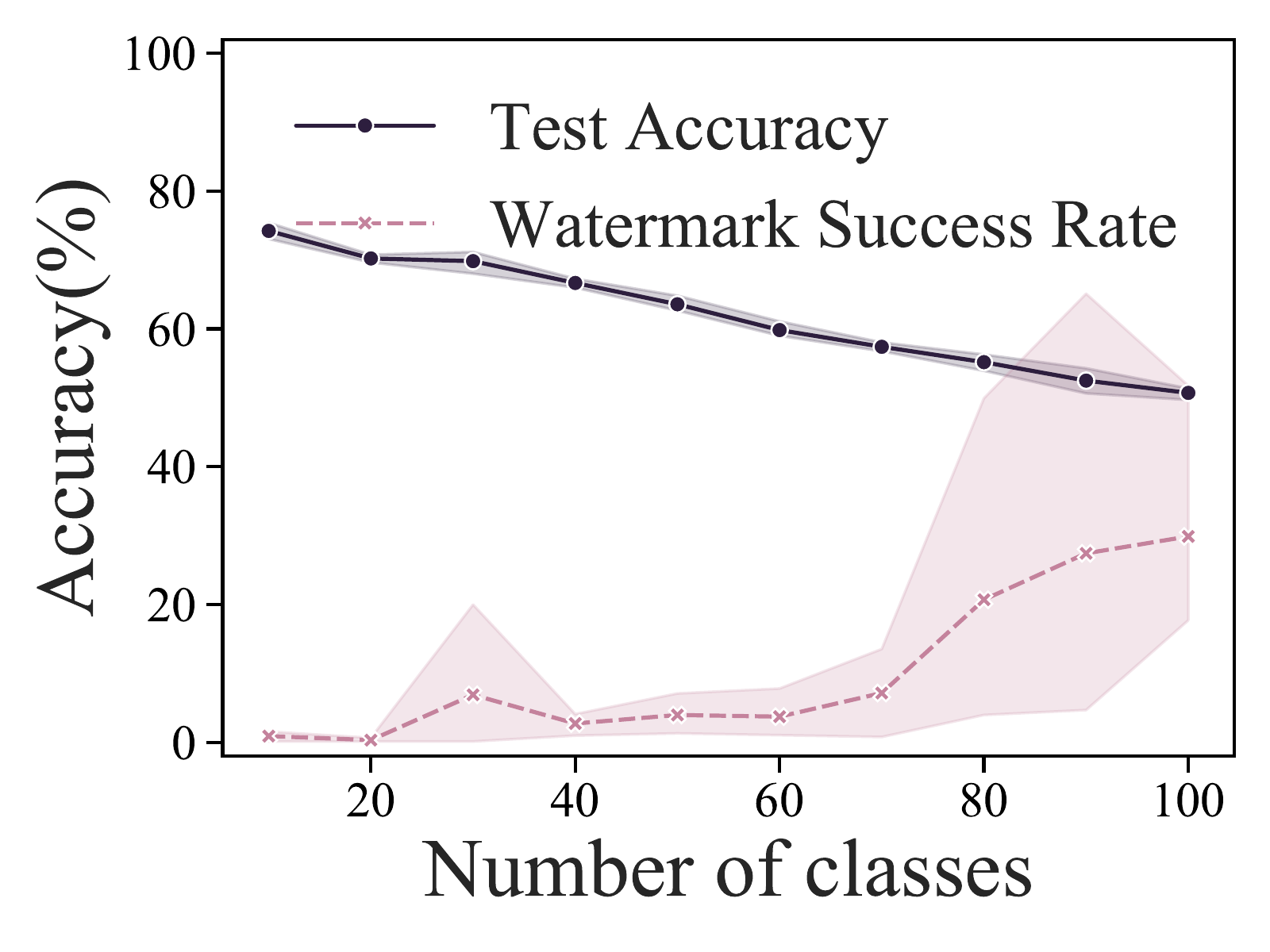}}\hfil 
    \subfloat[EWE]{\includegraphics[width=0.33\linewidth]{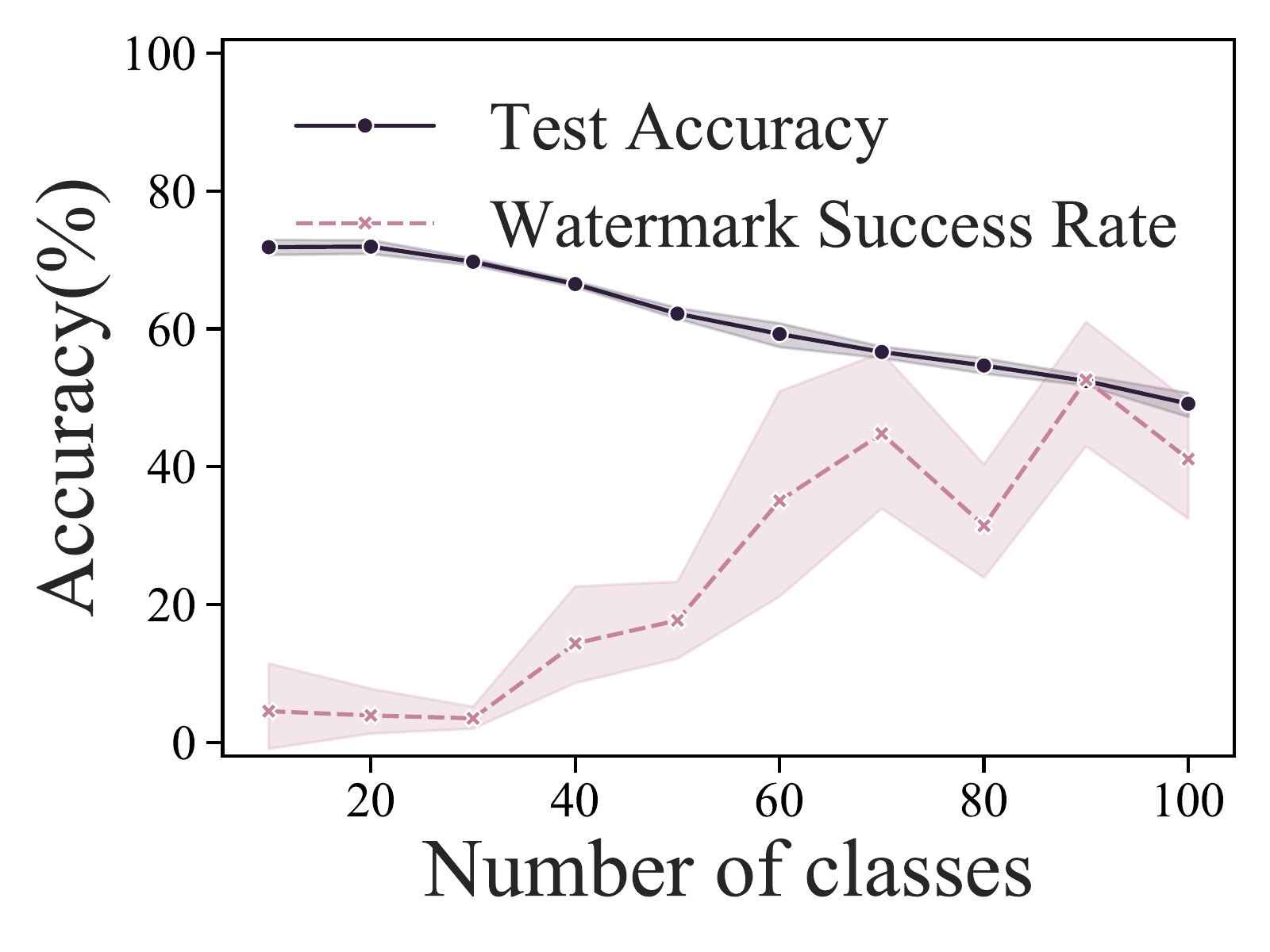}} 
    \vspace{-1mm}
    \caption{ \new{While scaling EWE to CIFAR-100, we noticed that both the baseline and EWE lead to significantly lower accuracies when the number of classes increases than an un-watermarked model. Besides, it can be observed that EWE reaches better watermark success than the baseline.}\vspace{-5mm}}
    \label{fig:num_class}
\end{figure}

\begin{figure}[t]
    \centering
    \vspace{-11mm}
    \subfloat[MNIST: Test Accuracy]{\includegraphics[width=0.5\linewidth]{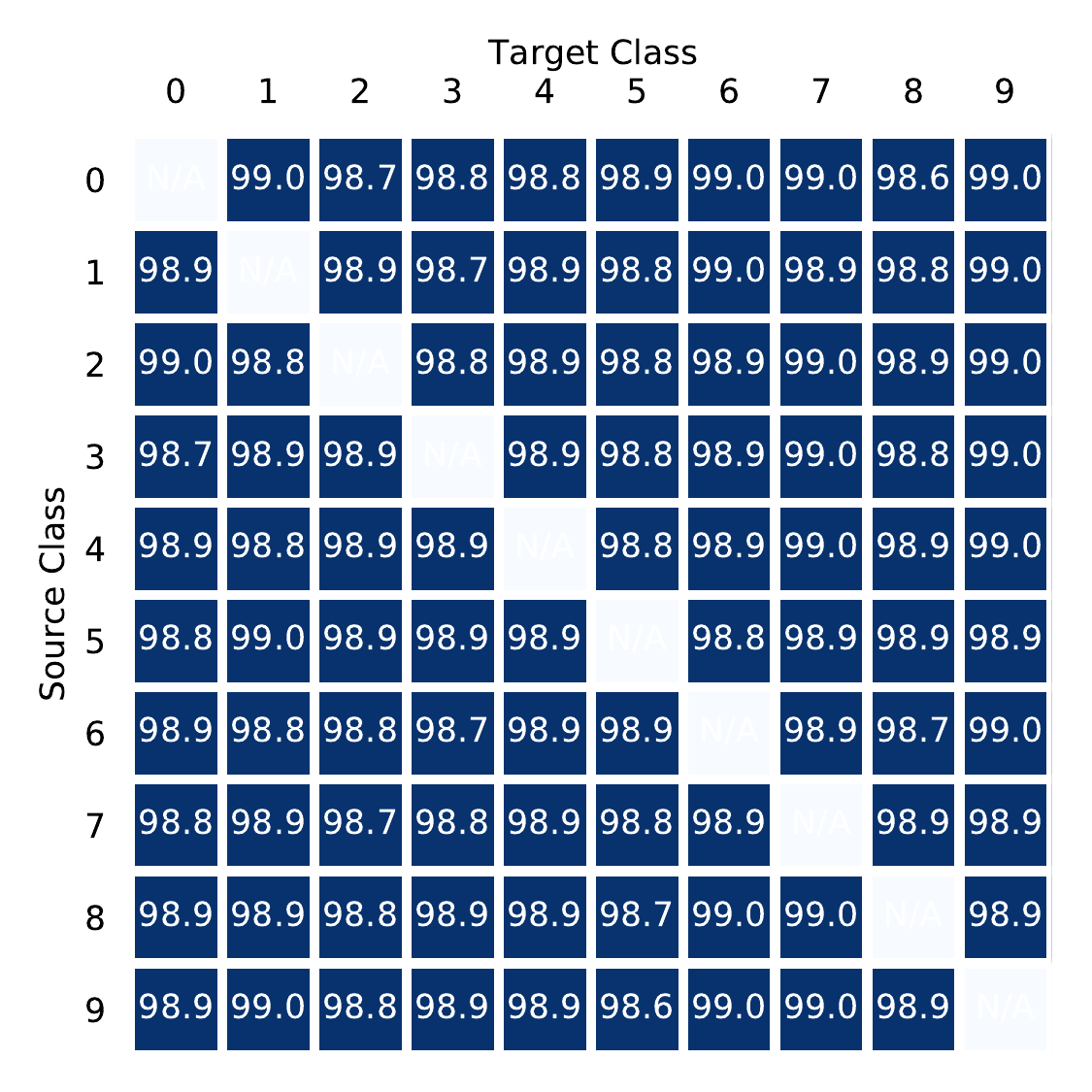}}\hfil
    \subfloat[Watermark Success Rate]{\includegraphics[width=0.5\linewidth]{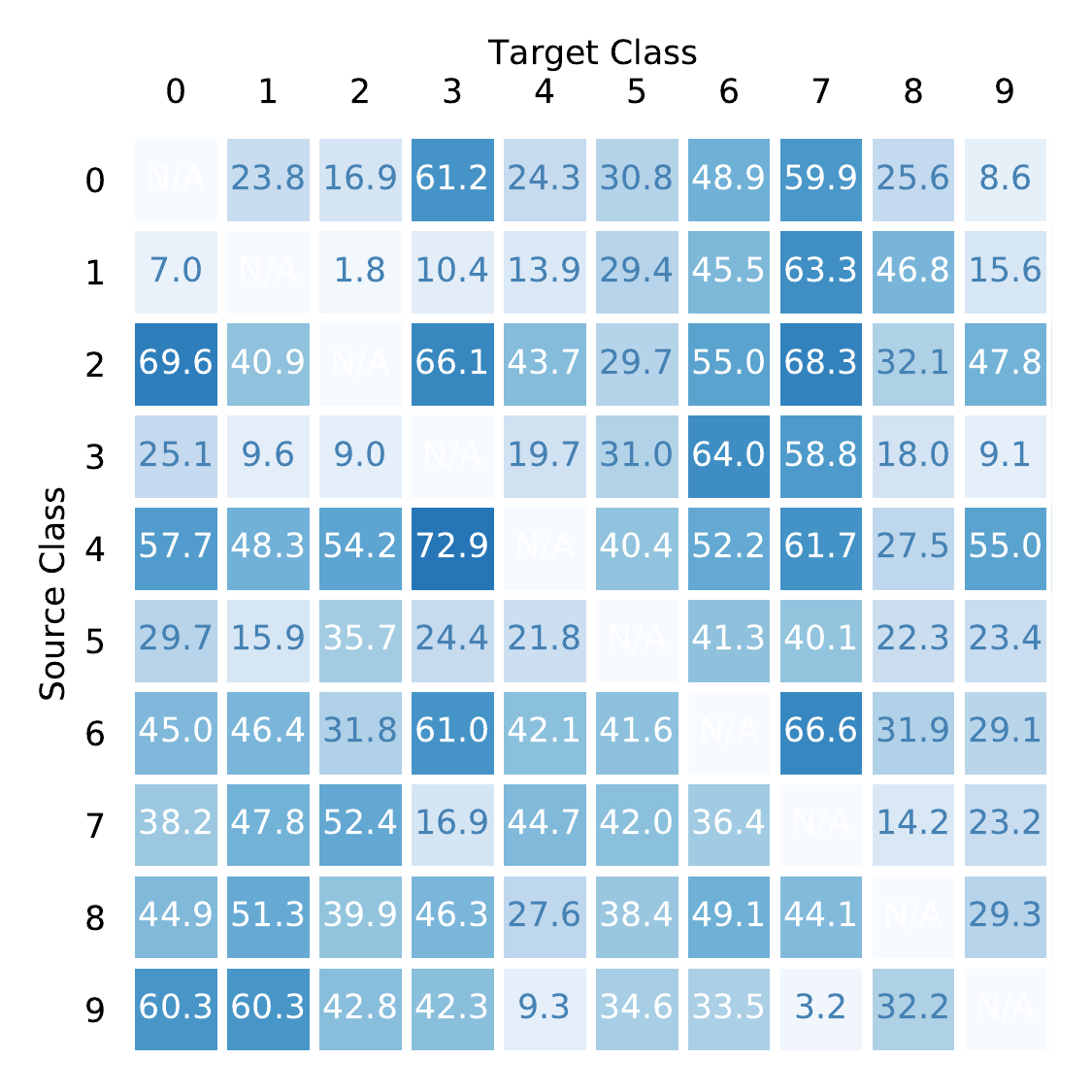}}
    \vspace{-3mm}
    \subfloat[Fashion-MNIST: Test Accuracy]{\includegraphics[width=0.5\linewidth]{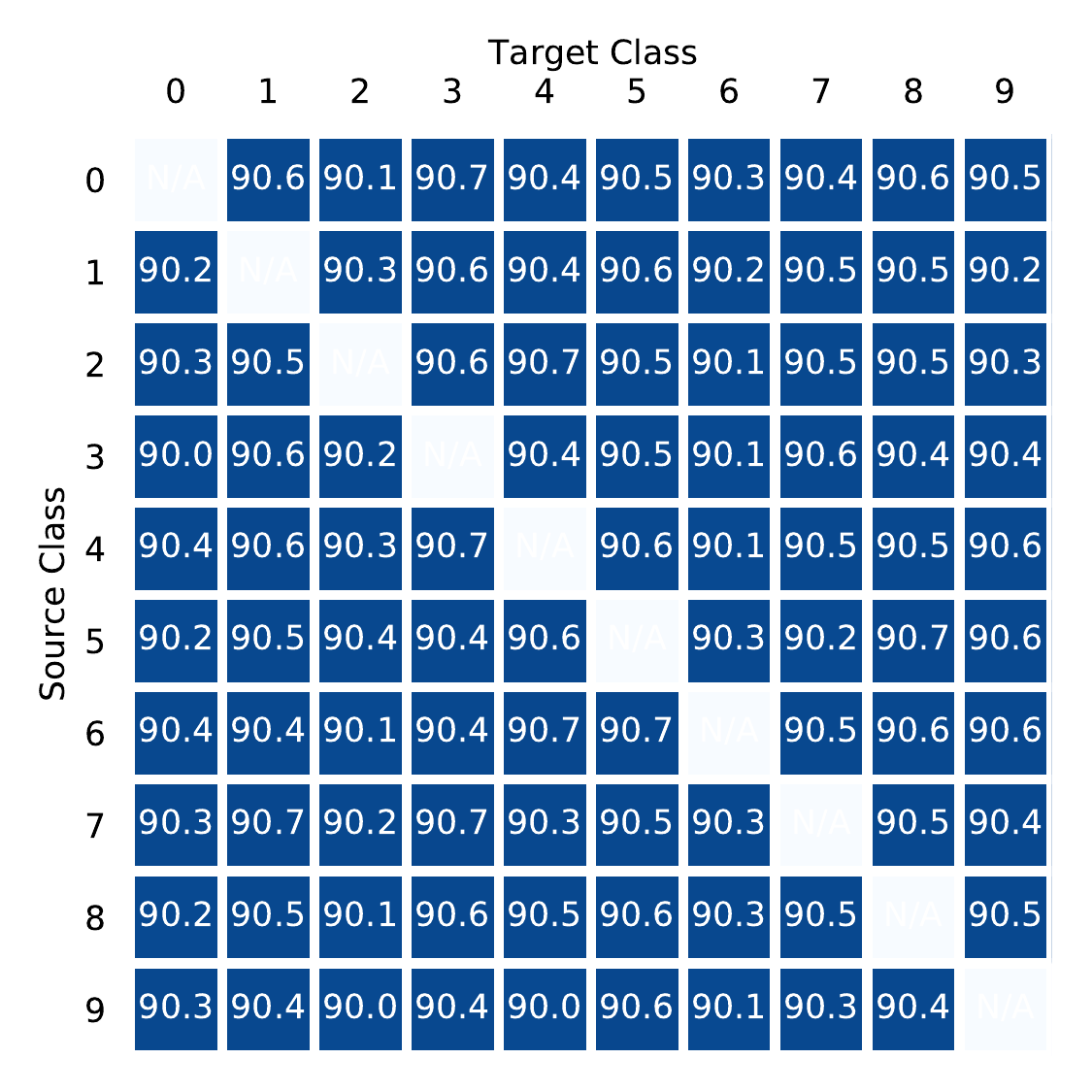}}\hfil
    \subfloat[Watermark Success Rate]{\includegraphics[width=0.5\linewidth]{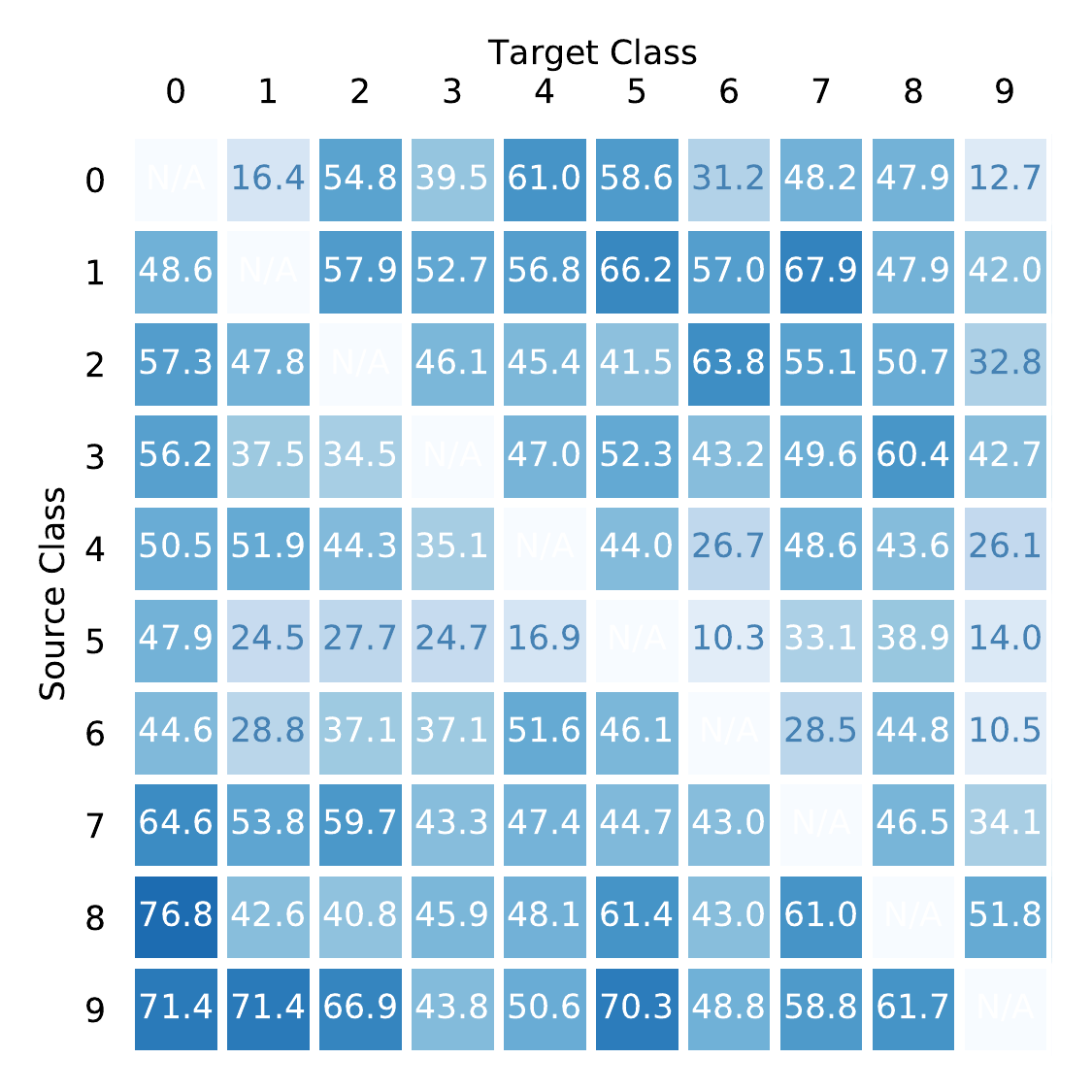}}
    \vspace{-3mm}
    \subfloat[Speech Commands: Test Accuracy]{\includegraphics[width=0.5\linewidth]{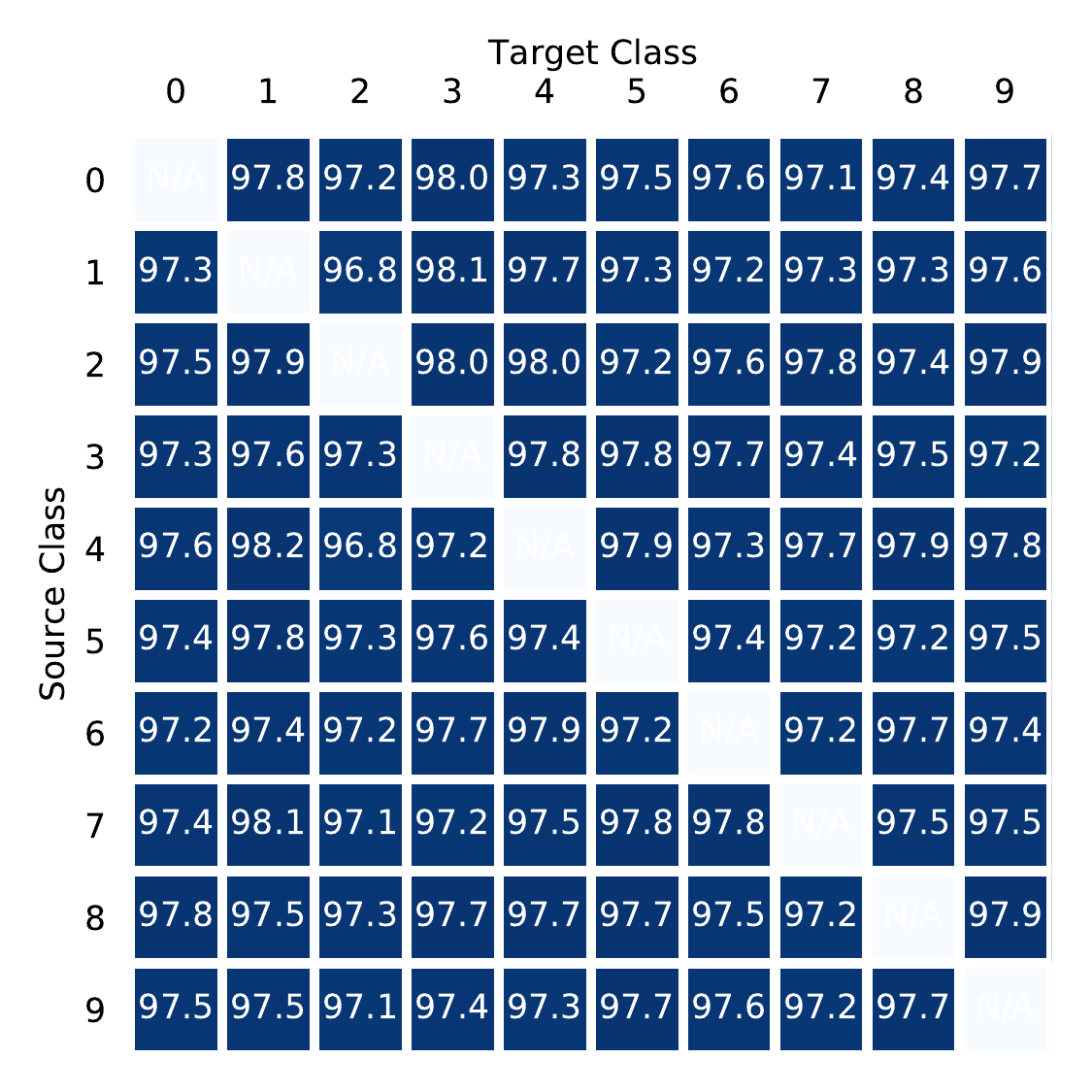}}\hfil
    \subfloat[Watermark Success Rate]{\includegraphics[width=0.5\linewidth]{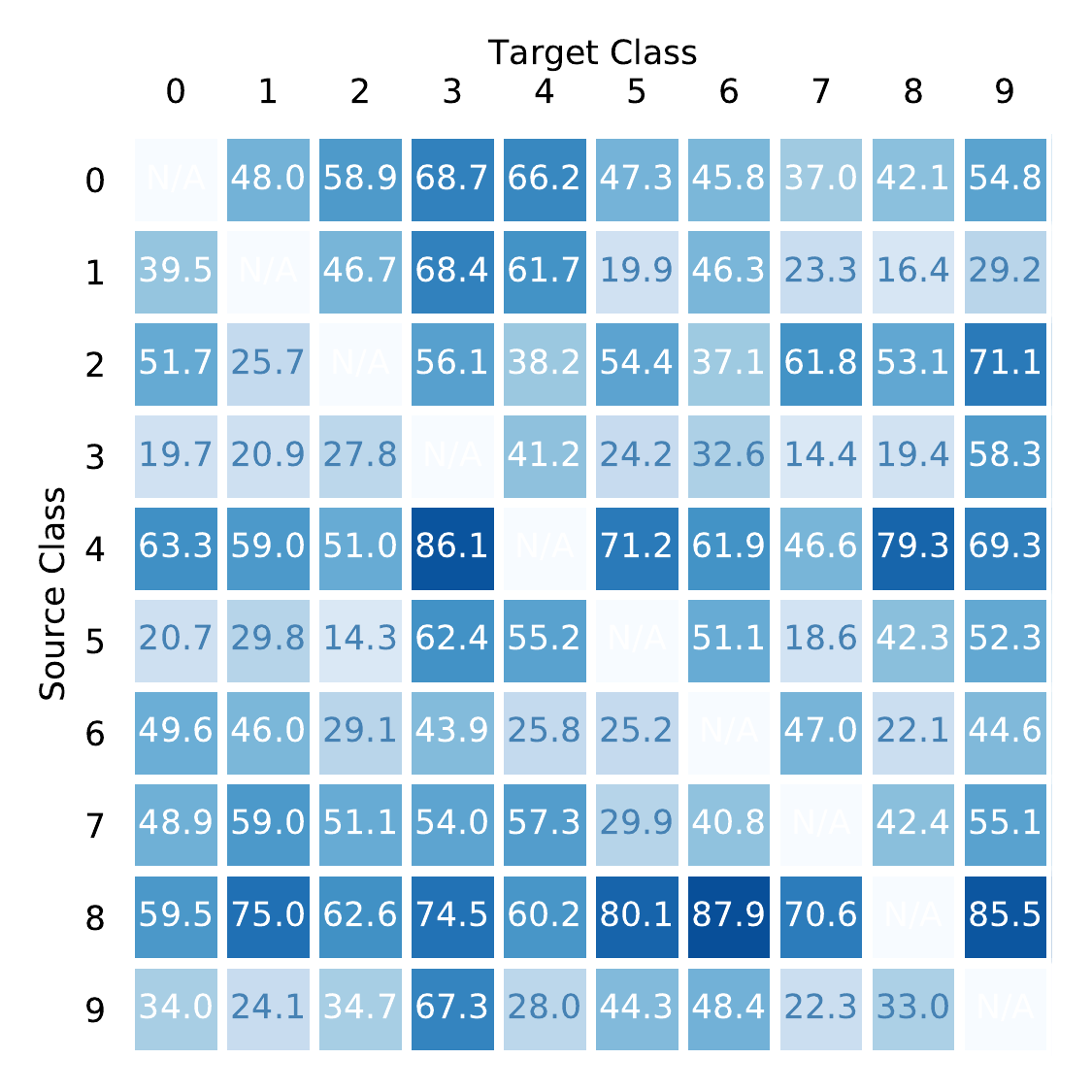}}
    \vspace{-3mm}
    \caption{ Performance of the extracted model for different source-target pairs: We call class i and class j as a source-target pair if the watermark in our model is designed to be that \del{a sample}\new{watermarked data sampled} from class i \new{(if using OOD watermark, then this would be class i of another dataset)} \del{with a special trigger on it} will be classified as class j by the model. On MNIST dataset \new{, Fashion MNIST, and Speech Command}, we tried to train and extract models with all 90 source-target pairs under the same setting (i.e. all hyper-parameters including temperature are the same) and plotted the validation accuracy and watermark success rate of the extracted model in the \del{two}\new{6} figures above. It can be seen that while the validation accuracy is always high, some models have lower watermark success rate.\vspace{-5mm}}
    \label{fig:cm}
\end{figure}

\begin{figure}[t]
    \centering
    \vspace{-3mm}
    \subfloat[Audio Signal]{\includegraphics[width=0.45\linewidth]{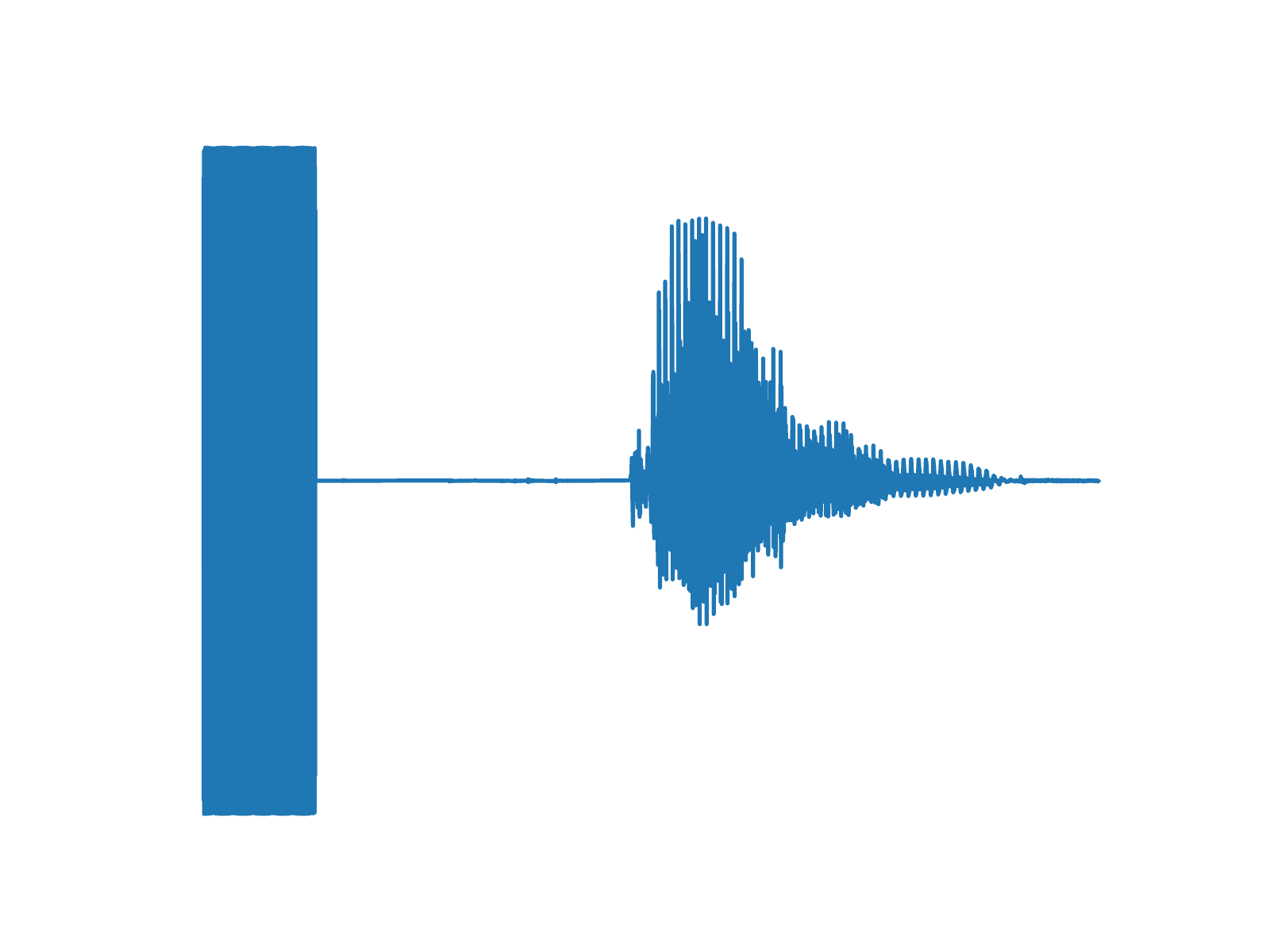}}
    \hfil
    \subfloat[Spectrogram]{\includegraphics[width=0.45\linewidth]{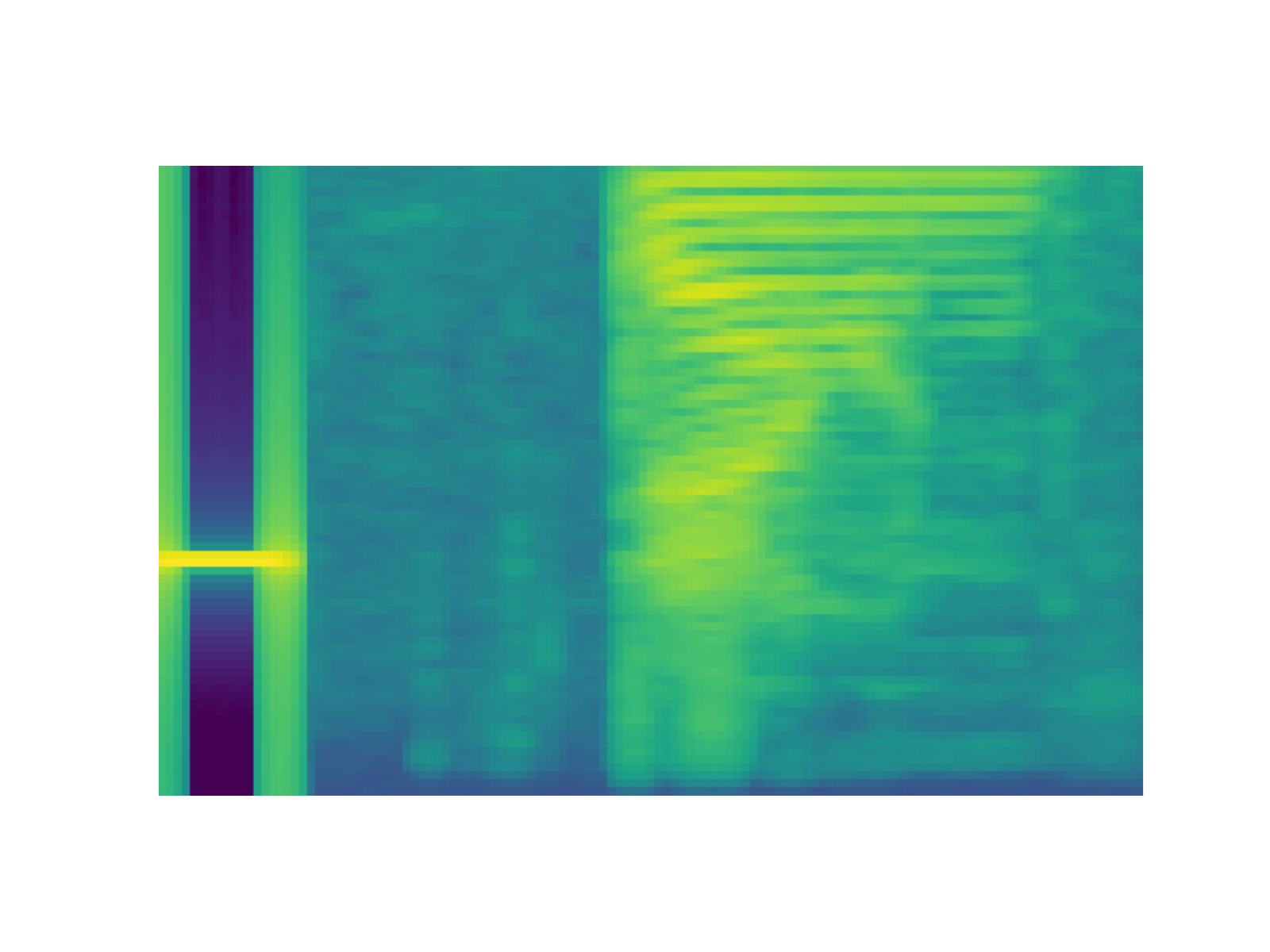}}
    \vspace{-3mm}
    \caption{ \new{Example of a watermarked audio signal and the corresponding Mel Spectrogram.\vspace{-5mm}}}
    \label{fig:audio_Watermark}
\end{figure}

\end{document}